\documentclass[iop]{emulateapj}

\usepackage{url}
\usepackage{hyperref}

\usepackage{graphicx}        
\usepackage{bm}              
\usepackage{natbib}
\usepackage{color}
\usepackage{amsmath}

\usepackage[mediumspace,Gray,squaren]{SIunits}

\shorttitle{ACT: SZ Selected Galaxy Clusters}
\shortauthors{Hasselfield, Hilton, Marriage et al.}

\newcommand{\arone}{148\,GHz}
\newcommand{\artwo}{218\,GHz}

\newcommand{\commentx}[1]{}

\renewcommand{\vec}[1]{\mbox{\boldmath$#1$}} 

\newcommand{\ra}[3]   
   {\makebox[1.5em][r]{#1}\makebox[1.5em][r]{#2} \makebox[2em][r]{#3}}

\newcommand{\hms}[3]  
   {${#1}^{\mathrm{h}}{#2}^{\mathrm{m}}{#3}^{\mathrm{s}}$}

\newcommand{\hmin}[2]  
   {\ensuremath{{#1}^{\mathrm{h}}{#2}^{\mathrm{m}}}}
   
\newcommand{\hours}[1]  
   {\ensuremath{{#1}^{\mathrm{h}}}}

\newcommand{\dms}[3]  
   {\ensuremath{{#1}\degree{#2}\arcminute{#3}\arcsecond}}

\newcommand{\dm}[2]  
   {\ensuremath{{#1}\degree{#2}\arcminute}}

\newcommand{\ukcmb}  
           {\ensuremath{\micro \kelvin_\mathrm{cmb}}}

\newcommand{\uk}  
           {\ensuremath{\micro \kelvin}}

\newcommand{\fdeg} 
           {\hbox{$.\!\!^{\circ}$}}

\usepackage{color}

\hypersetup{ colorlinks,
   linkcolor=blue,
   filecolor=blue,
   urlcolor=blue,
   citecolor=blue}

\newcommand\mytitle{}
\newcommand\mycaption{}

\begin{document}

\title{The Atacama Cosmology Telescope: Sunyaev-Zel'dovich Selected Galaxy
Clusters at \arone{} from Three Seasons of Data}

\author{
Matthew~Hasselfield\altaffilmark{1,2},
Matt~Hilton\altaffilmark{3,4},
Tobias~A.~Marriage\altaffilmark{5},
Graeme~E.~Addison\altaffilmark{2,6},
L.~Felipe~Barrientos\altaffilmark{7},
Nicholas~Battaglia\altaffilmark{8,9},
Elia~S.~Battistelli\altaffilmark{10,2},
J.~Richard~Bond\altaffilmark{9},
Devin~Crichton\altaffilmark{5},
Sudeep~Das\altaffilmark{11,12},
Mark~J.~Devlin\altaffilmark{13},
Simon~R.~Dicker\altaffilmark{13},
Joanna~Dunkley\altaffilmark{6},
Rolando~D\"{u}nner\altaffilmark{7},
Joseph~W.~Fowler\altaffilmark{14,15},
Megan~B.~Gralla\altaffilmark{5},
Amir~Hajian\altaffilmark{9},
Mark~Halpern\altaffilmark{2},
Adam~D.~Hincks\altaffilmark{9},
Ren\'ee~Hlozek\altaffilmark{1},
John~P.~Hughes\altaffilmark{16},
Leopoldo~Infante\altaffilmark{7},
Kent~D.~Irwin\altaffilmark{14},
Arthur~Kosowsky\altaffilmark{17},
Danica~Marsden\altaffilmark{18,13},
Felipe~Menanteau\altaffilmark{16},
Kavilan~Moodley\altaffilmark{3},
Michael~D.~Niemack\altaffilmark{14,15,19},
Michael~R.~Nolta\altaffilmark{9},
Lyman~A.~Page\altaffilmark{15},
Bruce~Partridge\altaffilmark{20},
Erik~D.~Reese\altaffilmark{13},
Benjamin~L.~Schmitt\altaffilmark{13},
Neelima~Sehgal\altaffilmark{21},
Blake~D.~Sherwin\altaffilmark{15},
Jon~Sievers\altaffilmark{1,9},
Crist\'obal~Sif\'on\altaffilmark{22},
David~N.~Spergel\altaffilmark{1},
Suzanne~T.~Staggs\altaffilmark{15},
Daniel~S.~Swetz\altaffilmark{14,13},
Eric~R.~Switzer\altaffilmark{9},
Robert~Thornton\altaffilmark{13,23},
Hy~Trac\altaffilmark{8},
Edward~J.~Wollack\altaffilmark{24}
}
\altaffiltext{1}{Department of Astrophysical Sciences, Peyton Hall, 
Princeton University, Princeton, NJ 08544, USA}
\altaffiltext{2}{Department of Physics and Astronomy, University of
British Columbia, Vancouver, BC, V6T 1Z4, Canada}
\altaffiltext{3}{Astrophysics and Cosmology Research Unit, School of
Mathematics, Statistics \& Computer Science, University of KwaZulu-Natal,
Durban, 4041, South Africa}
\altaffiltext{4}{Centre for Astronomy \& Particle Theory, School of Physics
\& Astronomy, University of Nottingham, Nottingham, NG7 2RD, UK}
\altaffiltext{5}{Dept. of Physics and Astronomy, The Johns Hopkins University, 3400 N. Charles St., Baltimore, MD 21218-2686, USA}
\altaffiltext{6}{Department of Astrophysics, Oxford University, Oxford, OX1
3RH, UK}
\altaffiltext{7}{Departamento de Astronom{\'{i}}a y Astrof{\'{i}}sica, 
Facultad de F{\'{i}}sica, Pontific\'{i}a Universidad Cat\'{o}lica,
Casilla 306, Santiago 22, Chile}
\altaffiltext{8}{Department of Physics, Carnegie Mellon University, Pittsburgh, PA 15213, USA}
\altaffiltext{9}{Canadian Institute for Theoretical Astrophysics, University of
Toronto, Toronto, ON, M5S 3H8, Canada}
\altaffiltext{10}{Department of Physics, University of Rome ``La Sapienza'', 
Piazzale Aldo Moro 5, I-00185 Rome, Italy}
\altaffiltext{11}{High Energy Physics Division, Argonne National Laboratory,
9700 S Cass Avenue, Lemont, IL 60439, USA}
\altaffiltext{12}{Berkeley Center for Cosmological Physics, LBL and
Department of Physics, University of California, Berkeley, CA 94720, USA}
\altaffiltext{13}{Department of Physics and Astronomy, University of
Pennsylvania, 209 South 33rd Street, Philadelphia, PA 19104, USA}
\altaffiltext{14}{NIST Quantum Devices Group, 325
Broadway Mailcode 817.03, Boulder, CO 80305, USA}
\altaffiltext{15}{Joseph Henry Laboratories of Physics, Jadwin Hall,
Princeton University, Princeton, NJ 08544, USA}
\altaffiltext{16}{Department of Physics and Astronomy, Rutgers, 
The State University of New Jersey, Piscataway, NJ 08854-8019, USA}
\altaffiltext{17}{Department of Physics and Astronomy, University of Pittsburgh, 
Pittsburgh, PA 15260, USA}
\altaffiltext{18}{Department of Physics, University of California Santa Barbara,
CA 93106, USA}
\altaffiltext{19}{Department of Physics, Cornell University, Ithaca, NY 14853, USA}
\altaffiltext{20}{Department of Physics and Astronomy, Haverford College,
Haverford, PA 19041, USA}
\altaffiltext{21}{Department of Physics and Astronomy, Stony Brook, NY 11794-3800, USA}
\altaffiltext{22}{Leiden Observatory, Leiden University, PO Box 9513, NL-2300 RA Leiden, Netherlands}
\altaffiltext{23}{Department of Physics, West Chester University 
of Pennsylvania, West Chester, PA 19383, USA}
\altaffiltext{24}{NASA/Goddard Space Flight Center,
Greenbelt, MD 20771, USA}


\newcommand\sigmaeUpp{0.786 \pm 0.013}
\newcommand\omegamUpp{0.250 \pm 0.012}
\newcommand\sigmaeBxii{0.824 \pm 0.014}
\newcommand\omegamBxii{0.285 \pm 0.014}
\newcommand\sigmaeDyn{0.829 \pm 0.024}
\newcommand\omegamDyn{0.292 \pm 0.025}

\begin{abstract}
We present a catalog of 68 galaxy clusters, of which 19 are new discoveries,
detected via the Sunyaev-Zel'dovich effect (SZ) at 148~GHz in the Atacama
Cosmology Telescope (ACT) survey on the celestial equator. With this addition,
the ACT collaboration has reported a total of 91 optically confirmed, SZ
detected clusters.  The 504 square degree survey region includes 270 square
degrees of overlap with SDSS Stripe 82, permitting the confirmation of SZ
cluster candidates in deep archival optical data.  The subsample of 48
clusters within Stripe 82 is estimated to be 90\% complete for $M_{500c} > 4.5
\times 10^{14} \textrm{M}_\odot$ and redshifts $0.15 < z < 0.8$.  While a full
suite of matched filters is used to detect the clusters, the sample is studied
further through a ``Profile Based Amplitude Analysis'' using a statistic
derived from a single filter at a fixed $\theta_{500}=5\farcm{9}$ angular
scale.
This new approach incorporates the cluster redshift along with prior
information on the cluster pressure profile to fix the relationship between
the cluster characteristic size ($R_{500}$) and the integrated Compton
parameter ($Y_{500}$).  We adopt a one-parameter family of ``Universal
Pressure Profiles'' (UPP) with associated scaling laws, derived from X-ray
measurements of nearby clusters, as a baseline model.
Three additional models of cluster physics are used to
investigate a range of scaling relations beyond the UPP prescription.
Assuming a concordance cosmology, the UPP scalings are found to be nearly
identical to an adiabatic model, while a model incorporating non-thermal
pressure better matches dynamical mass measurements and masses from the South
Pole Telescope. A high signal to noise ratio subsample of 15 ACT
clusters with complete optical follow-up is used to obtain cosmological
constraints.  We demonstrate, using 
fixed scaling relations, how the constraints depend on the assumed gas model
if only SZ measurements are used, and show that constraints from SZ data are
limited by uncertainty in the scaling relation parameters rather than sample
size or measurement uncertainty.  We next add in seven clusters from the ACT
Southern survey, including their dynamical mass measurements, which are based
on galaxy velocity dispersions and thus are independent of the gas physics.
In combination with WMAP7 these data simultaneously constrain the scaling
relation and cosmological parameters, yielding 68\% confidence ranges
described by $\sigma_8 = \sigmaeDyn$ and $\Omega_m = \omegamDyn$.  We consider
these results in the context of constraints from CMB and other cluster
studies.  The constraints arise mainly due to the inclusion of the dynamical
mass information and do not require strong priors on the SZ scaling relation
parameters.  The results include marginalization over a 15\% bias in dynamical
masses relative to the true halo mass.  In an extension to $\Lambda$CDM that
incorporates non-zero neutrino mass density, we combine our data with WMAP7,
Baryon Acoustic Oscillation data, and Hubble constant measurements to
constrain the sum of the neutrino mass species to be $\sum_\nu m_\nu <
0.29$~eV (95\% confidence limit).
\end{abstract}



\newcommand\Mpivot{M_{\rm pivot}}
\newcommand\hinv{h_{70}^{-1}}
\newcommand\Msun{{\rm M}_\odot}
\newcommand\Planck{\textit{Planck}}
\newcommand\Tcmb{T_\text{CMB}}

\section{Introduction}
\setcounter{footnote}{0} 

Galaxy clusters are sensitive tracers of the growth of structure in the
Universe.  The measurement of their evolving abundance with redshift has the
potential to provide constraints on cosmological parameters that are
complementary to other measurements, such as the angular power spectrum of the
cosmic microwave background \citep[CMB; e.g.,][]{hinshaw/etal:prep,
  dunkley/etal:2011, keisler/etal:prep, story/etal:prep}, Type Ia supernovae
\citep[e.g.,][]{hicken/etal:2009, lampeitl/etal:2010, suzuki/etal:prep}, or
baryon acoustic oscillations measured in galaxy correlation functions
\citep[e.g.,][]{percival/etal:2010}.

There is a long history of using optical \citep[e.g.,][]{abell/1958,
  lumsden/etal:1992, goto/etal:2002, lopes/etal:2004, miller/etal:2005,
  koester/etal:2007, hao/etal:2010, szabo/etal:2011, wen/etal:2009,
  wen/etal:2012} and X-ray \citep[e.g.,][]{henry/etal:1992,
  bohringer/etal:2004, burenin/etal:2007, mehrtens/etal:2012} surveys to
search for galaxy clusters. Data from such surveys offered an early indication
of an $\Omega_m < 1$ universe \citep[e.g.,][]{bahcall/etal:1992}.  Recent
results have demonstrated the power of modern optical and X-ray surveys for
constraining cosmology \citep[e.g.,][]{vikhlinin/etal:2009, mantz/etal:2010,
  rozo/etal:2010}. A promising method for both detecting clusters in optical
surveys and simultaneously providing mass estimates is to use weak
gravitational lensing shear selection, and the first such samples using this
technique have recently appeared \citep[e.g.,][]{wittmann/etal:2006,
  miyazaki/etal:2007}. Within the last few years, cluster surveys exploiting
the Sunyaev-Zel'dovich effect \citep[SZ;][]{sunyaev/zeldovich:1970} have also
begun to deliver cluster samples \citep[e.g.,][]{staniszewski/etal:2009,
  marriage/etal:2011, williamson/etal:2011, planck/esz:2011,
  reichardt/etal:2013} and constraints on cosmological parameters
\citep{vanderlinde/etal:2010, sehgal/etal:2011, benson/etal:2013,
  reichardt/etal:2013}.

The thermal SZ effect is the inverse Compton scattering of CMB photons by
electrons within the hot ($\sim 10^{7-8}$\,K) intracluster medium of galaxy
clusters. This leads to a spectral distortion in the direction of clusters,
with the size of the effect being proportional to the volume-integrated
thermal pressure and thus, in the adiabatic scenario, the total thermal energy
of the cluster gas.  Accordingly, this is correlated with cluster mass
\citep[e.g.,][]{bonamente/etal:2008, marrone/etal:prep, sifon/etal:inprep,
  planck/intermediateIII:2013}.  Since the SZ signal is not diminished due to
luminosity distance, it is nearly redshift independent; in principle SZ
surveys can detect all clusters in the Universe above a mass limit set by the
survey noise level \citep[e.g.,][]{birkinshaw:1999,
  carlstrom/holder/reese:2002}.

Although current SZ cluster samples are small in comparison to existing X-ray
and optical cluster catalogs, they provide very powerful complementary probes
because they are sensitive to the high mass, high redshift cluster population
\citep[e.g.,][]{brodwin/etal:2010, foley/etal:2011, planck/xxvi:2011,
  menanteau/etal:2012, stalder/etal:2012}. Many studies
\citep[e.g.,][]{hoyle/jimenez/verde:2011, mortonson/hu/huterer:2011,
  hotchkiss/2011, harrison/coles/2012} have noted that the discovery of a
sufficiently massive cluster at high redshift would be a challenge to
$\Lambda$CDM cosmology, and the approximately redshift independent,
mass-limited nature of SZ surveys means that they are well suited to reveal
such objects if they exist.

In this paper we describe the results of a search for galaxy clusters using
the SZ effect in maps of the celestial equator obtained by the Atacama
Cosmology Telescope \citep[ACT;][]{swetz/etal:2011}. ACT is a 6\,m telescope
located in northern Chile that observes the sky in three frequency bands
(centered at 148, 218, and 277\,GHz) simultaneously with arcminute
resolution. During 2008, ACT surveyed a 455 deg$^{2}$ patch of the Southern
sky, centered on $\delta = -55\deg$, detecting a number of SZ cluster
candidates of which 23 were optically confirmed as massive clusters
\citep{menanteau/etal:2010, marriage/etal:2011}.  The Equatorial survey area,
on which we report in this work, was chosen to overlap the deep ($r \approx
23.5$~mag) optical data from the Sloan Digital Sky Survey
\citep[SDSS;][]{abazajian/etal:2009} Stripe 82 region \citep[S82
  hereafter;][]{annis/etal:2011}. Optical confirmation of our SZ cluster
candidates is reported in \citet{menanteau/etal:2013}, using data from SDSS
and additional targeted optical and IR observations obtained at Apache Point
Observatory. All clusters have photometric redshifts, and most have
spectroscopic redshifts from a combination of SDSS and new observations at
Gemini South. X-ray fluxes from the ROSAT All Sky Survey confirm that this is
a massive cluster sample. The overlap of the ACT survey with SDSS has also
enabled stacking analyses which characterize the SZ-signal as a function of
halo mass from optically selected samples \citep{hand/etal:2011,
  sehgal/etal:2013}, as well as a first detection of the kinetic SZ effect
from the correlation of positions and redshifts of luminous red galaxies with
temperature in the ACT maps \citep{hand/etal:prep}.

The structure of this paper is as follows. In Section~\ref{sec:maps} we
describe the processing of the ACT data used in this work and the cluster
detection algorithm.  In Section~\ref{sec:recovery} we describe our approach
to relating the cluster signal in filtered SZ maps to cluster mass, and obtain
mass estimates for our cluster sample.  In Section~\ref{sec:catalog} we
compare our catalog to other SZ, optical, and X-ray selected cluster catalogs.
In Section~\ref{sec:cosmology} we obtain constraints on cosmological
parameters using the ACT cluster sample.  In an Appendix, we present analogous
SZ signal measurements and mass estimates for ACT's Southern field clusters,
using deeper data obtained over the course of the 2009-2010 observing seasons.

Where it is necessary to adopt a fiducial cosmology, we assume $\Omega_{\rm
  m}=0.3$, $\Omega_\Lambda=0.7$, and $H_0=70\,h_{70}$~km~s$^{-1}$~Mpc$^{-1}$
($h_{70} = 1$), unless stated otherwise. Throughout this paper, cluster mass
is measured within a characteristic radius with respect to the critical
density such that, e.g., $M_{500c}$ is defined as the mass measured within the
radius ($R_{500}$) at which the enclosed mean density is 500 times the
critical density at the cluster redshift. The function $E(z)$ denotes the
evolution of the Hubble parameter with redshift (i.e., $E(z) = [\Omega_{\rm
    m}(1+z)^3 + \Omega_{\Lambda}]^{1/2}$ for a universe with $\Omega_{\rm k} =
0$ and negligible radiation density).  Uncertainties and error bars are
specified at the 1-$\sigma$ level, and posterior distributions are summarized
in terms of their mean and standard deviation unless otherwise indicated.

\section{Maps and Cluster Detection}
\label{sec:maps}

In this section we discuss the detection of galaxy clusters in the ACT
Equatorial maps at 148\,GHz.  The maps are filtered to enhance structures
whose shape matches the Universal Pressure Profile of \cite{arnaud/etal:2010}.
The final cluster catalog consists of SZ candidates that have been confirmed
in optical or IR imaging.

\subsection{Equatorial Maps}

ACT's observations during the 2009 and 2010 seasons were concentrated on the
celestial equator.  For this study we make use of the 504 square degree deep,
contiguous region spanning from \hms{20}{16}{00} to \hms{3}{52}{24} in right
ascension and from $-2\degree07\arcmin$ to $2\degree18\arcmin$ in declination.
This region includes 270 square degrees of overlap with S82, which extends to
\hmin{20}{39} in R.A. and $\pm 1\degree15\arcmin$ in declination.  As shown in
Figure~\ref{fig:sensitivity}, the S82 region corresponds to the lowest-noise
region of the ACT Equatorial maps.

The bolometer time-stream data are acquired while scanning the telescope in
azimuth at fixed elevation.  Cross-linked data are obtained by observing the
same celestial region at two telescope pointings that produce approximately
orthogonal scan directions.  Because the scan strategy was optimized for
simultaneous observation with the \arone{} and \artwo{} arrays (the centers of
which are separated by approximately $33\arcmin$ when projected onto the sky),
the regions beyond declinations of $\pm 1\degree40\arcmin$ are not well
cross-linked.

The ACT data reduction pipeline and map-making procedure are described in
\citet{dunner/etal:2013}, in the context of ACT's 2008 data.  The time-stream
bolometer data are screened for pathologies and then combined to obtain a
maximum likelihood estimate of the microwave sky map (with $0.5\arcmin$
pixels) for each observing season.  As a result of the cross-linked scan
strategy, and the careful treatment of noise during map making, the ACT maps
are unbiased at angular scales $\ell > 300$.

Due to realignments of the primary and secondary mirrors, the telescope beams
vary slightly between seasons but are stable over the course of each
season. The telescope beams are determined from observations of Saturn, using
the method described in \cite{hincks/etal:2010}, but with additional
corrections to account for $\approx 6\arcsec$ RMS pointing variation between
observations made on different nights (Hasselfield et al., in prep.).  The
effective beam for the \arone{} array differs negligibly between the 2009 and
2010 seasons, with a FWHM of $1.4\arcmin$ and solid angle (including the
effects of pointing variation) of $224 \pm 2$~nsr.

Calibrations of ACT observations are based on frequent detector load curves,
and atmospheric opacity water vapor measurements \citep{dunner/etal:2013}.
Absolute calibration of the ACT maps is achieved by comparing the large
angular scale ($300 < \ell < 1100$) signal from the 2010 season maps to the
WMAP 95~GHz 7-year maps \citep{jarosik/etal:2011}.  Using a cross-correlation
technique as described in \cite{hajian/etal:2011}, an absolute calibration
uncertainty of $\approx 2\%$ in temperature is achieved \citep{das/etal:2013}.
The inter-calibration of 2009 and 2010 is measured through a similar
cross-correlation technique, with less than 2\% error.

\subsection{Gas Pressure Model}
\label{sec:pressure_model}

At several stages in the detection and analysis we will require a template for
the intracluster gas pressure profile.  To this end we adopt the ``Universal
Pressure Profile'' (UPP) of \citet[hereafter A10]{arnaud/etal:2010}, which
includes mass dependence in the profile shape and has been calibrated to X-ray
observations of nearby clusters.  In this section we review the form of the
UPP, and obtain several approximations that will be used in cluster detection
(Section~\ref{sec:detection}) and cluster property recovery
(Section~\ref{sec:recovery}).

In A10, the cluster electron pressure as a function of physical radius $r$ is
modeled with a generalized Navarro-Frenk-White (GNFW) profile
\citep*{nagai/kravtsov/vikhlinin:2007},
\begin{align}
  p(x) & = P_0~(c_{500}x)^{-\gamma}\left( 1+(c_{500}x)^\alpha
  \right)^{(\gamma-\beta)/ \alpha} \label{eqn:gnfw},
\end{align}
where $x = r/R_{500}$ and $P_0, c_{500}, \gamma, \alpha, \beta$ are fit
parameters.  The overall pressure normalization, under assumptions of
self-similarity (i.e., the case when gravity is the sole process responsible
for setting cluster properties), varies with mass and redshift according to
\begin{align}
  P_{500} & = \left[1.65 \times 10^{-3} h_{70}^2~\rm{keV~cm}^{-3} \right]
  m^{2/3} E^{8/3}(z),
\end{align}
where $E(z)$ is the ratio of the Hubble constant at redshift $z$ to its
present value, and
\begin{align}
  m & \equiv M_{500c}/(3\times10^{14}~\hinv\Msun)
\end{align}
is a convenient mass parameter.  Some deviation from strict self-similarity
may be encoded via an additional mass dependence in the shape of the profile,
yielding a form
\begin{align}
  P(r) & = P_{500} m^{\alpha_p(x)} p(x).\label{eqn:upp}
\end{align}
In this framework, A10 use X-ray observations of local ($z < 0.2$) clusters to
obtain best-fit GNFW parameters $[P_0, c_{500}, \gamma, \alpha, \beta] =$
[8.403 $h_{70}^{−3/2}$ , 1.177, 0.3081, 1.0510, 5.4905], and an additional
radial dependence described reasonably well by $\alpha_p(x) = 0.22/(1+8x^3)$.

Because hydrostatic mass estimates are used by A10 to assess the relationship
between cluster mass and the pressure profile, there may be systematic
differences when one makes use of an alternative mass proxy, such as weak
lensing or galaxy velocity dispersion.  Simulations suggest that hydrostatic
masses are under-estimates of the true cluster mass \citep[e.g.,
][]{nagai/kravtsov/vikhlinin:2007}. However, there is little consensus among
recent studies which compare X-ray hydrostatic and weak lensing mass
measurements. For example, \citet{mahdavi/etal:2013} find that hydrostatic
masses are lower than weak lensing masses by about 10\% at $R_{500}$;
\citet{zhang/etal:2010} and \citet{vikhlinin/etal:2009b} find reasonable
agreement; while \citet{planck/intermediateIII:2013} find hydrostatic masses
to be about 20\% larger than weak lensing masses. Therefore in our initial
treatment of the UPP we neglect this bias; later we will address this issue by
adding degrees of freedom to allow for changes in the normalization of the
pressure profile.

The thermal SZ signal is related to the optical depth for Compton scattering
along a given line of sight.  For our pressure profile, and in the absence of
relativistic effects, this Compton parameter at projected angle $\theta$ from
the cluster center is
\begin{align}
  y(\theta) & = \frac{\sigma_T}{m_{\rm e}c^2}\int ds~ P\left(
  \sqrt{s^2 + (R_{500}\theta/\theta_{500})^2}\right),\label{eqn:ytheta}
\end{align}
where $\theta_{500} = R_{500}/D_{\rm A}(z)$ with $D_{\rm A}(z)$ the angular
diameter distance to redshift $z$, $\sigma_T$ is the Thomson cross section,
$m_{\rm e}$ is the electron mass, and the integral in $s$ is along the line of
sight.  Relativistic effects change this picture somewhat, but for convenience
we will use the above definition of $y(\theta)$ and apply the relativistic
correction only when calculating the SZ signal associated with a particular
$y$.

To simplify the expression for the cluster pressure profile, we first consider
the mass parameter $m = 1$ and factor the expression in
equation~(\ref{eqn:ytheta}) to get
\begin{align}
  y(\theta,m=1) & = 10^{A_0} E(z)^2 \tau(\theta/\theta_{500})
  \label{eqn:uppcyl1}
\end{align}
where $\tau(x)$ is a dimensionless profile normalized to $\tau(0) = 1$, and
$10^{A_0} = 4.950 \times 10^{-5} h_{70}^{1/2}$ gives the normalization.  The
deviations from self-similarity are weak enough that we may model the changes
in the profile shape with mass as simple adjustments to the normalization and
angular scale of the profile.  For the masses of interest here ($1<m<10$) we
obtain
\begin{align}
  y(\theta, m) & \approx 10^{A_0} E(z)^2 m^{1+B_0}
  \tau(m^{C_0}\theta/\theta_{500})\label{eqn:uppcyl}
\end{align}
with $B_0 = 0.08$ and $C_0 = -0.025$.  This approximation reproduces the inner
signal shape extremely well, with deviations increasing to the 0.5\% level by
$\theta_{500}$.  For $0.1\theta_{500} < \theta < 3\theta_{500}$, the enclosed
signal ($\int_0^\theta d\theta'~2\pi \theta' y(\theta',m)$) differs by less
than 1\% from the results of the full computation.  This parametrization of
the cluster signal in terms of a normalization and dimensionless profile is
not used for cluster detection (Section~\ref{sec:detection}), but will
motivate the formulation of scaling relations and permit the estimation of
cluster masses (Section~\ref{sec:fixed_scale}).

\newcommand\fsz{f_{\rm SZ}}
\newcommand\frel{f_{\rm rel}}
\newcommand\ytilde{\widetilde{y}_0}

The observed signal due to the SZ effect is a change in radiation intensity,
expressed in units of CMB temperature:
\begin{align}
  \frac{\Delta T(\theta)}{\Tcmb} & = \fsz~ y(\theta).
\end{align}
In the non-relativistic limit, the factor $\fsz$ depends only on the observed
radiation frequency.  Integrating this non-relativistic SZ spectral response
over the nominal \arone{} array band-pass, we obtain an effective frequency of
146.9~GHz \citep{swetz/etal:2011}.  At this frequency, the formulae of
\cite{itoh/kohyama/nozawa:1998} provide a spectral factor, including
relativistic effects for gas temperature $T_e$, of $\fsz(t) = -0.992 \frel(t)$
where $t= k_BT_e/m_ec^2$ and $\frel(t) = 1 + 3.79t - 28.2t^2$.  This results
in a 6\% correction for a cluster with $T = 10$\,keV.  We use the scaling
relation of \cite{arnaud/pointecouteau/pratt:2005}, $t = -0.00848\times(m
E(z))^{-0.585}$, to express the mean temperature dependence in terms of the
cluster mass and redshift.  This yields a final form, $\fsz(m,z) = -0.992
\frel(m,z)$, which we use in all subsequent modeling of the SZ signal.  The
corrections for the ACT cluster sample range from roughly 3\% to 10\%.

\subsection{Galaxy Cluster Detection}
\label{sec:detection}

\newcommand\thetac{{\theta_c} }
\newcommand\thetaD{{\theta_{500}} }
\newcommand\thetaFixed{{5\farcm9}}
\newcommand\xv{\vec{x} } 
\newcommand\kv{\vec{k} } 

\begin{figure*}[ht]
\begin{center}
 \resizebox{\textwidth}{!}{
 \plotone{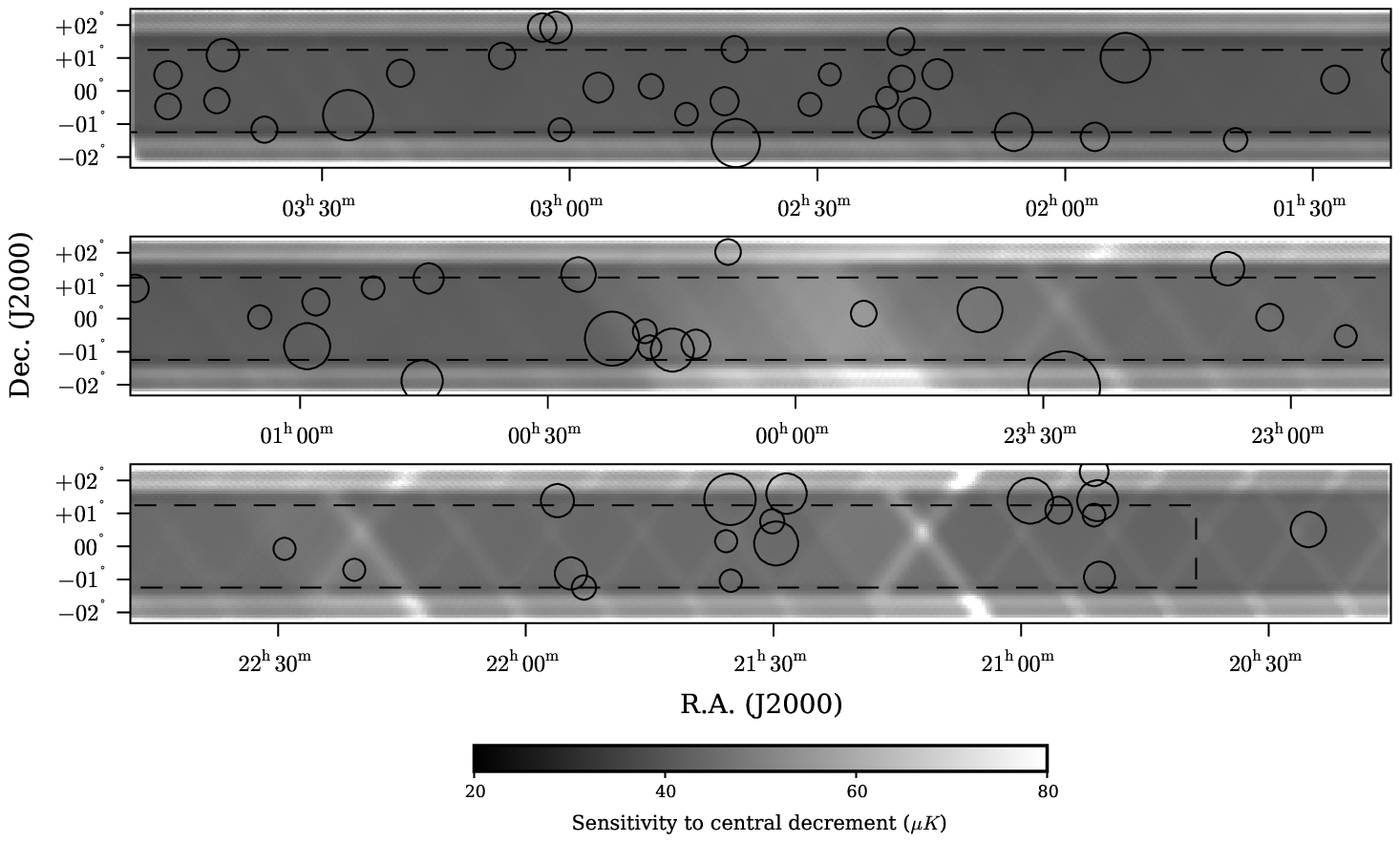}}
\caption{The portion of the ACT Equatorial survey region considered in this
  work.  It spans from \hms{20}{16}{00} to \hms{3}{52}{24} in R.A. and from
  $-2\degree07\arcmin$ to $2\degree18\arcmin$ in declination for a total of
  504 square degrees.  The overlap with Stripe 82 (dashed line) extends only
  to \hmin{20}{39} in R.A. and covers $\pm 1\degree15\arcmin$ in declination,
  for a total of 270 square degrees.  Circles identify the optically confirmed
  SZ-selected galaxy clusters, with radius proportional to the signal to noise
  ratio of the detection (which ranges from 4 to 13).  The gray-scale gives
  the sensitivity (in CMB\,\uk{}) to detection of galaxy clusters, \emph{after
    filtering}, for the matched filter with $\thetaD = \thetaFixed$ (see
  Section~\ref{sec:detection}).  Inside the Stripe 82 region the median noise
  level is 44~\uk{}, with one quarter of pixels having noise less
  (respectively, more) than 41~\uk{} (46 \uk{}).  Outside Stripe 82, the
  median level is 54 \uk{}, with one quarter of pixels having less (more) than
  47 (64) \uk{} noise.  The higher noise, \emph{X}-shaped regions are due to
  breaks in the scan for calibration operations.}
\label{fig:sensitivity}
\end{center}
\end{figure*}

In addition to the temperature decrements due to galaxy clusters, the ACT maps
at \arone{} contain contributions from the CMB, radio point sources, dusty
galaxies, and noise from atmospheric fluctuations and the detectors.  To
detect galaxy clusters in the ACT maps we make use of a set of matched
filters, with signal templates based on the UPP through the integrated profile
template $\tau(\theta/\theta_{500})$.

We consider signal templates $S_\thetaD(\theta) \equiv \tau(\theta/\thetaD)$
for $\thetaD = 1.\!\arcmin18$ to $27\arcmin$ in increments of $1.\!\arcmin18$.
Each fixed angular scale corresponds to a physical scale that varies with
redshift, but can be computed for a given cosmology.  For each signal template
we form an associated matched filter in Fourier space
\begin{align}
  \Psi_\thetaD(\kv) & = \frac{1}{\Sigma_\thetaD}
  \frac{B(\kv)S_\thetaD(k)}{N(\kv)}
\end{align}
where $B(\kv)$ is the product of the telescope beam response with the map
pixel window function, $N(\kv)$ is the (anisotropic) noise power spectrum of
the map, and $\Sigma_\thetaD$ is a normalization factor chosen so that, when
applied to a map containing a beam-convolved cluster signal $-\Delta T
[S_\thetaD \ast B](\theta)$ (in temperature units), the matched filter returns
the central decrement $-\Delta T$.

Since the total power from the galaxy cluster SZ signal is low compared to the
CMB, atmospheric noise, and white noise that contaminate the cluster signal,
we estimate the noise spectrum $N(\kv)$ from the map directly.  Bright (signal
to noise ratio greater than five) point sources are masked from the map, with
the masking radius ranging from $2\arcmin$ for the dimmest sources to
$1^\circ$ for the brightest source.  A plane is fit to the map signal
(weighting by the inverse number of samples in each pixel) and removed, and
the map is apodized within $0.2\degree$ of the map edges.

For ACT, the effect of the noise term in the matched filter is to strongly
suppress signal below $\ell \approx 3000$ (corresponding to scales larger than
7\arcmin).  When combined with the signal template (and beam), the filters
form band-passes centered at $\ell$ ranging from roughly 2500 to 5000.  The
angular scales probed by the filters are thus sufficiently small that
filtering artifacts near the map boundaries are mitigated by the map
apodization.  While the suppression of large angular scales disfavors the
detection of clusters with large angular sizes, we apply the full suite of
filters in order to maximize detection probability and to study the features
of inferred cluster properties as the assumed cluster scale is varied.

\begin{figure}
\begin{center}
\resizebox{3.5in}{!}{\plotone{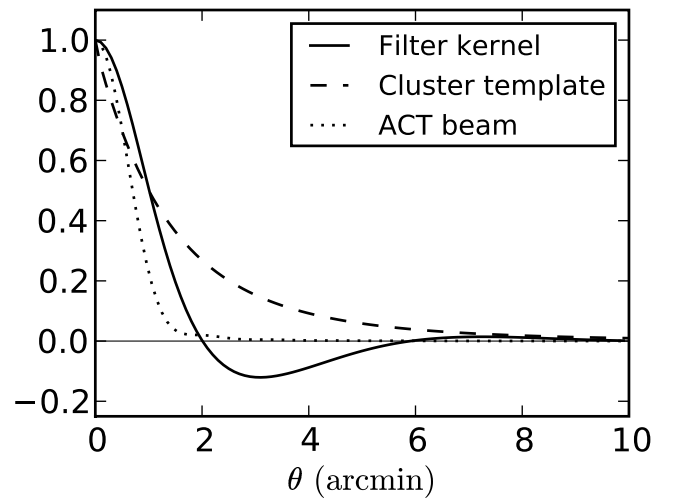}}
\caption{The azimuthally averaged real space matched filter kernel,
  proportional to $\Psi_{\thetaFixed}(\theta)$, for signal template with
  $\thetaD = \thetaFixed$.  Shown for reference are the ACT \arone{} beam, and
  the cluster signal template $S_{\thetaFixed}(\theta)$.  While filters tuned
  to many different angular scales are used for cluster detection
  (Section~\ref{sec:detection}), the $\thetaFixed$ filter is used for cluster
  characterization and cosmology (Section~\ref{sec:fixed_scale}). }
\label{fig:kernel}
\end{center}
\end{figure}

The azimuthally averaged real space filter kernel corresponding to
$\thetaD=\thetaFixed$ is shown in Figure~\ref{fig:kernel}, and compared to
both the ACT \arone{} beam and the signal template $S_\thetaD$.

The true noise spectrum may vary somewhat over the map due to variations in
atmospheric and detector noise levels, and thus the matched filter
$\Psi_\thetaD(\kv)$ might be said to be sub-optimal at any point.  The filter
remains \emph{unbiased}, however, and a reasonable estimate of the signal to
noise ratio (S/N) may still be obtained by recognizing that the local noise
level will be highly correlated with the number of observations contributing
to a given map pixel.

\begin{figure*}
\begin{center}
\resizebox{\textwidth}{!}{\plotone{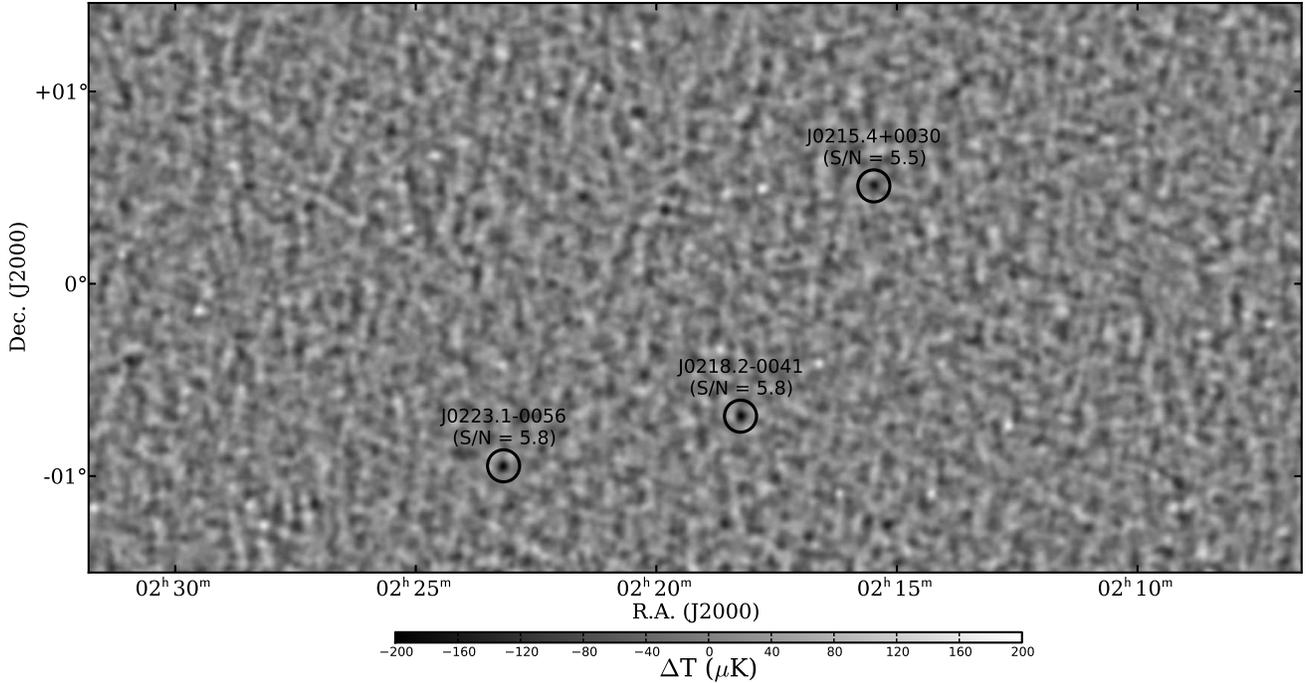}}
\caption{Section of the 148\,GHz map (covering 18.7 deg$^2$) match-filtered
  with a GNFW profile of scale $\thetaD = \thetaFixed$. Point sources are
  removed prior to filtering. Three optically confirmed clusters with S/N $>
  4.9$ are highlighted (see Table~\ref{tab:clusters}).  Within this area,
  there are an additional 11 candidates ($4 <$ S/N $< 4.9$) which are not
  confirmed as clusters in the SDSS data (and thus may be spurious detections
  or high-redshift clusters). }
\label{fig:mapsection}
\end{center}
\end{figure*}

Prior to matched filtering, the ACT maps are conditioned in the same way as
for noise estimation, except that the point sources are subtracted from the
maps instead of being masked.  (Subsequent analysis disregards regions near
those point sources, amounting to 1\% of the map area.)  With the application
of each filter we obtain a map of $\Delta T$ values.  A section of a filtered
map (for $\thetaD = \thetaFixed$) is shown in Figure~\ref{fig:mapsection}.

We characterize the noise in each filtered map by modeling the variance at
position $\xv$ as $\sigma^2(\xv) = \sigma_0^2 + \sigma_{\rm hits}^2/n_{\rm
  hits}(\xv)$, where $n_{\rm hits}(\xv)$ is the number of detector samples
falling in the pixel at $\xv$.  We fit constants $\sigma_0$ and $\sigma_{\rm
  hits}$ by binning in small ranges of $1/n_{\rm hits}$.  The fit is iterated
after excluding regions near pixels that are strong outliers to the noise
model.  Typically such pixels are near eventual galaxy cluster
candidates. Figure~\ref{fig:sensitivity} shows the noise map for the $\thetaD
= \thetaFixed$ filter.

After forming the signal to noise ratio map $-\Delta T(\xv)/\sigma(\xv)$,
cluster candidates are identified as all pixels with values exceeding four.
The catalog of cluster candidates contains positions, central decrements
($\Delta T$), and the local map noise level.  Candidates seen at multiple
filter scales are cross-identified if the detection positions are within
$1\arcmin$; the cluster candidate positions that we list come from the map
where the cluster was most significantly detected. We adopt the largest S/N
value obtained over the range of filter scales as the detection significance
for each candidate.

\begin{figure}
\resizebox{3.5in}{!}{\plotone{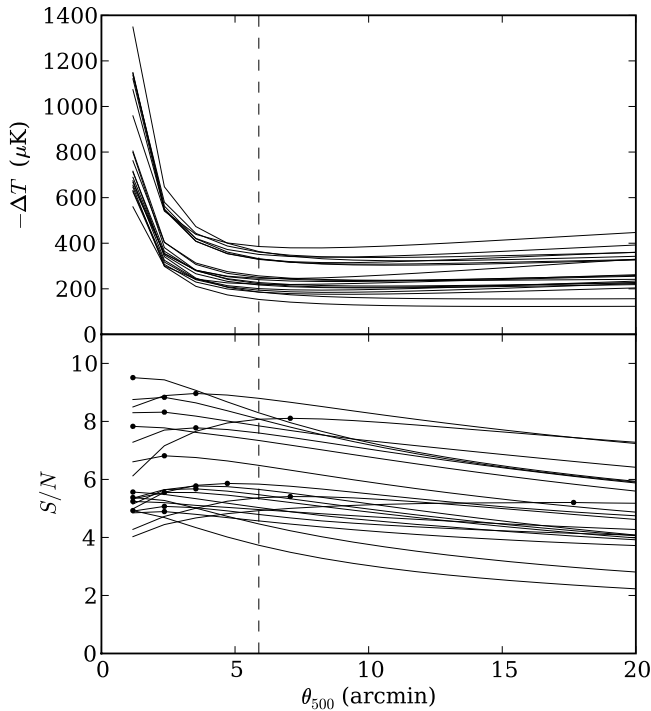}}
\caption{Central decrement and signal to noise ratio as a function of filter
  scale for the 20 clusters in S82 detected with peak S/N $> 5$. \emph{Top
    panel}: Although the central decrement is a model-dependent quantity, the
  value tends to be stable for filter scales of $\thetaD > 3\arcmin{}$.
  \emph{Bottom panel}: On each curve, the circular point identifies the filter
  scale at which the peak S/N was observed.  The vertical dashed line shows
  the angular scale chosen for cluster property and cosmology analysis,
  $\theta_{500} = \thetaFixed$.  Despite the apparent gap near S/N $\approx$
  6, the clusters shown represent a single population. }
\label{fig:decrements}
\end{figure}

For a given candidate, the S/N tends to vary only weakly with the filter
scale.  The reconstructed central decrement $\Delta T$ varies weakly above
filter scales of $\thetaD \approx 3\arcmin$, as may be seen for the most
significantly detected clusters in Figure~\ref{fig:decrements}.  This
stabilization occurs when the assumed cluster size is larger than the true
cluster size, because the filter is optimized to return the difference in the
level of the signal at the cluster position and the level of the signal away
from the cluster center.  The filter interprets the signal at the cluster
position as being due to the convolution of the telescope beam with the
cluster signal.  The inferred central decrement thus rises rapidly as the
assumed $\thetaD$ decreases, since total SZ flux scales as $\Delta T
\thetaD^2$.  As is discussed in Section~\ref{sec:fixed_scale}, only the
results from the $\thetaD = \thetaFixed$ matched filter are used for inferring
masses, scaling relations, and cosmological results.  The corresponding
physical scale may be determined, as a function of redshift, based on cluster
distance.

\subsection{Galaxy Cluster Confirmation}
\label{sec:confirmation}

The cluster candidates obtained from the \arone{} map analysis are confirmed
using optical and infrared imaging.  A complete discussion of this process may
be found in \cite{menanteau/etal:2013}.  For the purposes of this work, we
briefly summarize the confirmation process and the redshift limits of the
sample (which must be understood in order to derive cosmological constraints).
These limits differ according to the depth of the optical imaging available
over a given part of the map.

Most cluster candidates are confirmed through the analysis of SDSS imaging.
The ACT Equatorial survey is almost entirely covered by SDSS archival data
\citep{abazajian/etal:2009}, with a central strip designed to overlap with the
deep optical data ($r\approx 23.5$) in the S82 region \citep{annis/etal:2011},
as shown in Figure~\ref{fig:sensitivity}.  For each ACT cluster candidate with
peak S/N~$> 4$, SDSS images are studied using an iterative photometric
analysis to identify a brightest cluster galaxy (BCG) and an associated red
sequence of member galaxies.  A minimum richness of $N_{\rm gal}$ = 15,
evaluated within a projected 1 $h^{-1}$ Mpc of the nominal cluster center and
within $0.045(1+z_c)$ of the nominal cluster redshift $z_c$, is required for
the candidate to be confirmed as a cluster.  The redshifts of confirmed
clusters are obtained from either a photometric analysis of the
images, from SDSS spectroscopy of bright cluster members, or from targeted
multi-object spectroscopic follow-up.  The redshift limit of cluster
confirmation using SDSS data alone is estimated to be $z \approx 0.8$ within
S82 and $z \approx 0.5$ outside of S82.  Cluster candidates that are not
confirmed in SDSS imaging are targeted, in an on-going follow-up
campaign, under the assumption that they may be high redshift clusters.

Within the S82 region, 49 of 155 candidates are confirmed, with 44 of these
confirmations resulting from analysis of SDSS data only.  Targeted follow-up
of the high S/N candidates was pursued at the Apache Point Observatory,
yielding five more confirmations, all at $z > 0.9$.  All cluster
candidates with S/N $> 5.1$ were confirmed as clusters.  This is consistent
with our estimate of 1.8 false detections in this region,
based on filtering of simulated noise.  The follow-up in the S82
region is deemed complete to a S/N of 5.1, in the sense that all SZ candidates
with ACT S/N $> 5.1$ have been targeted.  It is thus this sample, and this
region, that are considered for the cosmological analysis (in addition to a
subset of the \citet{marriage/etal:2011} sample; see
Section~\ref{sec:cosmology}).
The completeness within S82, as a function of mass and redshift, is estimated in
Section~\ref{sec:completeness}.

Outside of S82, 19 clusters are confirmed using SDSS DR8 data.  High
significance SZ detections in this region that are not confirmed in the DR8
data constitute good candidates for high redshift galaxy clusters and are
being investigated in a targeted follow-up campaign.

The confirmed cluster sample may contain a small number of false positives,
due to chance superposition of a low mass cluster at the location of an
otherwise spurious SZ candidate.  Most of our confirmed clusters are
associated with rich optical counterparts, and thus are truly massive
clusters. However, our search was carried out over considerable sky area in
the $\approx 150$ regions around SZ candidates.  Assigning an effective area of
13 square arcminutes to each of these fields yields a total area of
approximately 0.5 square degrees. From the maxBCG catalog
\citep{koester/etal:2007}, which includes optical richness measurements for
clusters with $0.1 < z < 0.3$, we expect that the density of clusters
satisfying our richness criteria in the range $0.1 < z < 0.8$ is approximately
6 per square degree. We conclude that roughly three of our low
richness confirmed clusters could potentially be spurious associations. Such
contamination is not likely to affect the high significance (S/N $> 5.1$)
sample, where the lowest richness is $\approx 30$.

\begin{figure*}
\begin{center}
\resizebox{\textwidth}{!}{\plotone{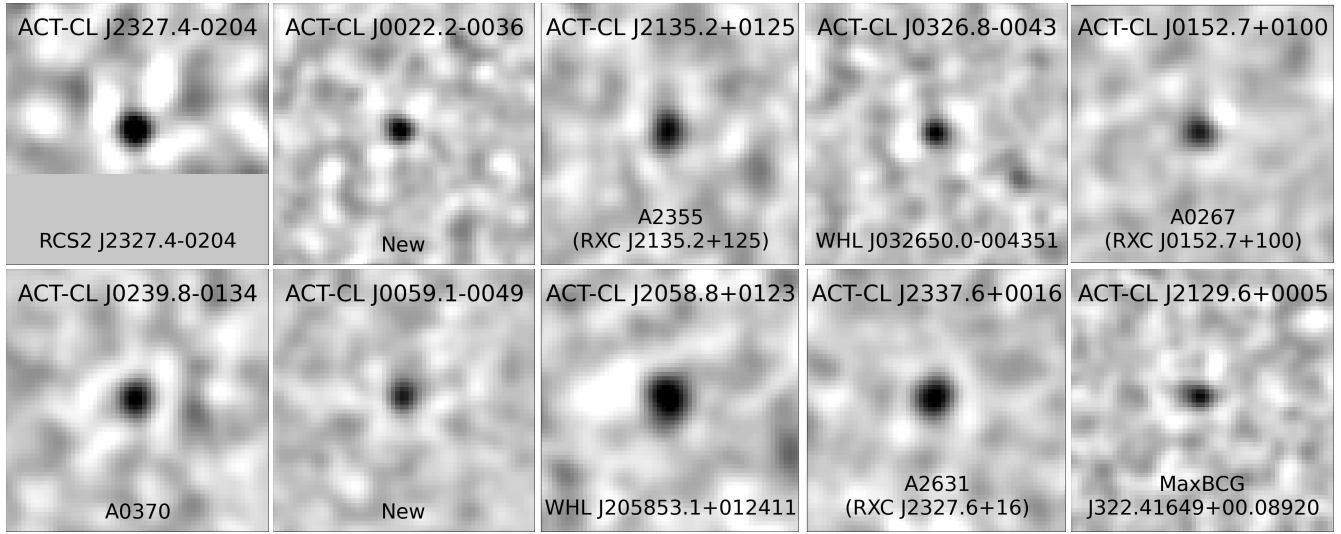}}
\caption{Postage stamp images ($30\arcmin$ on a side) for the 10 highest S/N
  detections in the catalog (see Table~\ref{tab:clusters}), taken from the
  filtered ACT maps.  The clusters are ordered by detection S/N, from top left
  to bottom right, and each postage stamp shown is filtered at the scale which
  optimizes the detection S/N. Note that J2327.4$-$0204 is at the edge of the
  map. The greyscale is linear and runs from -350$\,\mu$K (black) to
  +100$\,\mu$K (white).}
\label{fig:montage}
\end{center}
\end{figure*}

In Table~\ref{tab:clusters} we present the catalog of 68 confirmed clusters.
For each object we list its coordinates, redshift \citep[see][for
  details]{menanteau/etal:2013}, S/N of the detection (we adopt the maximum
S/N across the range of filters used), and SZ
properties. Figure~\ref{fig:montage} shows postage stamp images of some
high-significance clusters, taken from the filtered ACT 148\,GHz maps.

\section{Recovered Cluster Properties}
\label{sec:recovery}

\newcommand\MethodName{Profile Based Amplitude Analysis}
\newcommand\PBAA{PBAA}

In this section we develop a relationship between cluster mass and the
expected signal in the ACT filtered maps.  The form of the scaling
relationship between the SZ observable and the cluster mass is based on the
UPP, and parameters of that relationship are studied using models of cluster
physics and dynamical mass measurements.  We obtain masses for the ACT
Equatorial clusters assuming a representative set of parameters.

\subsection{\MethodName{}}
\label{sec:fixed_scale}

Scaling relations between cluster mass and cluster SZ signal strength are
often expressed in terms of bulk integrated Compton quantities, such as
$Y_{500}$, which are expected to be correlated to mass with low intrinsic
scatter \citep[e.g.,][]{motl/etal:2005,reid/spergel:2006}.  Due to projection
effects, and the current levels of telescope resolution and survey depth,
measurements of $Y_{500}$ for individual clusters can be obtained only by
comparing the microwave data to a simple, parametrized model for the cluster
pressure profile.  Such fits may be done directly, or indirectly as part of
the cluster detection process through the application of one or more matched
filters (where the filters are ``matched'' in the sense of being tuned to a
particular angular scale).  In such comparisons, the inferred values of
$Y_{500}$ are very sensitive to the assumed cluster scale (i.e.,
$\theta_{500}$\footnote{$M_{500c} = (4\pi/3)\times 500\rho_c(z) R_{500}^3$;
  $\theta_{500} = R_{500}/D_A(z)$.}), and this scale is poorly
constrained by microwave data alone.

Recent microwave survey instruments make use of spatial filters to both detect
and characterize their cluster samples, coping with $\theta_{500}$ uncertainty
in different ways.  For example, the \Planck{} team uses X-ray luminosity
based masses \citep{planck/esz:2011} as well as more detailed X-ray and weak
lensing studies \citep{planck/intermediateIII:2013} to constrain $R_{500}$,
and obtain $Y_{500}$ measurements assuming profile shapes described by the
UPP.

In cases where suitable X-ray or optical constraints on the cluster scale are
not available, authors have constructed empirical scaling relations based on
alternative SZ statistics, such as the amplitude returned by some particular
filter \citep{sehgal/etal:2011}, or the maximum S/N over some ensemble of
filters \citep{vanderlinde/etal:2010}.  Recognizing that the cluster angular
scale is poorly constrained by the filter ensemble, recent work from the South
Pole Telescope has included a marginalization over the results returned by the
ensemble of filters \citep[e.g.,][]{story/etal:2011,reichardt/etal:2013}.
Such approaches rely on simulated maps to guide the interpretation of their
results.

For the purposes of using the SZ signal to understand scaling relations and to
obtain cosmological constraints, we develop an approach in which the cluster
SZ signal is parametrized by a single statistic, obtained from the ACT map
that has been filtered using $\Psi_\thetaFixed(\kv)$.  Instead of using
simulations to inform our interpretation of the data, we develop a framework
where the SZ observable is expressed in terms of the parameters of some
underlying model for the cluster pressure profile.  In particular, we model
the clusters as being well described, up to some overall adjustments to the
normalization and mass dependence, by the UPP (see
Section~\ref{sec:pressure_model}).

An estimate of the cluster central Compton parameter, based only on the
non-relativistic SZ treatment, is given by
\begin{align}
  \ytilde & \equiv \frac{\Delta T}{\Tcmb}\fsz^{-1}(m=0,z=0), \label{eqn:ytil}
\end{align}
where $\fsz(m=0,z=0) = -0.992$ as explained in
Section~\ref{sec:pressure_model}.  This ``uncorrected'' central Compton
parameter is used in place of $\Delta T$ to develop an interpretation of the
SZ signal. This quantity is uncorrected in the sense that it is associated
with the fixed angular scale filter and does not include a relativistic
correction.

For a cluster with SZ signal described by equation (\ref{eqn:uppcyl}), the
value of $\ytilde$ that we would expect to observe by applying the filter
$\Psi_\thetaFixed$ to the beam-convolved map is
\begin{align}
  \ytilde & = 10^{A_0}E(z)^2m^{1+B_0}Q(\theta_{500}/m^{C_0})\frel(m,z)
     \label{eqn:uppcompton}
\end{align}
where
\begin{align}
  Q(\theta) & = \int \frac{d^2k}{(2\pi)^2} \Psi_\thetaFixed(\kv) B(\kv) \int d^2\theta'~
    e^{i\vec{\theta'}\cdot\kv} \tau(\theta'/\theta).\label{eqn:Q}
\end{align}
is the spatial convolution of the filter, the beam, and the cluster's
unit-normalized integrated pressure profile.  We note that in this formalism,
$\theta_{500} = R_{500}/D_A(z)$ is determined by the cluster mass and the
cosmology (rather than being some independent parameter describing the angular
scale of the pressure profile).

\begin{figure}
\includegraphics[width=8.5cm]{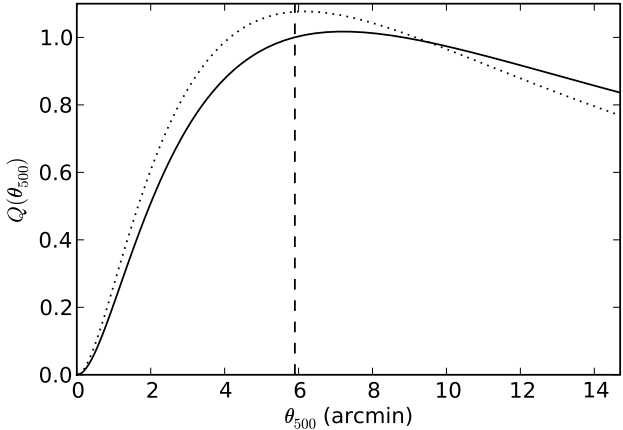}
\caption{The response function used to reconstruct the cluster central
  decrement as a function of cluster angular size (solid line).  At $\thetaD =
  \thetaFixed$, the filter is perfectly matched and $Q=1$.  At scales slightly
  above $\thetaFixed$, $Q > 1$ because such profiles have high in-band signal
  despite being an imperfect match, overall, to the template profile.  For the
  definition of $Q$, see Section~\ref{sec:fixed_scale}. Dotted line shows
  analogous function computed under the assumption that the cluster signal is
  described by the \Planck{} Pressure Profile (see
  Section~\ref{sec:planck_profile}).}
\label{fig:Qeq}
\end{figure}

The response function $Q(\theta)$ for the Equatorial clusters is shown in
Figure~\ref{fig:Qeq}.  It encapsulates the bias incurred in the central
decrement estimate due to a mismatch between the true cluster size and the
size encoded in the filter, for the family of clusters described by the UPP.
While this bias is in some cases substantial ($Q \approx 0.3$ for clusters
with $\theta_{500} \approx 1.5\arcmin$), the function
$Q(\theta)$ is not strongly sensitive to the details of the assumed pressure
profile (as demonstrated in Section~\ref{sec:planck_profile}), and the
assumptions underlying this approach are not a significant departure from
other analyses that rely on a family of cluster templates to extract a cluster
observable.

\begin{figure}
\includegraphics[width=8.5cm]{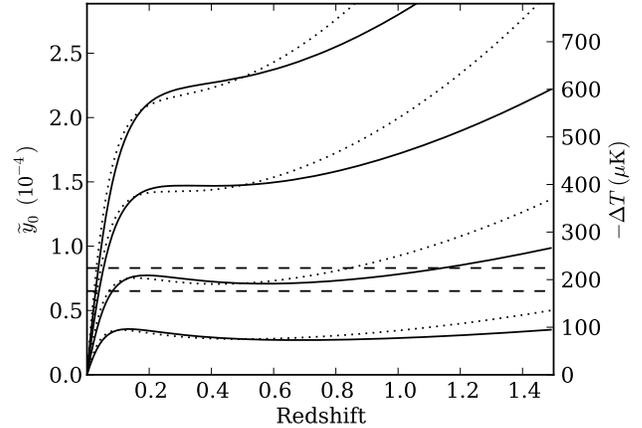}
\caption{Prediction, based on the UPP, for cluster signal in a map
  match-filtered with $\thetaD = \thetaFixed$, in units of uncorrected central
  Compton parameter $\ytilde$ and apparent temperature decrement $-\Delta T$
  at \arone{} (Section~\ref{sec:fixed_scale}).  Solid lines trace constant
  masses of, from top to bottom, $M_{500c} = 10^{15}, 7 \times 10^{14},
  4\times 10^{14}$, and $2 \times 10^{14}~\hinv\Msun$.  Dotted lines are for
  the same masses, but with the scaling relation parameter $C=0.5$ to show the
  redshift sensitivity to this parameter.  Above $z \approx 0.5$, the scaling
  behavior of the observable $\ytilde$ with redshift is stronger for higher
  masses because their angular size is a better match to the cluster template
  and the redshift dependence in $Q$ does not attenuate the scaling of the
  central decrement, $y_0 \propto E(z)^2$, as much as it does for lower
  masses.  The dashed lines correspond to $S/N > 4$ and $S/N > 5.1$, based on
  the median noise level in the S82 region.}
\label{fig:ymz}
\end{figure}

Equation~(\ref{eqn:uppcompton}) thus relates $\ytilde$ to cluster mass and
redshift while accounting for the impact of the filter on clusters whose
angular size is determined by their mass and redshift.  This relationship can
be seen in Figure~\ref{fig:ymz}.

The essence of our approach, then, is to filter the maps with $\Psi_\thetaFixed(\kv)$
and for each confirmed cluster obtain $\Delta T$ and its error.  This is
equivalent to measuring $\ytilde$, which can then be compared to the right
hand side of equation~(\ref{eqn:uppcompton}).  If the cluster redshift is also
known, then for a given cosmology the only free parameter in the expression
for $\ytilde$ is the mass parameter, $m$.\footnote{ With $\ytilde$ and $z$
  measurements in hand, one could certainly proceed to solve
  equation~(\ref{eqn:uppcompton}) to obtain a mass for each cluster.  Because
  we are treating mass as one of the independent variables, however, such an
  approach would produce biased mass estimates; see
  Section~\ref{sec:corrected_masses}.}

We refer to this alternative approach, where a family of pressure profiles is used to
model the amplitude of a source in a filtered map, as ``\MethodName{}''
(\PBAA{}).  While we have applied a filter tuned to a particular angular
scale, the effects of angular diameter distance, telescope beam, and the
spatial filtering are modeled in a way that accounts for the (mass and
redshift dependent) cluster angular scale.  For a given cosmology, and having
computed $Q(\theta)$ based on the UPP, the parameters ($A_0$, $B_0$, and
$C_0$) of the scaling relation between $\ytilde$ and mass have a physical
significance and can be verified through measurements of $\ytilde$, redshift,
and mass for a suitable set of clusters.

While the usage of a single filter clearly simplifies data processing, the
most compelling advantage is that one does not suffer from inter-filter noise
bias.  For example, when optimizing filter scale, a CMB cold spot near a
cluster candidate will draw the preferred filter to larger angular scales than
would the isolated cluster signal.  The preferred filter scale is thus driven
by the amplitude of local noise excursions as much as it is driven by the
cluster signal.  In a single filter context, a CMB cold spot affects the
amplitude measurement by contributing spurious signal to the apparent cluster
decrement; but if CMB hot and cold spots are equally likely, and uncorrelated
with cluster positions, then the CMB as a whole acts as a Gaussian noise
contribution to cluster signal.  The effects of coherent noise on large scales
are thus somewhat better behaved if we do not permit the re-weighting of
angular scales to maximize the apparent signal.

To achieve the goal of detecting as many clusters as possible, one should
certainly explore a variety of candidate cluster profiles and apply an
ensemble of matched filters.  However, for cosmological studies, or when
trying to understand the relationship between observables in samples that are
selected based on one of the observables under study, it is critical to
understand the selection function that describes how the population of objects
in the sample relates to the broader population of objects in the universe.
While we sacrifice a certain amount of signal when choosing a single filter
scale to use for cosmological and scaling relation analysis, we benefit from
having a simpler selection function.

Much of our approach can be simply generalized so that a suite of filters are
used, but with each filter intended to correspond to a particular redshift
interval.  The redshift-dependent angular scales might be selected to match
clusters of a particular mass, for example.  Such an approach benefits from
the lack of inter-scale noise bias, because there is no data-based
optimization over angular scale.  However, interpretation of the signal is
then complicated by the need to consider the impact of the full suite of
filters on the cluster signal and noise models.  Such an approach is
tractable, but is not considered in this work.

We also note that $\thetaD = \thetaFixed$ is chosen because it lies in a
regime of $\thetaD$ where the measured $\ytilde$ statistic for our high
significance clusters is approximately constant.  Our approach does not
require this, however, and could instead have used a filter corresponding to
some smaller $\thetaD$, where signal to noise ratios are, on average, slightly
higher.

In order to compare the predictions of the UPP based formalism to models and
other data sets, we introduce a more general relationship relating cluster
mass to the uncorrected central Compton parameter.  We allow for variations in
the normalization, mass dependence, and scale evolution through parameters
$A$, $B$, and $C$ and model $\ytilde$ as
\begin{align}
  \ytilde & = 10^{A_0+A} E(z)^2
      (M/\Mpivot)^{1+B_0+B} \times\nonumber\\
 & ~~~~~~ Q\left[\left(\frac{1+z}{1.5}\right)^C\theta_{500}/m^{C_0}\right]
  \frel(m,z). \label{eqn:ymmodelmod}
\end{align}
To abbreviate the argument to $Q(\theta)$, we will often simply write
$Q(m,z)$.  The exponents $(A_0, B_0, C_0)$ remain fixed to the UPP model
values of equations~(\ref{eqn:uppcyl1}) and (\ref{eqn:uppcyl}), except where
otherwise noted.  For a given data set or model, $\Mpivot$ will be chosen to
reduce covariance in the fit values of $A$ and $B$.  In
Table~\ref{tab:abcfits} we present the fit parameters for various models and
data sets discussed in subsequent sections.  In order to compare fits from
data sets with different $\Mpivot$, we also compute the normalization exponent
$A_m$ associated with $\Mpivot = 3 \times10^{14}~\hinv\Msun$ for each data
set.  In these terms, the UPP model described by
equation~(\ref{eqn:uppcompton}) corresponds to $(A_m,B,C)$ = $(0,0,0)$.

In cases where independent surveys each measure $\ytilde$ values for a cluster
based on following the algorithm described here, the $\ytilde$ measurements
should not, in general, be compared directly.  This is because the filter
$\Psi_\thetaD$ and the resulting bias factor $Q$ depend on the telescope beam
and the noise spectra of the resulting maps.  However, it is possible to
filter one set of maps in a way that matches the beam and filtering of a
preceding analysis.  In such cases an independent measurement of $\ytilde$ is
obtained, which may be compared between experiments.  Such comparisons are
likely to be most interesting in cases where two telescopes have similar
resolution.

Alternatively, $\ytilde$ measurements and redshifts may be converted, for some
particular values of the scaling relation parameters, into physical parameters
such as $M_{500c}$, $Y_{500}$, or the corrected $y_0$.  Such derived
quantities may be compared between experiments that probe different angular
scales.  The physical parameters can be updated as one's understanding of the
scaling relation parameters is improved.  The use of $\ytilde$ thus
facilitates the re-use of the data in analyses that explore different models
for the cluster signal.

The uncorrected central Compton parameter measurements ($\ytilde$) for the ACT
Equatorial clusters are presented in Table~\ref{tab:clusters}.  They are
used in subsequent sections to estimate cluster properties (such as corrected
SZ quantities and mass) and to constrain cosmological parameters.  For the
Southern cluster sample, analogous measurements are presented in the
Appendix. Between the Equatorial and Southern cluster samples, the ACT
collaboration has reported a total of 91 optically confirmed, SZ detected
clusters.

\subsection{Cluster Mass and SZ Quantity Estimates}
\label{sec:corrected_masses}

Given measurements of cluster $\ytilde$ and redshift, one cannot naively
invert Equations~(\ref{eqn:uppcompton}) or (\ref{eqn:ymmodelmod}) to obtain a
mass estimate.  Because of intrinsic scatter, measurement noise, and the
non-trivial (very steep) cluster mass function, the mean mass at fixed SZ
signal $\ytilde$ will be lower than the mass whose mean predicted SZ signal is
$\ytilde$.  The bias due to noise is often referred to as ``flux boosting" and
can be corrected in a Bayesian analysis that accounts for the underlying
distribution of flux densities \citep{coppin/etal:2005}.  The bias due to
intrinisic scatter, however, is not restricted to the low significance
measurements.  Considering the population of clusters (at fixed redshift) in
the $(\log m, \log \ytilde)$ plane, the locus $\langle \log m | \log \ytilde
\rangle$ (i.e., the expectation value of the log of the mass for a given
central Compton parameter) lies at lower mass than $\langle \log \ytilde| \log
m \rangle$.  This phenomenon has been discussed in the context of galaxy
cluster surveys by, e.g., \citet[][see also the review by
  \citealt*{allen/etal:2011}]{mantz/etal:2010b}.

\newcommand\ytil{\widetilde{y}}
\newcommand\ytrue{\widetilde{y}_{0}^{\rm tr}}
\newcommand\yobs{\widetilde{y}_{0}^{\rm ob}}

The mass of a cluster, however, can be estimated if one has an expression for
the cluster mass function.  We adapt the Bayesian framework of
\cite{mantz/etal:2010b} to this purpose.  The posterior probability of the
mass parameter $m$ given the observation $\yobs$ is
\begin{align}
 P(m|\yobs) & \propto P(\yobs | m) P(m)\nonumber\\
  & = \left( \int d\ytrue~P(\yobs|\ytrue) P(\ytrue | m) \right) P(m)
 \label{eqn:P_mass}
\end{align}
where $\ytrue$ represents the ``true'' SZ signal in the absence of noise,
$P(\yobs|\ytrue)$ is the distribution of $\yobs$ given $\ytrue$ and the
observed noise $\delta \yobs$, and $P(m)$ is proportional to the distribution
of cluster masses at the cluster redshift.  The distribution $P(\ytrue|m)$ of
the noise-free cluster signal $\ytrue$ is assumed to be log-normal about the
mean relation given by Equation~(\ref{eqn:ymmodelmod}), i.e.,
\begin{align}
  \log \ytrue & \sim N(\log \ytilde(m,z); \sigma_{\rm int}^2)
  \label{eqn:ymmodelscatter}
\end{align}
with $\sigma_{\rm int}$ denoting the intrinsic scatter.

We use the results of \cite{tinker/etal:2008} to compute the cluster mass
function, assuming the fiducial $\Lambda$CDM cosmology, with $\sigma_8 = 0.8$.
Scaling the mass function by the comoving volume element at fixed solid angle,
we obtain $dN(<m,z)/dz$, the number of clusters, per unit solid angle and per
unit redshift, that have mass less than $m$.  The probability of a cluster in
this light cone having mass $m$ and redshift $z$ may then be taken as $P(m,z)
\propto d^2N(<m,z)/dz~dm$.  We account for redshift uncertainty by
marginalizing the cluster mass function $P(m,z)$ over the cluster's redshift
error to obtain an effective $P(m)$ at the observed cluster redshift.

The marginalized masses obtained using Equation (\ref{eqn:P_mass}) are
presented in Table~\ref{tab:masses}.  For the ACT Southern cluster sample,
these masses are presented in the Appendix.  In each case, masses are
presented for the UPP scaling relation as well as for scaling relation
parameters fit to SZ signal models (see Section~\ref{sec:calsim}) or dynamical
mass data (see Section~\ref{sec:caldyn}).

A similar approach may be taken to estimate the true values of SZ quantities,
given the observed quantities.  In this case we are effectively only undoing
the noise bias, while intrinsic scatter affects the underlying population
function.  We are interested in
\begin{align}
 P(\ytrue|\yobs) & \propto P(\yobs|\ytrue) P(\ytrue)\nonumber\\
  & = P(\yobs|\ytrue) \int dm~P(\ytrue|m) P(m).
 \label{eqn:P_y}
\end{align}
The resulting probability distribution is used to obtain marginalized
estimates of $y_0$, $\theta_{500}$ (which should be interpreted as giving the
scale of the pressure profile rather than the scale of the mass density
profile), $Y_{500}$ (estimated within the SZ-inferred $\theta_{500}$) and $Q$.
These quantities are presented in Table~\ref{tab:masses}, for the UPP scaling
relation parameters.

\begin{figure}
\includegraphics[width=8.5cm]{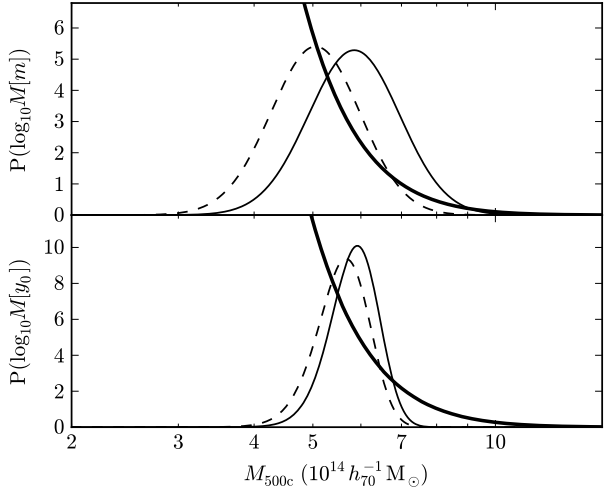}
\caption{Example probability distributions for cluster mass (upper panel) and
  SZ signal strength parametrized as a mass according to
  equation~(\ref{eqn:ymmodelmod}) (lower panel).  The solid line PDF is the
  result of a direct inversion of the scaling relation described by
  equation~(\ref{eqn:ymmodelmod}).  The corrected PDF (dashed line) is
  obtained by accounting for the underlying population distribution (bold
  line; arbitrary normalization).  The correction is computed according to
  equation (\ref{eqn:P_mass}) for the upper panel, and according to equation
  (\ref{eqn:P_y}) for the lower panel.  Curves shown correspond to ACT--CL
  J0022.2$-$0036.  }
\label{fig:mass_post}
\end{figure}

In Figure~\ref{fig:mass_post} we demonstrate the impact of the steep mass
function on the inferred mass and SZ quantities.  As the measurement noise
decreases, the $\ytilde$ measurements are less biased; but any intrinsic
scatter in the $\ytilde$--$M$ relation will lead to bias in the naively
estimated mass.

\subsection{The \Planck{} Pressure Profile}
\label{sec:planck_profile}

In \cite{planck/intermediateV:prep}, data for 62 massive clusters from the
\Planck{} all-sky Early Sunyaev-Zel'dovich cluster sample
\citep{planck/esz:2011} are analyzed to obtain a ``\Planck{} pressure
profile'' (PPP) based on measurements of the SZ signal.  Integrating the PPP
along lines of sight, the central pressure is 20\% lower than the UPP but is
higher than the UPP outside of $0.5R_{500}$.  \Planck{} finds overall
consistency between results obtained with the UPP and the PPP.

We assess the difference in inferred mass due to this alternative pressure
profile by re-analyzing the ACT $\ytilde$ using the PPP.  A bias function $Q$
is computed as in Equation~(\ref{eqn:Q}), but with the normalized Compton
profile $\tau$ associated to the PPP (see Figure~\ref{fig:Qeq}).  We
additionally determine scaling parameters, compatible with
Equation~(\ref{eqn:uppcyl}), of $(10^{A_0}, B_0, C_0)$ = $(4.153 \times
10^{-5} h_{70}^{1/2}, 0.12, 0)$.  Note that while the bias function shows
increased sensitivity, compared to the UPP, for $\theta_{500} < 9\arcmin$,
this is compensated for by the lower normalization factor $10^{A_0}$.  For the
Equatorial cluster sample, we find the PPP masses to be well described by a
simple mean shift of $M_{500c}^{\rm{PPP}} = 1.015~M_{500c}^{\rm{UPP}}$, with
3\% RMS scatter.  Note that this is only a statement about the dependence of
the ACT results on the assumed pressure profile; experiments that probe
different angular scales may be more or less sensitive to such a change.

While the change in inferred masses is in this case negligible, we reiterate
that our fully parametrized relationship between SZ signal and mass (Equation
(\ref{eqn:ymmodelmod})) allows for freedom in the normalization, mass
dependence, and evolution of cluster concentration with redshift.  Mass or
cosmological parameter estimation can be computed after fixing these
parameters based on any chosen pressure profile, model, simulation, or data
set; all that is required is to compensate for the mismatch of our assumed
pressure profile to the true mean pressure profile.

\subsection{Scaling Relation Calibration from SZ Models}
\label{sec:calsim}

The previous sections have described a general approach that relates cluster
mass and redshift to SZ signal in a filtered map, given values for the scaling
relation parameters.  In this section we obtain scaling relation parameters
based on three models for cluster gas physics.  While the ACT data will be
interpreted using each of these results, we do not yet consider any ACT data
explicitly.

Current models for the SZ signal from clusters include contributions from
non-thermal pressure support, star formation, and energy feedback and are
calibrated to match detailed hydrodynamical studies and X-ray or optical
observations \citep{shaw/etal:2010, bode/ostriker/cen/trac:2012}.  Such models
provide a useful testing ground for the assumptions and methodology of our
approach to predicting SZ signal based on cluster mass.  While models may
suffer from incomplete modeling of relevant physical effects, they are less
vulnerable to some measurement biases (e.g., by providing a cluster mass and
alleviating the need for secondary mass proxies).  In order to explore the
current uncertainty in the SZ--mass scaling relation, we consider simulated
sky maps based on three models of cluster SZ signal that include different
treatments of cluster physics.

Our study will center on maps of SZ signal produced from the SZ models and
structure formation simulations of \citet[ hereafter
  B12]{bode/ostriker/cen/trac:2012}.  The N-body simulations
\citep{bode/ostriker:2003} are obtained in a Tree-Particle-Mesh framework, in
which dark matter halos have been identified by a friends-of-friends
algorithm.  The intracluster medium (ICM) of massive halos is subsequently
added, following a hydrostatic equilibrium prescription, and calibrated to
X-ray and optical data \citep*{bode/ostriker/vikhlinin:2009}.  The density and
temperature of the ICM of lower mass halos and the IGM are modeled as a
virialized ideal gas with density (assuming cosmic baryon fraction
$\Omega_b/\Omega_m = 0.167$) and kinematics that follow the dark matter.

To complement the model of B12, we also consider the Adiabatic and
Nonthermal20 models described in \cite{trac/bode/ostriker:2011}, which make
use of the same N-body results as B12.  In the Adiabatic model the absence of
feedback and star formation leads to a higher gas fraction than in the B12
model.  In the Nonthermal20 model, 20\% of the hydrostatic pressure is assumed
to be nonthermal, leading to substantially less SZ signal compared to the B12
model.  The SZ-mass relations derived from these two models are thus
interpreted as, respectively, upper and lower bounds on the SZ signal.

The model of B12 differs from those in \citet{trac/bode/ostriker:2011} through
a more detailed handling of non-thermal pressure support, which is tied to the
dynamical state of the cluster and is allowed to vary over the cluster extent.
Both B12 and the similar treatment of \citet{shaw/etal:2010} make use of
hydrodynamic simulations \citep{nagai/kravtsov/vikhlinin:2007,
  lau/kravtsov/nagai:2009, battaglia/etal:2012} to understand these
non-thermal contributions.

\begin{figure}
\includegraphics[width=8.5cm]{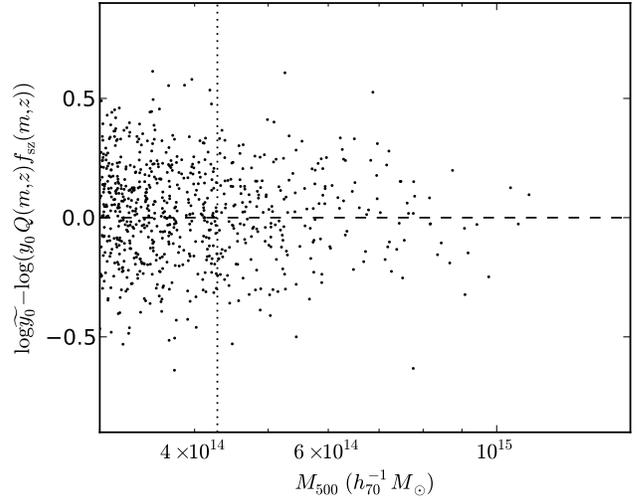}
\caption{Residuals of the scaling relation fit for the B12 model
  (Section~\ref{sec:calsim}).  Only clusters with $0.2 < z < 1.4$ and
  $M_{500c} > 4.3\times 10^{14}~\hinv\Msun$ (indicated by dotted line) are
  used for the fit.  The scatter in the relation is measured from the RMS of
  the residuals.}
\label{fig:simCalibration}
\end{figure}

To calibrate our scaling relation approach to these models, we make use of
light-cone integrated maps of the thermal and kinetic SZ at 145~GHz
\citep[constructed as in ][]{sehgal/etal:2010}, and the associated catalog of
cluster positions and masses.  A set of 192 non-overlapping patches of area
18.2\,deg$^2$ each are extracted from the simulated map, and convolved with
the ACT \arone{} beam to simulate observation with the telescope.  The maps
are then filtered with the same filter $\Psi_\thetaFixed(\kv)$ that was used
for the ACT 
Equatorial clusters. Because the filtering is a linear operation, it is
counter-productive to the purpose of calibration and intrinsic scatter
estimation to add noise (CMB, detector noise) to the simulated signal map, and
so we do not.  The uncorrected central decrements are extracted and used to
constrain the parameters of Equation~(\ref{eqn:ymmodelmod}).  To probe the
high-mass regime, only the 257 clusters having $M_{500c} > 4.3\times
10^{14}~\hinv\Msun$ and $0.2 < z < 1.4$ are considered; the fit is performed
around $\Mpivot = 5.5 \times 10^{14}~\hinv\Msun$.  The intrinsic scatter of
the relation is also obtained from the RMS of the residuals.  For the B12
model, the residuals of the fit are plotted against mass in
Figure~\ref{fig:simCalibration}.

For each of the three models, fit parameters are presented in
Table~\ref{tab:abcfits}.  The mass dependence is consistent, in all cases,
with the UPP prediction ($B \approx 0$), and additional redshift dependence is
only present in the Nonthermal20 model.  Only the Adiabatic model is
consistent in its normalization with the UPP value.  This is despite the
explicit calibration, in B12, of the mean pressure profile to the UPP at
$R_{500}$. The origin of this inconsistency is due to the relative shallowness
of the mean pressure profile in B12 compared to the UPP. Thus, the profiles in
B12 have less total signal within $R_{500}$, where ACT is sensitive.

The scaling relation parameters obtained for the Adiabatic model are
sufficiently close to zero (i.e., to the UPP scaling prediction), that we drop
them from further consideration.  While the B12 normalization lies somewhat
below the UPP prediction, we note that even lower normalizations (such as that
found in the Nonthermal20 model) are favored by recent measurements of the SZ
contribution to the CMB angular power spectrum
\citep{dunkley/etal:2011,reichardt/etal:2012}.

We thus proceed to consider quantities derived from each of the UPP, B12, and
Nonthermal20 scaling relation parameter sets.  Mass estimates for the B12 and
Nonthermal20 models are computed for the ACT Equatorial cluster sample as
described in Section~\ref{sec:corrected_masses}, and are presented alongside
the UPP estimates in Table~\ref{tab:masses}.

\renewcommand\mytitle{Scaling relation parameters}
\renewcommand\mycaption{
  Scaling relation parameters, fit to: (i) various SZ
  models (see Section~\ref{sec:calsim}); (ii) the dynamical mass data of
  \cite{sifon/etal:inprep} (Section~\ref{sec:caldyn}); (iii) a cosmological
  MCMC including WMAP data along with the ACT Southern and Equatorial cluster
  samples and dynamical mass
  data (Section~\ref{sec:parameters_dyn}).  Scaling relation parameters $A$,
  $B$, and $C$ are defined as in equation (\ref{eqn:ymmodelmod}), with
  $M_\text{pivot}$ chosen to yield uncorrelated $A$ and $B$.  $A_m$ is the
  normalization parameter corresponding to $\Mpivot = 3\times
  10^{14}~\hinv\Msun$ and may be compared among rows.  Parameters $A_m, B$ and
  $C$ indicate the level of deviation from the predictions based on the
  Universal Pressure Profile of \cite{arnaud/etal:2010} (equations
  (\ref{eqn:uppcyl1}) and (\ref{eqn:uppcyl}); shown for reference).  The
  intrinsic scatter $\sigma_{\rm int}$ is defined as the square root of the
  variance of the observed $\log \ytilde$, in the absence of noise, relative
  to the mean relation defined by equation~\ref{eqn:ymmodelmod}.  The
  parameter $C$ is fixed to 0 when fitting scaling relations to dynamical
  masses.}
\begin{deluxetable*}{l c c c c c c}\tablecaption{\mytitle }

\tablehead{
Description & $M_{\text{pivot}}$ & $A$ & $A_m$ & $B$ & $C$ & $\sigma_\text{int}$\\
 & $(10^{14} h_{70}^{-1}M_\odot)$ &  &  &  &  & 
}
\startdata
Universal Pressure Profile (UPP) & -- & -- & 0 & 0 & 0 & \phantom{0}0.20\tablenotemark{a}\\
Models (\S \ref{sec:calsim}) &  &  &  &  &  & \\
\quad B12 & 5.5 & $\phantom{+}0.111 \pm 0.021$ & $-0.17 \pm 0.06$ & $-0.00 \pm 0.20$ & $-0.04 \pm 0.37$ & $0.20$\\
\quad Nonthermal20 & 5.5 & $-0.003 \pm 0.020$ & $-0.29 \pm 0.06$ & $\phantom{+}0.00 \pm 0.20$ & $\phantom{+}0.67 \pm 0.47$ & $0.21$\\
\quad Adiabatic & 5.5 & $\phantom{+}0.241 \pm 0.020$ & $-0.02 \pm 0.06$ & $-0.08 \pm 0.20$ & $\phantom{+}0.10 \pm 0.43$ & $0.21$\\
Dynamical mass data (\S \ref{sec:caldyn}) &  &  &  &  &  & \\
\quad All clusters & 7.5 & $\phantom{+}0.237 \pm 0.060$ & $-0.21 \pm 0.21$ & $\phantom{+}0.03 \pm 0.51$ & $0$ & $0.31 \pm 0.13$\\
\quad Excluding J0102 & 7.5 & $\phantom{+}0.205 \pm 0.045$ & $-0.11 \pm 0.15$ & $-0.28 \pm 0.35$ & $0$ & $0.19 \pm 0.10$\\
Full cosmological MCMC (\S \ref{sec:parameters_dyn}) &  &  &  &  &  & \\
\quad $\Lambda$CDM model & 7.0 & $\phantom{+}0.079 \pm 0.135$ & $-0.45 \pm 0.19$ & $\phantom{+}0.36 \pm 0.36$ & $\phantom{+}0.43 \pm 0.62$ & $0.42 \pm 0.19$\\
\quad $w$CDM model & 7.0 & $\phantom{+}0.065 \pm 0.153$ & $-0.46 \pm 0.21$ & $\phantom{+}0.36 \pm 0.35$ & $\phantom{+}0.34 \pm 0.65$ & $0.45 \pm 0.20$
\enddata
\tablenotetext{a}{This value, based on the B12 model value, is used for results computed for the UPP scaling relation parameters that also require a value for the intrinsic scatter.}
\tablecomments{\mycaption}
\label{tab:abcfits}\end{deluxetable*}

\subsection{Scaling Relation Calibration from Dynamical Masses}
\label{sec:caldyn}

\newcommand\bdyn{\beta^{\rm dyn}}

\citet[hereafter S12]{sifon/etal:inprep} measure galaxy velocity dispersions
to obtain mass estimates for clusters in ACT's Southern field.  S12 also
present the uncorrected central Compton parameter measurements $\ytilde$ and
the corrected versions $y_0$ obtained as described in
Section~\ref{sec:corrected_masses} and presented in the Appendix.  S12 perform
power law fits of both the uncorrected ($\ytilde$) and corrected ($y_0$)
central Compton parameters to the dynamical masses to establish scaling
relations for those cluster observables.

Here, we fit the full scaling relation of equation (\ref{eqn:ymmodelmod}) to
the dynamical mass data for the 16 $z > 0.3$ clusters from the Southern field
that were detected by ACT and observed by S12. We do not use the scaling
relation as estimated in S12, because the parametrization of the scaling
relation in that study is different. Also, the linear regression that is used
in S12 (the bisector algorithm of \cite{akritas/bershady:1996}) is not suited
to predicting the SZ signal given only the mass, which is the aim in
formulating the cluster abundance likelihood in
Section~\ref{sec:cosmology}.\footnote{The cluster abundance likelihood assumes
  a scaling relation where $\ytilde$ is the dependent variable and takes full
  account of the mass function; in this section we will use the
  likelihood-based approach of \cite{kelly:2007} which includes iterative
  estimation of the distribution of the independent variable.  }

Dynamical masses are estimated in S12 for each cluster based on an average of
60 member galaxy spectroscopic redshifts.  For each cluster, the galaxy
velocity dispersion $S_{\rm BI}$ is interpreted according to the simulation
based results of \cite{evrard/etal:2008}, who find that the dark matter
velocity dispersion $\sigma_{\rm DM}$ is related to the halo mass $M_{200c}$
by
\begin{align}
  \sigma_{\rm DM} & = \sigma_{15} \left( \frac{0.7\times E(z)~M_{200c}}{
    10^{15}~\hinv\Msun} \right)^{\alpha},\label{eqn:mdyndm} 
\end{align}
where $\sigma_{15} = 1082.9 \pm 4~\rm{km~s}^{-1}$ and $\alpha = 0.3361 \pm
0.0026$.  By inverting equation~(\ref{eqn:mdyndm}) and assuming that $S_{\rm
  BI} = \sigma_{\rm DM}$, S12 obtain dynamical estimates, which we will denote
by $M_{200c}^{\rm dyn}$, of the halo mass.  As discussed in S12 and
\cite{evrard/etal:2008}, the systematic bias between galaxy and dark matter
velocity dispersions, $b_v \equiv S_{\rm BI}/\sigma_{\rm DM}$, is believed to
be within 5\% of unity.  To account for this, and any other potential
systematic biases in the dynamical mass estimates, we introduce the parameter
\begin{align}
 \bdyn{} & \equiv \left\langle \frac{M_{200c}^{\rm dyn}}{M_{200c}} \right\rangle.
\end{align}
Based on a velocity dispersion bias of $b_v = 1.00 \pm 0.05$, the equivalent
mass bias is $\bdyn = 1.00 \pm 0.15$.  For the present discussion, we
disregard this bias in order to distinguish its effects from other calibration
issues.  However, in the cosmological parameter analysis of
Section~\ref{sec:cosmology} we include $\bdyn$ as a nuisance parameter and
discuss its impact on the cosmological parameter constraints.

The $\ytilde$ measurements associated with the Southern sample of clusters are
obtained using a filter matched to the noise power spectrum of the Southern
field maps used in \cite{sifon/etal:inprep}.  Thus, while the signal template
is the same, the full form of the filter $\Psi$ and the associated response
function $Q$ differ slightly from the ones used on the Equatorial data.  We
apply the same correction for selection bias that was used by S12, and denote
the corrected values as $\ytilde^{\rm corr}$.

To convert the dynamical masses to $M_{500c}$ values, we model the cluster
halo with a Navarro-Frenk-White profile \citep{nfw:1995} with concentration
parameters and uncertainties obtained from the fits of \cite{duffy/etal:2008}.
For the fit we use a pivot mass of $7.5 \times 10^{14}~\hinv\Msun$, and fix
the parameter $C$ to 0 (otherwise the fit is poorly constrained).  We use the
likelihood-based approach of \cite{kelly:2007} to fit for the intrinsic
scatter along with the parameters $A$ and $B$, given measurement errors on
both independent and dependent variables.  The scatter is modeled, as before,
as an additional Gaussian random contribution to $\log \ytilde$ relative to
the mean relation $\langle \ytilde | m,z\rangle$.

The fit parameters are presented in Table~\ref{tab:abcfits}.  A substantial
contribution to the scatter in the dynamical mass fits comes from the
exceptional, merging cluster ACT-CL J0102$-$4915 \citep[``El Gordo,''
][]{menanteau/etal:2012}: when this cluster is excluded from the fit, the
scatter drops to $0.19 \pm 0.10$, which is more consistent with fits based
solely on models.

\begin{figure}
\includegraphics[width=8.5cm]{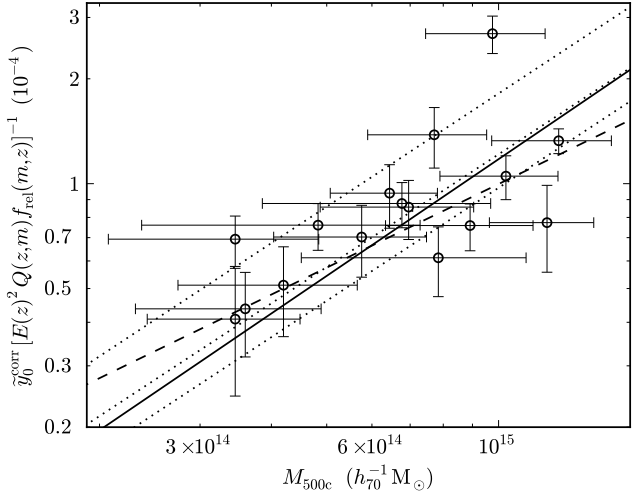}
\caption{Corrected central Compton parameter vs. dynamical mass for the 16
  ACT-detected clusters presented in \cite{sifon/etal:inprep} for the Southern
  sample.  Values on $y$-axis include factor of $E(z)^{-2}$, which arises in
  the derivation of $y_0$ in self-similar models.  The high signal outlier is
  ``El Gordo'' \citep[ACT-CL J0102$-$4915,][]{menanteau/etal:2012}, an
  exceptional, merging system.  The solid line represents the best fit of
  equation (\ref{eqn:ymmodelmod}) with $M_\text{pivot} = 7.5 \times 10^{14}
  ~\hinv\Msun$.  The dashed line is for the fit with J0102$-$4915 excluded.
  Dotted lines, from top to bottom, are computed for scaling relation
  parameters corresponding to the UPP, B12 and Nonthermal20 ($z=0.5$) models.
}
\label{fig:dynCalibration}
\end{figure}

In Figure~\ref{fig:dynCalibration} we plot the cluster SZ measurements against
the dynamical masses, along with the best fit scaling relation.  The scaling
relations from the UPP and from the parameters fit to the B12 and Nonthermal20
models are also shown.  While the fit parameters for the dynamical mass data
are consistent with either the B12 or Nonthermal20 models, the dynamical mass
data lie well below the mean scaling relation predicted by the UPP.  These
results reinforce the need to consider a broad range of possible scaling
relation parameters, within our framework based on the UPP.  We note, however,
that the possibility of a systematic difference between dynamical masses and
other mass proxies must be considered when comparing the parameters obtained
in this section to other results.

\subsection{Completeness Estimate}
\label{sec:completeness}

In this section we estimate the mass, as a function of redshift, above which
the ACT cluster sample within SDSS Stripe 82 (which we will refer to as the
S82 sample) is 90\% complete.  We consider the S82 sample as a whole ($S/N >
4$), and also consider the subsample that has complete high redshift follow-up
($S/N > 5.1$).

For a cluster of a given mass and redshift, we use the formalism of
Section~\ref{sec:fixed_scale} to predict its SZ signal and to infer the
amplitude $\ytil_{\thetaD}$ that we would expect to measure in a map to which
filter $\Psi_{\thetaD}$ has been applied, in the absence of noise and
intrinsic scatter.  We then assume that the cluster occupies a map pixel with
a particular noise level, and consider all possible realizations of the noise
(assumed to be Gaussian) and intrinsic scatter, to get the probability
distribution of observed $\ytil_{\thetaD}$ values.  Applying the sample
selection criteria, we thus obtain the probability of detection for this mass,
redshift, filter scale, and map noise level.

We obtain the total probability that the cluster will be detected at a given
filter scale by averaging over the distribution of noise levels in the
corresponding filtered map.  The distribution of noise levels in the real
filtered maps is used to perform this computation.  To obtain a total
detection probability for the cluster, we take the maximum of the detection
probabilities over the ensemble of filters.  This assumes that noise and
intrinsic scatter are strongly covariant between the filter scales, so that a
cluster that is not detected in the optimal filter is very unlikely to be
detected in a sub-optimal filter.  This assumption may lead to a slight
underestimate of the total detection probability.  The calculation is repeated
to obtain the detection efficiency as a function of mass and redshift.

At redshift $z$, the completeness at mass level $M$ is the average fraction of
all existing clusters with mass greater than $M$ that we would expect to
detect.  The total number of clusters is obtained by integrating the
\cite{tinker/etal:2008} mass function at our fiducial cosmology; the average
number of detected clusters is obtained by integrating the mass function
scaled by the detection efficiency.  Such computations are used to obtain the
mass, as a function of redshift, at which the completeness level is 90\%.

\begin{figure}
\includegraphics[width=8.5cm]{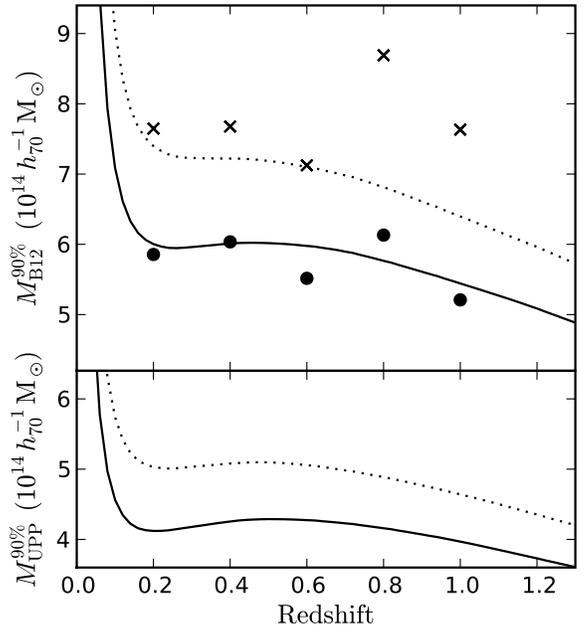}
\caption{Estimate of the mass ($M_{500c}$) above which the ACT cluster sample
  within S82 is 90\% complete (see Section~\ref{sec:completeness}).  Lower
  panel assumes a UPP-based scaling relation with 20\% intrinsic scatter;
  solid line is for $S/N > 4$ (full S82 sample, valid to $z<0.8$), dotted line
  is for the $S/N > 5.1$ subsample (valid to $z \approx 1.4$).  The upper
  panel shows analogous limits, but assuming scaling relation parameters
  obtained for the B12 model (Section~\ref{sec:calsim}).  Circles (crosses)
  are based on filtering and analysis of B12 model clusters for the $S/N > $ 4
  (5.1) cut.  The completeness threshold decreases steadily above $z \approx
  0.6$ because clusters at this mass are easily resolved and the total SZ
  signal, at constant mass, increases with redshift.}
\label{fig:completeness}
\end{figure}

The completeness mass levels are shown in Figure~\ref{fig:completeness}.  Note
that we also show results obtained for the B12 scaling relation parameters.
In this case we also obtained completeness estimates based in part on the
filtered simulated maps.  The central Compton parameters were measured in the
filtered maps, and the S82 noise model was applied to generate a detection
probability for each simulated cluster.  Because of the small number of
sufficiently high mass clusters in the model simulations, we have compensated
for sample variance by reweighting the contribution of each cluster to
correspond to the \cite{tinker/etal:2008} mass function.

In summary, the S82 sample with S/N $> 4$, for which optical confirmation
should be 100\% complete for $z < 0.8$, is estimated to have SZ detection
completeness of 90\% above masses of $M_{500c} \approx 4.5\times
10^{14}~\hinv\Msun$ for $z > 0.2$.  The S82 sample having $S/N > 5.1$, for
which optical confirmation is 100\% complete for $z < 1.4$, is estimated to
have SZ detection completeness of 90\% above masses of $M_{500c} \approx 5.1
\times 10^{14}~\hinv\Msun$ for $z > 0.2$.  (Note in the latter case, however,
that the mass threshold falls steadily beyond redshift of 0.5.)

\subsection{Redshift distribution}

\begin{figure}
\includegraphics[width=8.5cm]{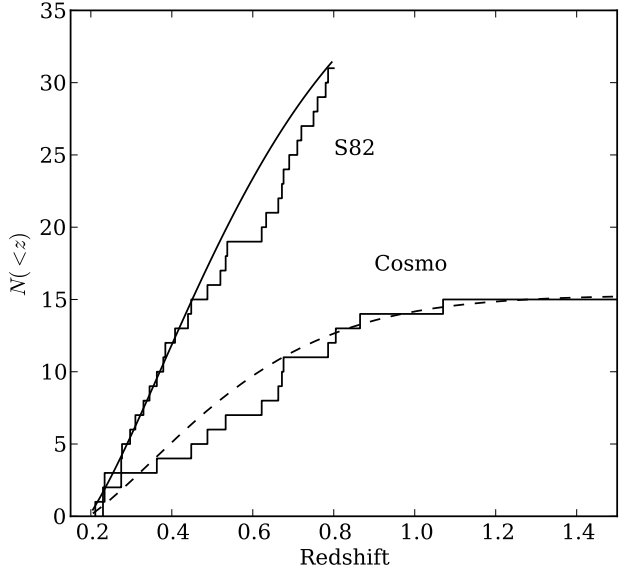}
\caption{Cumulative number counts for two subsamples of the full cluster
  catalog for which confirmation is complete.  The upper lines are data and
  model counts for the S82 sample of clusters having $\ytilde/\delta\ytilde >
  4$ and $0.2 < z < 0.8$.  The lower lines represent the cosmological sample
  of 15 clusters with fixed-scale $\ytilde/\delta\ytilde > 5.1$ and $z > 0.2$.
  The model for the counts is obtained from a maximum likelihood fit, with
  only $\sigma_8$ as a free parameter.  The model includes a full treatment of
  selection effects for the sample under consideration. }
\label{fig:redshift_dist}
\end{figure}

While a full cosmological analysis will be undertaken in
Section~\ref{sec:cosmology}, we briefly confirm the consistency of our cluster
redshift distribution with expectations.  As in the cosmological analysis, we
will select our samples based on the signal to noise ratio of the uncorrected
central decrement $\ytilde \pm \delta\ytilde$ obtained for each cluster using
the filter corresponding to $\thetaD = \thetaFixed$.  We first consider the
S82 clusters that have $\ytilde/\delta\ytilde > 4$, over the redshift range
$0.2 < z < 0.8$.  Secondly we consider the ``cosmological'' sample of
clusters, consisting of 15 clusters with $\ytilde/\delta\ytilde > 5.1$ and $z
> 0.2$.  The cumulative number density as a function of redshift is shown in
Figure~\ref{fig:redshift_dist}.  For each of the two subsamples, we bin the
clusters into redshift bins of width 0.1 and perform a maximum likelihood fit
(assuming Poisson statistics in each bin) to estimate $\sigma_8$.  To
facilitate comparison with cosmological results presented in
Section~\ref{sec:cosmology}, we assume a flat $\Lambda$CDM cosmology with
$\Omega_{\rm m} = 0.25$ and $n_s=0.96$, and fix the scaling relation parameters to
the values associated with the UPP.  Cluster count predictions are obtained starting
from the cluster mass function of \cite{tinker/etal:2008}, and include all
selection effects (intrinsic scatter, noise, and $\ytilde/\delta\ytilde$ cut).
The fits yield $\sigma_8 = 0.782$ for the cosmological sample, and $\sigma_8 =
0.789$ for the S82 sample.  Both of these are consistent with the result of
the full cosmological analysis for the UPP scaling relation.
The best-fit model is a good fit to the data in the sense that the likelihood
score of the data, given the best-fit model, lies near the median of the
likelihood scores for all samples drawn from the best-fit model that have the same
total cluster count as the data.  For the S82 (respectively, cosmological)
sample, 55\% (59\%) of such random samples are less likely.  Each of these
samples is dominated by clusters with spectroscopic redshift estimates, and
thus any features in the distribution cannot be attributed to redshift error.

\section{Comparison with Other Catalogs}
\label{sec:catalog}

In this section we compare the Equatorial cluster catalog and the SZ derived
cluster properties to those obtained by other studies in microwave, X-ray, and
optical wavelengths.  While optical studies have good overlap with our sample
in S82 to $z<0.6$, previous X-ray and SZ survey data include only a small
fraction of the clusters in our sample.  We also examine the question of radio
contamination of cluster decrements through a comparison of extrapolated
fluxes near our cluster positions relative to random positions in the field.

\subsection{Comparison to \Planck{} Early SZ Sample}
\label{sec:compare_planck}

We compare our catalog and our derived cluster properties, to the catalog
presented in the \Planck{} all-sky Early Sunyaev-Zel'dovich cluster sample
\citep[ESZ;][]{planck/esz:2011}.  The ESZ presents 189 clusters, of which 4
lie within the ACT Equatorial footprint, and of which 2 are detected by ACT.
The two clusters detected by both \Planck{} and ACT consist of two of the
three clusters having ACT $Y_{500}$ exceeding the 50\% completeness level of
the ESZ.  (The third, not matched to the ESZ, is RCS2~J2327.4$-$0204.)

The two clusters not detected by ACT are low redshift clusters: Abell 2440 at
z = 0.091 and Abell 119 at z = 0.044.  Based on their integrated X-ray gas
temperature measurements of $3.88\pm0.14$ and $5.62\pm0.12$~keV
\citep{white:2000}, we estimate masses of $\approx 4\times 10^{14}$ and $7
\times 10^{14}~\hinv\Msun$, respectively; these are well below our 90\%
completeness level (Section~\ref{sec:completeness}) at these redshifts.

For the two clusters detected by both \Planck{} and ACT, a summary comparison
of measured cluster properties may be found in Table~\ref{tab:planck}.
MACS~J2135.2$-$0102 is detected by ACT, at low significance, inside S82.
Abell 2355 (ACT-CL~J2135.2+0125) is detected by ACT at high significance
($S/N=9.3$) just outside the S82 region.  The specifics of each case are
discussed below.

\begin{deluxetable*}{l c c c c c c c c}\tablecaption{Comparison of \emph{Planck} and ACT cluster measurements}

\tablehead{
Cluster ID & Redshift & \multicolumn{3}{c}{\Planck} &  & \multicolumn{3}{c}{ACT}\\
 &  & $\theta_{500}$ & $Y_{500}$ & $M_{500}$ &  & $\theta_{500}$ & $Y_{500}$ & $M_{500}$\\
\cline{3-5} \cline{7-9}
 &  & (arcmin) & ($10^{-4}~\textrm{arcmin}^2)$ & ($10^{14}~\hinv~\Msun$) &  & (arcmin) & ($10^{-4}~\textrm{arcmin}^2$) & ($10^{14}~\hinv~\Msun$)
}
\startdata
Early SZ clusters &  &  &  &  &   &  &  & \\
\quad MACS J2135.2-0102 & $0.325$ & 1.6 $\pm$ 1.0 & 9.8 $\pm$ 1.9 & 0.8 $\pm$ 0.8 &   & 3.1 $\pm$ 0.4 & 2.5 $\pm$ 1.2 & 2.7 $\pm$ 1.0\\
\quad Abell 2355 & $0.231$ & 5.1 & 15 $\pm$ 4 & 5.2 &   & 5.3 $\pm$ 0.2 & 14.3 $\pm$ 2.4 & 6.3 $\pm$ 1.3\\
 &  &  &  &  &  &  &  & \\
Intermediate Results &  &  &  &  &   &  &  & \\
\quad Abell 267 & 0.235 & 4.5 $\pm$ 0.2 & 9.3 $\pm$ 2.3 & 3.6 $\pm$ 0.5 &   & 5.4 $\pm$ 0.2 & 13.1 $\pm$ 2.4 & 5.6 $\pm$ 1.2\\
\quad RXC J2129.6+0005 & 0.234 & 4.7 $\pm$ 0.2 & 7.8 $\pm$ 2.0 & 4.3 $\pm$ 0.5 &   & 5.2 $\pm$ 0.2 & 11.4 $\pm$ 2.5 & 5.2 $\pm$ 1.2\\
\quad Abell 2631 & 0.275 & 5.4 $\pm$ 0.6 & 15.5 $\pm$ 2.3 & 9.8 $\pm$ 3.3 &   & 4.8 $\pm$ 0.2 & 11.5 $\pm$ 2.2 & 6.0 $\pm$ 1.3
\enddata
\tablecomments{
Comparison of cluster properties as determined by \Planck{} and by
ACT, for two clusters from the ESZ
\citep[Section~\ref{sec:compare_planck}]{planck/esz:2011} and three
clusters from Intermediate Results
\citep[Section~\ref{sec:compare_weak_lensing}]{planck/intermediateIII:2013}.
For \Planck{}, $\theta_{500}$ is derived from X-ray measurements of
$M_{500}$ for all clusters except MACS J2135.2$-$0102, for which the
angular scale was determined from the SZ data only.  \Planck{} values
for Abell 2355 are corrected to redshift 0.231 as discussed in the text;
$M_{500}$ for this cluster is estimated from X-ray luminosity and thus
carries a large ($\approx$ 50\%) uncertainty.
For the ACT measurements, the angular scale, mass and $Y_{500}$ of
each cluster are obtained simultaneously from the \arone{} data
assuming that the cluster pressure profile is described by the
Standard UPP.
}
\label{tab:planck}\end{deluxetable*}

For ease of comparison, we convert the \Planck{} measurement of the SZ signal
within $5R_{500}$ through the ESZ-provided conversion factor $Y_{500} =
Y_{5R_{500}}/1.81$.  We also use the ESZ value for $\theta_{500}$ (which is
either determined from X-ray luminosity measurements, or from the SZ signal
alone) to obtain an approximate value for $M_{500c}$.  The ESZ analysis makes
use of the ``Standard'' version of the Universal Pressure Profile, which
assumes a self-similar scaling relation \citep[see the Appendix
  of][]{arnaud/etal:2010}.  We thus re-analyze the ACT $\ytilde$ measurements
using the profiles and scaling relation of the Standard UPP to estimate
$\theta_{500}$, $Y_{500}$, and $M_{500c}$.  The results of this analysis
differ only slightly from the results obtained using the full UPP
(Table~\ref{tab:masses}).

For MACS J2135.2$-$0102 ($z=0.329$), X-ray luminosity data was not available
and the ESZ presents the angular scale of the cluster based on SZ data alone.
The scale, $\theta_{500} = 1.6 \pm 1.0\arcmin{}$ is the smallest
$\theta_{500}$ in the ESZ, and corresponds to a very low mass ($\approx
1\times10^{14}$).  Such a mass seems inconsistent with the SZ signal observed
by either \Planck{} or ACT.  Lensed submillimeter galaxies have been observed
near this cluster \citep{ivison/etal:2010}, and the ESZ notes include a
reference to possible point source contamination.  The ACT measurement is
likely to be less contaminated by such emission, since we use only \arone{}
data where dusty sources are comparatively dim.  This may explain the large
difference in $Y_{500}$ inferred by the two telescopes.  Overall this cluster is a
peculiar case, and it is difficult to draw any useful conclusions from the
disagreement of ACT and Planck measurements without more detailed X-ray
information.

Abell 2355 (ACT--CL J2135.2$+$0125) is one of the most significant ACT
detections, and one for which the ACT analysis implies a very high mass.  A
spectroscopic redshift of 0.1244 for this cluster has been obtained by
\cite{kowalski/ulmer/cruddace:1983}, and has subsequently been adopted in both
the MCXC and the \Planck{} ESZ.  However, \cite{sarazin/rood/struble:1982}
identify $z=0.231$ as a more probable spectroscopic redshift, and this value
is adopted by \citet{menanteau/etal:2013}, who find it to be much more
consistent with their photometric estimate of $z = 0.25 \pm 0.01$.
Furthermore, the NED\footnote{NASA Extragalactic Database;
  http://ned.ipac.caltech.edu/.  Retrieved July 15, 2012.} entry for this
cluster refers to an unpublished spectroscopic redshift $z=0.228$ obtained
from 3 galaxies.  In order to compare the \Planck{} and ACT measurements, we
correct the \Planck{} SZ measurements to $z=0.231$.

The cluster angular scale used by \Planck{} is obtained from X-ray luminosity
based masses in the MCXC.  We compute a new mass estimate using the MCXC
scaling relations, correcting the X-ray luminosity for the changes in
luminosity distance and K-correction \citep[according to the $T=5~\rm{keV}$
  tabulation of][]{bohringer/etal:2004}.  The resulting inferred mass is more
than double the estimate obtained for $z=0.1244$.  The corresponding $\theta_{500}$ is
slightly smaller, and so we obtain a crude estimate of the $Y_{500}$ that
\Planck{} might have measured if they had used this angular scale.  From the
inspection of Figure~9 of \cite{planck/esz:2011}, the axis of degeneracy for
the scale and signal measurements lies along $Y_{500} \propto
\theta_{500}^\alpha$ with $\alpha$ in the range of 0.75 (for resolved
clusters) to 1.5 (unresolved clusters).  We compute the new value assuming
$\alpha=1$, and add 20\% error to account for the uncertainty in $\alpha$.
This gives $Y_{500} = (15 \pm 4) \times 10^{-4}~\text{arcmin}^2$.  The ACT
mass and $Y_{500}$ are in good agreement with the X-ray mass and our estimate
of the resulting \Planck{} SZ signal.

Comparison to SZ measurements of three more clusters detected by \Planck{} may
be found in the next section.

\subsection{Comparison to Weak Lensing Masses}
\label{sec:compare_weak_lensing}

In this section we examine weak lensing mass measurements of 4 clusters in the
ACT Equatorial sample.  While one is a high redshift cluster discovered by
ACT, the other three are well-known moderate redshift clusters that have been
observed by \emph{XMM-Newton} and \Planck{}.

\hyphenation{Su-ba-ru}

Weak lensing measurements of ACT-CL J0022.2$-$0036 ($z = 0.81$) are presented
in \citet{miyatake/etal:2013}.  Subaru imaging is analyzed and radial
profiles of tangential shear are fit with an NFW profile.  They obtain a mass
estimate of $M_{500c} = 8.4^{+3.3}_{-3.0} \times 10^{14}~\hinv \Msun$.  While
consistent with our SZ masses for any of the three model scaling relations,
this mass is higher than the one deduced from the UPP scaling relation
parameters and more consistent with the B12 and Nonthermal20 models. The SZ
and lensing mass estimates are also consistent with the dynamical mass
reported for ACT-CL J0022.2$-$0036 in \citet{menanteau/etal:2013}.

The ACT Equatorial sample also includes 3 clusters (A267, A2361, and
RXCJ2129.6+0005) treated by \citet[hereafter PI3]{planck/intermediateIII:2013}
in a comparison of weak lensing mass, X-ray mass proxies, and SZ signal.
X-ray data are obtained from the \emph{XMM-Newton} archive, and weak lensing
masses for the clusters we consider here originate in \citet{okabe/etal:2010}.
Each of these clusters is detected by ACT inside S82 with $S/N > 8$.

Comparing the ACT UPP based SZ masses to the weak lensing masses for these
three clusters we obtain a mean weighted mass ratio of $M_{\rm ACT}^{\rm UPP}
/ M_{\rm WL} = 1.3 \pm 0.2$.  This is consistent with the mass ratio found by
PI3 between X-ray masses and weak lensing masses for their full sample of 17
objects.  For the B12 scaling relation parameters the ratio is $M_{\rm
  ACT}^{\rm B12} / M_{\rm WL} = 1.8 \pm 0.3$.

PI3 also report $Y_{500}$, measured from \Planck{}'s multi-band data using a
matched filter, with the scale ($\theta_{500}$) of the cluster template fixed
using either the X-ray or the weak lensing mass.  For each object we see
agreement in the $Y_{500}$ measurements at the 1 to $2-\sigma$ level, and our
mean ratio is consistent with unity.  For $\theta_{500}$ determined from X-ray
(weak lensing) mass we find weighted mean ratio $Y_{500,\rm ACT}^{UPP} /
Y_{500,Planck} = 0.90 \pm 0.16$ ($0.98 \pm 0.16$).

The angular scales, $Y_{500}$ and $M_{500c}$ obtained by \Planck{} and ACT are
provided in Table~\ref{tab:planck}.  The ACT values are computed for the
Standard version of the UPP, following our treatment of the ESZ clusters in
section \ref{sec:compare_planck}; these results are almost indistinguishable
from those obtained with the full UPP treatment.

\subsection{Comparison to SZA Measurements}

Higher resolution SZ data can be obtained through interferometric
observations.  \citet{reese/etal:2012} present Sunyaev-Zel'dovich Array (SZA)
observations of two ACT Equatorial clusters at 30~GHz.  For the high redshift,
newly discovered cluster ACT-CL~J0022-0036, a GNFW profile is fit to the
30~GHz SZ signal, and is used to infer the cluster mass assuming an NFW
density profile and virialization of the gas.  This yields a mass estimate of
$M_{500c} = 7.3^{+1.0}_{-1.0} \times 10^{14}~ \hinv\Msun$.  For Abell 2631
(ACT-CL~J2337.6+0016), X-ray and SZ data are both used to constrain the
density profile, producing a mass estimate of $M_{500c} = 9.4^{+4.8}_{-2.4}
\times 10^{14}~\hinv\Msun$.  These masses are somewhat higher than the ACT
results for the UPP scaling relation, and are more consistent with the masses
arising from the B12 scaling relation parameters.

\subsection{Optical Cluster Catalogs}

The extensive overlap of the ACT observations described in this work with the
SDSS means that there are a number of existing optically selected cluster
catalogs with which the ACT SZ selected cluster sample can be compared
\citep[see also][]{menanteau/etal:2013}. For all the comparisons described
below, we matched each catalog to the ACT cluster sample using a 0.5\,Mpc
matching radius, evaluated at the ACT cluster redshift.

Several optical cluster catalogues have been extracted from the SDSS legacy
survey \citep[e.g.,][]{goto/etal:2002, miller/etal:2005, koester/etal:2007,
  szabo/etal:2011, wen/etal:2009, wen/etal:2012}. For the purposes of this
comparison, we focus on the MaxBCG catalog \citep{koester/etal:2007} and its
successor the GMBCG catalog \citep{hao/etal:2010}. Both of these catalogs make
use of the color-magnitude red-sequence characteristic of the cluster early
type galaxy population, plus the presence of a Brightest Cluster Galaxy (BCG),
to identify clusters. The MaxBCG catalog contains 13,823 clusters, of which
492 fall within the footprint of the 148\,GHz map used in this work, and is
thought to be $> 90$\% complete and $>90\%$ pure for clusters with $N_{\rm
  gal} > 20$ over its entire redshift range ($0.1 < z < 0.3$). The GMBCG
catalog builds on this work using the entire SDSS DR7 survey area, and is
thought to have $> 95$\% completeness and purity for clusters with richness $>
20$ galaxies and $z < 0.48$. A total of 1903 of the 55,424 GMBCG clusters fall
within the ACT footprint.

We find that the ACT cluster catalog contains 8 clusters in common with
MaxBCG, and 16 clusters in common with the GMBCG catalog. There are no ACT
clusters at $z < 0.3$ in the SDSS DR7 footprint that are not cross matched
with MaxBCG objects; however, there are 2 ACT clusters at $z < 0.48$
(ACT-CL~J0348.6$-$0028 and ACT-CL~J0230.9$-$0024) that are not cross matched
with objects with richness $> 20$ in the GMBCG catalog within the common
area\footnote{To determine the overlap between the ACT maps and various SDSS
  data releases, we make use of the angular selection function from
  \url{http://space.mit.edu/\textasciitilde molly/mangle/download/data.html}
  \citep[see][for details]{swanson/etal:2008, hamilton/tegmark:2004,
    blanton/etal:2005}.}  between the two surveys. ACT-CL~J0348.6$-$0028 ($z =
0.29$) is an optically rich system ($N_{\rm gal} = 56.9 \pm 7.5$; as measured
by \citealt{menanteau/etal:2013}), while ACT-CL~J0230.9$-$0024 is optically
fairly poor ($N_{\rm gal} = 19.9 \pm 4.5$).  In both cases,
\citet{menanteau/etal:2013} find the BCG to have a small offset from the SZ
position ($0.1$\,Mpc for J0348.6 and $0.16$\,Mpc for J0230.9). However,
neither of these objects has a plausible cross-match in the full GMBCG
catalog, although we note that there is a GMBCG cluster
(J057.14850$-$00.43348) at $z = 0.31$ located within a projected distance of
0.65\,Mpc of ACT-CL~J0348.6$-$0028.

We also compared the ACT catalog with that of \citet*{geach/etal:2011}. This
catalog is constructed using the Overdense Red-sequence Cluster Algorithm
\citep*[ORCA;][]{murphy/etal:2011} and is the first optical cluster catalog
available based on the deep ($r \approx 23.5$~mag) SDSS S82 region. It reaches
to higher redshift ($z \approx 0.6$) than the catalogs based on the SDSS
legacy survey data (such as GMBCG), and all of the MaxBCG clusters within a
7\,deg$^2$ test area are re-detected \citep{murphy/etal:2011}. We find that 26
$z < 0.6$ ACT clusters are cross-matched with objects in the
\citeauthor{geach/etal:2011} catalog.  However, there are 24 ACT clusters,
which we have optically confirmed using the S82 data, that were not detected
by \citeauthor{geach/etal:2011}, and most of these (19 objects) are at $z >
0.6$. We also confirmed a further 4 objects at $z > 1$ in the S82 region with
the addition of $K_s$-band imaging obtained at the Apache Point Observatory.
This suggests that the ACT cluster catalog has a higher level of completeness
for massive clusters at high redshift compared to current optical surveys.

\subsection{X-ray Cluster Catalogs}
\label{sec:compare_xray}

Clusters are detected in X-rays through thermal bremsstrahlung emission from
the intracluster gas, and so X-ray selected cluster surveys are
complementary to SZ searches. However, current large area X-ray surveys are
relatively shallow. The REFLEX cluster catalog \citep{bohringer/etal:2004} is
derived from the ROSAT All Sky Survey data \citep{voges/etal:1999} and
overlaps completely with the ACT survey. The full catalog contains a total of
448 clusters reaching to $z = 0.45$, from which 17 objects fall within the
footprint of the ACT Equatorial maps. We find that only five of these clusters
are cross-matched with ACT cluster detections. The detected objects are all
luminous systems, with $L_{\rm X} > 3.3 \times 10^{44}$\,erg\,s$^{-1}$. The
undetected objects are all lower luminosity (hence lower mass) and are at $z <
0.1$, where SZ completeness is low.

As discussed in Section~\ref{sec:compare_planck}, the \Planck{} ESZ relies on
the $M_{500c}$ values presented in the MCXC (derived from the $L_{\rm X}-M$
relation in \citet{arnaud/etal:2010}, which has intrinsic scatter $\approx
50$\%) to constrain the angular scale of detected clusters, and reduce the
uncertainty in $Y_{5R_{500}}$.  The MCXC catalog includes 9 clusters from the
ACT Equatorial cluster sample, and 6 clusters from the ACT Southern cluster
sample.

Comparing the masses we derive from the UPP scaling relation to the MCXC
masses, we obtain a mean ratio $M_{500c}^{\rm{ACT,UPP}} / M_{500c}^{\rm{MCXC}}
= 1.03 \pm 0.19$.  Adding in the Southern clusters (see Appendix) we find a
ratio of $0.83 \pm 0.13$.

\begin{figure}
\includegraphics[width=8.5cm]{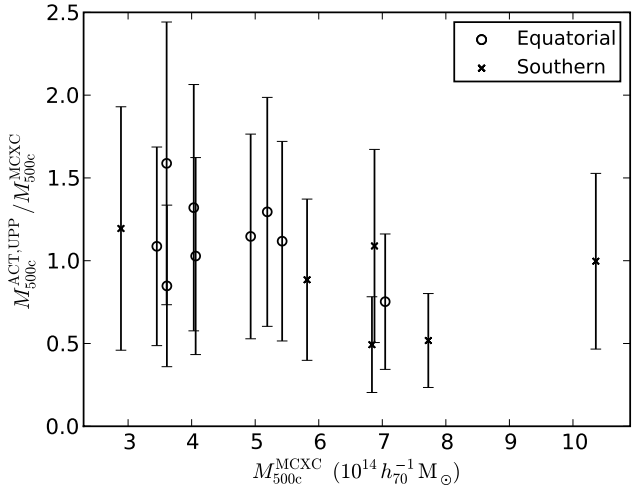}
\caption{Ratio of ACT SZ determined masses to X-ray luminosity based masses
  from the MCXC.  ACT masses assume the UPP scaling relation parameters.
  Error bars on mass include uncertainty from the ACT SZ measurements, and
  50\% uncertainty on MCXC masses.  The weighted mean ratio is $1.03 \pm 0.19$
  for the Equatorial clusters and $0.83 \pm 0.13$ for the full sample.}
\label{fig:mcxcMasses}
\end{figure}

While the Equatorial cluster result is consistent with the UPP scaling
relation, the full sample prefers higher masses.  For the B12 scaling relation
parameters and the full sample we find $M_{500c}^{\rm{ACT,B12}} /
M_{500c}^{\rm{MCXC}} = 1.12 \pm 0.17$.  For the Equatorial sample alone, this
ratio is $1.41 \pm 0.27$.  The mass ratios are shown in
Figure~\ref{fig:mcxcMasses}.

\subsection{Radio Point Sources}

Relative to the field, galaxy clusters are observed to contain an excess of
radio point sources \citep[e.g., ][]{cooray/etal:1998}. Studies of potential
SZ signal contamination by radio sources have been carried out at frequencies
below 50 GHz \cite[e.g., ][]{lin/etal:2009}.  There are, however, few such
studies at 150 GHz.  Recently \cite{sayers/etal:2013} used Bolocam at 140 GHz
to study possible radio source contamination in 45 massive clusters. They
found that the SZ signals from only 25\% of the sample were contaminated at a
level greater than 1\%.  The largest contamination observed was 20\% of the SZ
signal.  We study this phenomenon in the ACT cluster sample, using the FIRST
catalog of flux densities at 1.4\,GHz \citep{white/etal:1997}.  The lower flux
density limit of the catalog is at most 1\,mJy, and overlaps with the ACT
Equatorial field between R.A.  of \hmin{21}{20} and \hmin{3}{20}.

For each source in the FIRST catalog, we extrapolate the flux density to
\arone{} using flux-dependent spectral indices computed based on a stacking
analysis in the ACT \arone{} and \artwo{} maps. 
For the sources of interest here, the spectral index ranges from approximately
-0.5 (at $S_{1.4} = 10$\,mJy) to -0.8 (at $S_{1.4} = 100$\,mJy).
While a single spectral index is inadequate for describing the spectral behavior
of radio sources over a decade in frequency, the extrapolation based on stacked 
ACT data is a reasonable technique for the purposes of predicting flux densities 
at ACT frequencies. Given the few sources found per cluster, the error in the 
extrapolation will be dominated by intrinsic scatter in \arone{} flux densities
($\le 1$\,mJy) corresponding to a given $S_{1.4}$ range. This analysis does not
take into account the probability of source orientation dependence that results in
significantly greater observed flux density, as in the case of blazars. 
For each of the 63 galaxy clusters lying in the FIRST survey area, we take all sources
within 2\arcmin{} of the cluster position and sum the predicted flux at
\arone{}.  We find a mean flux density of $S_{148} = 0.94$\,mJy, and the 9
most potentially contaminated clusters have flux density between 2.5 and
4.2\,mJy.  Converting these to a peak brightness in CMB temperature units
using the \arone{} beam, we obtain contamination of 11\,\uk{} on average and
50\,\uk{} at worst.  These levels are substantially higher than would be
expected if total radio flux were not correlated with cluster position.  For
random positions, the contamination is only $S_{148} = 0.2$\,mJy on average,
and 1\% of locations have total flux exceeding 2.4\,mJy.

For the purposes of inferring masses, we are interested in the impact the
point sources may have in the measurement of the uncorrected central
decrement.  To explore this, we create a simulated ACT map of the sources
using their inferred flux densities at \arone{} and the ACT beam shape.  This
map is then filtered with the same matched filter $\Psi_\thetaFixed$ described in
Section~\ref{sec:fixed_scale}.  We obtain a prediction for the contamination
level of each cluster by taking the maximum value in the filtered map within
2\arcmin{} of the cluster position.  More than half of the clusters have
predicted contamination less than 13\,\uk{} and 90\% have less than 30\,\uk{}
(which is smaller than the typical measurement uncertainty).  The worst
contamination prediction is associated with ACT-CL J0104.8$+$0002, at a level
of 59\,\uk{} (corresponding to 35\% of the signal strength, and larger than
the cluster noise level of 41\,\uk{}).

We emphasize that the contamination levels given here are extrapolations
based on 1.4 GHz flux, and thus the contamination at \arone{} for any
particular source associated with a cluster may vary somewhat from the values
stated.
Since such contamination is difficult to model in detail without knowing the
spectral indices of individual sources, and since the rate of significant
contamination seems to be quite low, we make no attempt to correct for this
effect.  This may introduce a small, redshift dependent bias into scaling
relation parameters obtained from these SZ data.

\section{Cosmological Interpretation}
\label{sec:cosmology}

Cluster count statistics, such as the number density of clusters above some
limiting mass, are particularly sensitive to the total matter density
($\Omega_{\rm m}$) and the amplitude of density fluctuations (as parametrized
by, e.g., $\sigma_8$).  Because SZ selected cluster samples can reach to
arbitrarily high redshift, they also probe parameters, such as the dark energy
equation of state parameter $w$, that describe the recent expansion history.

In order to constrain cosmological parameters, we incorporate our sample into
a Bayesian analysis and compute the posterior likelihood of cosmological
parameters given the cluster data.  We begin by outlining the formalism used
to model the probability of our data given values of cosmological and scaling
relation parameters.  We then demonstrate the constraints achieved by
combining the ACT cluster data with other data sets.

\subsection{Likelihood Formalism}
\label{sec:like_formalism}

\newcommand\true{{\rm tr}}
\newcommand\yob{\widetilde{y_0}}
\newcommand\ytr{\yob^\true}
\newcommand\mob{\widetilde{m}}
\newcommand\mtr{\mob^\true}
\newcommand\mdyn{\mob^{\rm dyn}}
\newcommand\ztr{z^\true}

\newcommand\vvect{\mathbf{x} }
\newcommand\Ntot{N_{\rm tot} }

\newcommand\thetav{\boldsymbol\theta}
\newcommand\psiv{\boldsymbol\psi}

In this section we outline a formalism for determining the Bayesian likelihood
of cosmological and scaling relation parameters given the ACT cluster
measurements, including redshift, SZ, and dynamical mass information, when
available.  Our approach follows previous work in developing an expression for
the probability of the cluster measurements based on the application of
Poisson statistics to finely spaced bins in the multi-dimensional space of
cluster observables.  Such approaches naturally support non-trivial sample
selection functions, the self-consistent calibration of scaling relations
between object properties (i.e., between mass, SZ signal, and redshift), and
missing data (i.e., the absence of independent mass measurements for some
detected clusters).  This approach to the comparison of a sample of detected
objects to a number density predicted by a model is described by
\citet{cash:1979}; \cite{mantz/etal:2010} present a useful general formalism
for dealing with cluster data, applying it to X-ray mass and luminosity
measurements to obtain cosmological constraints while calibrating the
mass--luminosity scaling relation.  The SZ studies of \cite{sehgal/etal:2011}
and SPT \cite[e.g., ][]{benson/etal:2013} have applied similar techniques to
SZ and X-ray data.  Here, we develop an approach that pays particular
attention to the uncertainties in the observed quantities, and in which the SZ
signal is interpreted through the \PBAA{} approach
(Section~\ref{sec:fixed_scale}).

We assume that the cluster data consist of a sample of confirmed clusters, and
that for each cluster an uncorrected central Compton parameter $\yob \pm
\delta\yob$ and redshift $z \pm \delta z$ have been measured.  The $\yob$ are
obtained from maps filtered with $\Psi_\thetaFixed$.  In some cases, clusters
may also have dynamical mass measurements, $M_{500c}^{\rm dyn}$ (see
Section~\ref{sec:caldyn}).  To parametrize these measurements in a way that is
independent of cosmology, we define the observed dynamical mass parameter,
$\mob \equiv E(z)M_{500c}^{\rm dyn} / (10^{14}~\hinv\Msun)$.

To compare the observed sample to predicted cluster number densities, we
consider all clusters to possess an intrinsic uncorrected central Compton
parameter $\ytr$, redshift $\ztr$, and mass parameter $\mtr$.  These true
intrinsic quantities represent the values one would measure in the absence of
any instrumental noise or astrophysical contamination (from, e.g., the CMB).
For the SZ signal, $\ytr$ should be thought of as the measurement we would
make if we applied our filter $\Psi_\thetaFixed$ (i.e., the fixed-scale filter
matched to the noise spectrum of our maps) to a map from which all noise and
astrophysical contamination had been removed.  The true cluster mass
parameter, $\mtr \equiv M_{500c}~h_{70} E(\ztr)$ with $E(\ztr)$ computed for
the true cosmology, is representative of the halo mass rather than the mass
inferred from an observational proxy (such as galaxy velocity dispersion).

We proceed by obtaining the number density of galaxy clusters in the space of
true cluster properties $\ytr$, $\mtr$, and $\ztr$.  As described in
Section~\ref{sec:corrected_masses}, we make use of the cluster mass function
of \cite{tinker/etal:2008} to predict, for cosmological parameters $\thetav$,
the number of clusters per unit redshift and unit mass within the area of the
survey:
\begin{align}
 n(\mtr,\ztr|\thetav) & = d^2N(<\mtr,\ztr)/d\ztr d\mtr.
\end{align}
Here we use the notation $n(\alpha|\beta)$, more commonly used with
probability densities, to indicate the conditional distribution of clusters
with respect to variables $\alpha$ when variables $\beta$ are held fixed.

Given $\mtr$, $\ztr$, and scaling relation parameters $\psiv =
(A_m,B,C,\sigma_{\rm int})$, the conditional distribution of Compton parameter values,
$P(\ytr|\mtr,\ztr,\psiv)$, is specified by equations (\ref{eqn:ymmodelmod})
and (\ref{eqn:ymmodelscatter}).  Summarizing these equations in our current
notation, $\log \ytr$ is normally distributed,
\begin{align}
  \log \ytr & \sim N(\log \ytilde(\mtr,\ztr,\psiv{}); \sigma_{\rm int}^2),
\end{align} 
and the mean relation is described by
\begin{align}
  \ytr(\mtr,\ztr,\psiv{}) & = 10^{A_0+A_m} E^2(\ztr) (\mtr)^{1+B_0+B} \times\nonumber\\
 & ~~~~~~
  Q\left[\left(\frac{1+\ztr}{1.5}\right)^C\theta_{500}/(\mtr)^{C_0}\right]
  \times\nonumber\\
 & ~~~~~~  \frel(\mtr,\ztr).
\end{align}

This may be used to compute the number
density of clusters in the full space of true cluster properties:
\begin{align}
  n(\ytr,\mtr,\ztr|\thetav,\psiv) & = 
    P(\ytr|\mtr,\ztr,\psiv) \times \nonumber\\
    & ~~~~ n(\mtr,\ztr|\thetav).
\end{align}

In order to compare our observed sample to the model, it is necessary to
properly account for the sample selection function, and for the effects of
measurement uncertainty.  This is especially important because our selection
function depends explicitly on $\delta \yob$, and the mass function (and thus
the cluster density as a function of $\yob$) is very steep.  This full
treatment of uncertainty allows us to include, in the same analysis, regions
of the map that have quite different noise levels.

In general, we imagine that the cluster observables $\vvect = (\yob,\mob,z,
\delta \yob, \delta \mob, \delta z)$ are related to the true cluster
properties $\ytr, \mtr,\ztr$ by some probability distribution $P(\vvect |\ytr,
\mtr, \ztr,\bdyn)$.  In addition to describing the scatter of each variable
about its true value, and accounting for the dynamical mass bias through the
parameter $\bdyn$ (which is defined in Section~\ref{sec:caldyn}), this
distribution also includes a description of what measurement uncertainties we
are likely to encounter for given values of the true cluster properties.  For
example, while most of our sample have spectroscopic redshifts for which the
measurement uncertainty is negligible, some clusters (particularly at high
redshift) have photometric redshift estimates with relatively large
uncertainties ($\delta z \approx 0.06$).  Although it may seem awkward to
worry about measurement uncertainty for clusters that have not been detected,
it is necessary, formally, to account for the distribution of errors if the
sample is defined based on observed cluster properties.  In our particular
case, the full probability distribution factors to:
\begin{align}
P(\vvect |\ytr, \mtr, \ztr, \bdyn)
& = P(\yob|\ytr,\delta \yob)P(\delta \yob) \times \nonumber\\
& ~~ P(\mob|\mtr, \delta \mob, \bdyn) P(\delta \mob) \times \nonumber\\
& ~~ P(z|\ztr, \delta z) P(\delta z | \ztr).
\end{align}

The expression above encodes the following properties of the ACT observations:
\begin{itemize}
\item For a given cluster, the measurements of $\yob$, $\mob$, and $z$ are
  independent; we do not expect any covariance in the errors.  In practice we
  assume cluster observables are normally distributed about their true values
  (with the exception of $\mob$; see next point).
\item The probability distribution of $\mob$ includes the effects of the
  dynamical mass bias parameter, $\bdyn$; specifically the measured dynamical
  mass parameter $\mob$ is expected to be normally distributed about mean
  $\bdyn \mtr$ with standard deviation $\delta \mob$.
\item The distribution of $\yob$ errors, $P(\delta \yob)$, is independent of
  all true cluster properties.  In practice $P(\delta \yob)$ is obtained from
  the histogram of the noise map.
\item The distribution of $\mob$ errors, $P(\delta \mob)$, is independent of
  true cluster properties.  In practice the uncertainty in the dynamical
  masses is related to the number of galaxies used for the velocity dispersion
  measurements.  In any case, we do not need to understand this distribution
  in detail, because observed mass is not a factor in sample selection.
\item The distribution of $z$ errors, $P(\delta z | \ztr)$, may depend on the
  true cluster redshift.  While spectroscopically measured redshifts are
  available for many sample clusters, high redshift clusters are more likely
  to have only photometric redshift estimates.  In practice, this distribution
  only enters when computing the prediction for the \emph{total} number of
  clusters observed within some volume.  For suitably chosen sample redshift
  limits, the details will not matter (see discussion below).
\end{itemize}

For a sample selected based on signal to noise ratio threshold $s$ and
observed redshift range $[z_A,z_B]$, we define the selection function
$S(\yob,z,\delta \yob)$ to take value unity when $\yob/\delta \yob > s$ and $z
\in [z_A,z_B]$, and to take value zero otherwise.  Then the predicted number
density in the 6-dimensional space of observables $\vvect$ is
\begin{align}
  n(\vvect | \thetav, \psiv, \bdyn) & = S(\yob,z,\delta \yob)\times
  \nonumber\\
& \int d\ytr d\mtr d\ztr P(\vvect|\ytr,\mtr,\ztr, \bdyn) \times
  \nonumber\\
& ~~~~ n(\ytr,\mtr,\ztr|\thetav,\psiv).
\label{eqn:nxthetapsi}
\end{align}

This cluster density function may be used to evaluate the extent to which the
observed cluster data are consistent with the model $\thetav,\psiv,\bdyn$.
This is achieved, as in \cite{cash:1979} by imagining a very fine binning in
the space of observables.  We take bins indexed by $\alpha$ centered at
$\vvect_\alpha$ and having (6-dimensional) volume $V_\alpha$.  In the limit of
very fine bins, the total predicted counts in bin $\alpha$ is
well-approximated by $N_\alpha(\thetav, \psiv,\bdyn) \approx V_\alpha
n(\vvect_\alpha | \thetav, \psiv, \bdyn)$.  Furthermore, the number of
observed clusters in bin $\alpha$, denoted $c_\alpha$, is either 0 or 1.

Letting $D$ denote the set of bins in which a cluster has been observed (i.e.,
the $\alpha$ where $c_\alpha = 1$), we assume Poisson statistics in each bin,
and obtain the probability of the data given the model parameters:
\begin{align}
  P(\{\vvect_i\} | \thetav,\psiv,\bdyn)
  & = P(\left\{c_\alpha\right\}|\thetav,\psiv,\bdyn) \nonumber\\
  & = \prod_\alpha e^{-N_\alpha(\thetav,\psiv,\bdyn)}
  N_\alpha(\thetav,\psiv,\bdyn)^{c_\alpha}\nonumber \\
 & = e^{-\Ntot(\thetav,\psiv,\bdyn)} \prod_{\alpha \in D} V_\alpha~
    n(\vvect_\alpha).~~~
    \label{eqn:Pcthetapsi}
\end{align}
We have defined
\begin{align}
  \Ntot(\thetav,\psiv,\bdyn) & =
  \sum_\alpha N_\alpha(\thetav,\psiv,\bdyn)\nonumber\\
  & = \int d^6 x ~n(\vvect|\thetav,\psiv,\bdyn),
\end{align}
the total number of clusters that the model predicts will be detected.

In equation (\ref{eqn:Pcthetapsi}), the product over occupied bins is only
sensitive to the values of the density function at the locations of the
detected clusters.  The volume elements $V_\alpha$ depend on the data and the
binning, but not on the cosmological or scaling relation parameters.  They
will thus cancel exactly in any ratio of probabilities comparing different
models.  So we may write the likelihood of parameters $\thetav,\psiv,\bdyn$
given the cluster data $\{\vvect_i\}$ as
\begin{align}
  L(\thetav,\psiv,\bdyn|\{\vvect_i\}) & =
  P(\{\vvect_i\}|\thetav,\psiv,\bdyn)\nonumber\\
  & \propto e^{-\Ntot(\thetav,\psiv,\bdyn)} \times \nonumber\\
  & ~~~~~~ \prod_{i} n(\vvect_i| \thetav,\psiv,\bdyn),
  \label{eqn:Lthetapsic}
\end{align}
where $i$ indexes the clusters in the sample.

When evaluating this expression in practice, we face the two related problems
of computing the total cluster count prediction $\Ntot$, and of computing the
number density $n(\vvect_i|\thetav,\psiv,\bdyn)$ for each cluster in the
sample.  Certain simplifications make possible the efficient computation of
these quantities.

When computing $n(\vvect_i|\thetav,\psiv,\bdyn)$, we approximate $P(\delta
z|\ztr)$ as being constant over the range of $\ztr$ under consideration.  This
is acceptable because the integral over $\ztr$ is restricted to the vicinity
of the observed cluster redshift $z_i$ by the distribution $P(z_i|\ztr,\delta
z)$.  We thus replace $P(\delta z_i|\ztr)$ with $p_i \equiv P(\delta
z_i|\ztr)|_{\ztr=z_i}$.  The $p_i$ can then be factored out of the integral
over true cluster properties in equation~(\ref{eqn:nxthetapsi}).  When
evaluating the likelihood in equation~(\ref{eqn:Lthetapsic}), these $p_i$
contribute a constant multiplicative factor that is independent of parameters
$\thetav,\psiv,\bdyn$.  So their contribution is irrelevant, and like the
$V_\alpha$ the $p_i$ may be dropped from the likelihood expression.

When computing $\Ntot$, the procedure is simplified by first integrating over
$\mtr$ to obtain the distribution $n(\ytr,\ztr|\thetav,\psiv)$.  The integrals
over $\mob$ and $\delta \mob$ are trivial to perform (independent of the form
of $P(\delta m)$ and the value of $\bdyn$), because the selection function
does not depend on $\mob$ or $\delta \mob$.  We may then write
$\Ntot(\thetav,\psiv)$, dropping the dependence on $\bdyn$.  The integral over
$\delta \yob$ can be accomplished with $P(\delta \yob)$ based on the noise
map.  Note that this is essential to properly predict the total cluster count
in cases where the map area under consideration includes a variety of local
noise levels.

It is necessary to consider the impact of $P(\delta z|\ztr)$ in the evaluation
of $\Ntot$.  The number of clusters within an observed redshift range $z \in
[z_A,z_B]$ will be approximately equal to the number of clusters with true
redshift $\ztr \in [z_A,z_B]$.  The difference between these two numbers may
be interpreted as the result of clusters ``scattering'' over the redshift
boundary due to measurement uncertainty.  The magnitude of this effect is
related to the noise level $\delta z$ and the steepness of the distribution
$n(z)$ at the boundaries.

Two properties of the ACT cluster sample allow us to avoid dealing with the
details of $P(\delta z|\ztr)$.  The first is that we have spectroscopic
redshift measurements for all but a small number of high redshift clusters.
This means that $P(\delta z|\ztr)$ strongly favors the case of $\delta z
\approx 0$ at the low redshift sample boundary.  Second, our upper redshift
limit, which arises based on the depth of optical confirmation observations,
is at $z=1.4$.  Clusters at $z=1.4$ are sufficiently rare that uncertainty in
$\delta z$ does not greatly affect the total number of clusters within the
full survey volume.  For example, at $z = 1.4$ the predicted cluster number
density has fallen off significantly and the effects of redshift uncertainty
can only affect the total predicted counts by less than 1\%.  This is well
below the error due to sample variance.  So provided we use a cluster sample
for which SZ candidate follow-up (optical/IR confirmation) is complete out to
at least $z = 1.4$, we may disregard redshift uncertainty when computing
$\Ntot$, and integrate over $\ztr$ instead of $z$ and $\delta z$.

For a more complicated data set (involving a large number of photometrically
obtained redshifts), the distribution $P(\delta z|\ztr)$ could be estimated
based on the redshift error data in hand.

Finally, in the case that mass data are not available for some or all of the
clusters, we can simply integrate over our ignorance of $\mob$.  The model and
scaling relation parameters give cluster density prediction, marginalized over
the missing data, of
\begin{align}
  n(\yob,z,\delta \yob, \delta z|\thetav, \psiv) & = 
  \int d(\delta \mob) \int dm ~n(\vvect|\thetav,\psiv,\bdyn).
\end{align}
The likelihood expression for the mixed case is
\begin{align}
  L(\thetav,\psiv,\bdyn|\{\vvect_i\})
  & \propto e^{-\Ntot(\thetav,\psiv)} \times \nonumber\\
  & ~~~~\prod_{i \in M} n(\vvect_i| \thetav,\psiv,\bdyn) \times \nonumber\\
  & ~~~~\prod_{i \notin M} n(\yob_i,z_i,\delta \yob_i, \delta z_i|\thetav,
  \psiv),
  \label{eqn:Lthetapsimy}
\end{align}
where $M$ denotes the subset of clusters that have mass measurements.

\subsection{Parameter Constraints for Fixed Scaling Relations}
\label{sec:parameters_fixed}

\newcommand\LCDM{$\Lambda$CDM}
\newcommand\wCDM{$w$CDM}
\newcommand\ACTCL[1]{ACTcl#1}
\newcommand\Omegam{\Omega_{\rm m}}
\newcommand\Omegab{\Omega_{\rm b}}
\newcommand\Omegac{\Omega_{\rm c}}

In this section we obtain cosmological parameter constraints by combining the
Equatorial cluster sample with various external data sets, with the SZ scaling
relation parameters fixed to values indicated by the UPP prescription, and by
the B12 and Nonthermal20 models.  For each case, we fix the scaling relation
parameters $(A_m,B,C,\sigma_{\rm int})$ to the values given in
Table~\ref{tab:abcfits}, and do not account for any uncertainty in these
parameters.  These results should thus be viewed as illustrating the potential
constraint achievable from these cluster data, should the uncertainty on the
scaling relations be reduced.  The constraints also demonstrate the
sensitivity of cosmological parameter estimates to the different physical
assumptions entering each model.

The posterior distributions of cosmological parameters are obtained through
Markov Chain Monte Carlo (MCMC) sampling, driven by the CosmoMC software
\citep{lewis/bridle:2002}, with matter power spectra and CMB angular power
spectra computed using the CAMB software \citep{lewis/etal:2000}.  The ACT
Equatorial data contribute to the likelihood according to
equation~(\ref{eqn:Lthetapsimy}).  The cluster sample includes the 15 objects
(identified in Table~\ref{tab:clusters}) that lie inside sample boundaries
defined by redshift range $0.2 < z < 1.4$ and signal to noise ratio cut
$\ytilde/\delta \ytilde > 5.1$.  As discussed in
Section~\ref{sec:confirmation}, the cut on $\ytilde/\delta \ytilde$
corresponds to the level above which the optical confirmation campaign has
successfully confirmed or falsified all SZ detections.

Posterior distributions of parameters for data sets that do not include ACT
cluster information are obtained from chains released with the WMAP seven-year
results \citep[WMAP7;][]{komatsu/etal:2011}.  In some cases, chains also
incorporate data from Baryon Acoustic Oscillation experiments
\citep[BAO;][]{percival/etal:2010}, and Type Ia supernovae
\citep[SNe;][]{hicken/etal:2009}.  When not constrained by CMB information,
ACT cluster data are combined with results from big bang nucleosynthesis
\citep[BBN;][]{hamann/etal:2008} to constrain the baryon fraction and with
Hubble constant measurements \citep[denoted H0;][]{riess/etal:2009}.  The H0
measurements are also used to restrict the parameter space of \wCDM{} model
studies, emphasizing the role clusters can play in such cases.

We first consider a $\Lambda$CDM model with 7 free parameters representing the
baryon and cold dark matter densities ($\Omega_{\rm b}$ and $\Omega_{\rm c}$),
the angular scale of the sound horizon ($\theta_A$), the normalization ($A_s$)
and spectral index ($n_s$) of the matter power spectrum, the optical depth of
reionization ($\tau$), and the SZ spectral amplitude ($A_{\rm SZ}$).  For
simplicity, the SZ spectral amplitude is not explicitly tied to the cluster
scaling relation parameters.

\begin{figure}
\includegraphics[width=8.5cm]{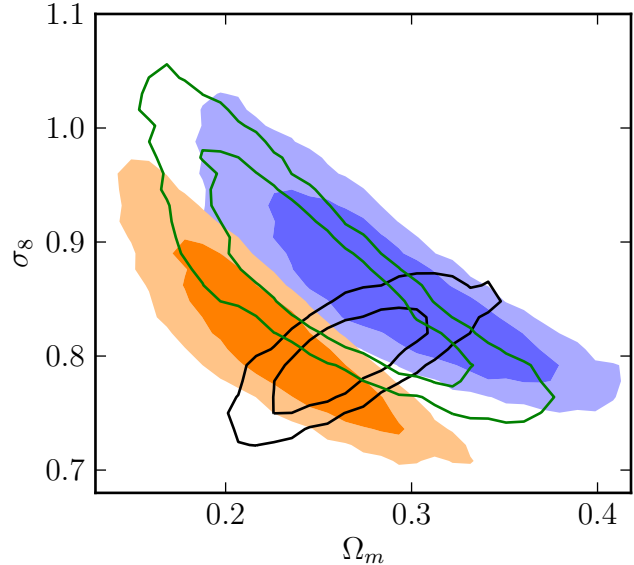}
\caption{Constraints on \LCDM{} cosmological parameters from WMAP7 (black line
  contours) and \ACTCL{}+BBN+H0 (without CMB information).  Contours indicate
  68 and 95\% confidence regions.  ACT results are shown for three scaling
  relations: UPP (orange contours), B12 (green lines), Nonthermal20 (violet
  contours).  While any one scaling relation provides an interesting
  complement to CMB information, the results from the three different scaling
  relations span the range of parameter values allowed by WMAP measurements. }
\label{fig:params_solo2}
\end{figure}

The potential for our cluster data to constrain $\Omegam$ and $\sigma_8$ is
demonstrated in Figure~\ref{fig:params_solo2}.  In order to emphasize the
impact of cluster studies, we have computed the parameter likelihood for the
ACT cluster data in combination with BBN and H0 only (these constrain the
baryon density to $\Omegab h^2 = 0.022 \pm 0.002$ and the Hubble constant to
$H_0 = 73.9 \pm 3.6~\textrm{km~s~Mpc}^{-1}$).  We plot the marginalized
two-dimensional distribution of parameters $\sigma_8$ and $\Omegam$, with
contours showing the 68\% and 95\% confidence regions.  The ACT cluster
constraints, without CMB information, are seen to nicely complement the
results from WMAP7.  But the variation in the constraints between models shows
the degree to which uncertainty in the cluster physics diminishes the
constraining power.  The marginalized parameter values are provided in
Table~\ref{tab:params_lcdm}.

\begin{deluxetable*}{l c c c c c}\tablecaption{Cosmological parameter constraints for the $\Lambda$CDM model.}

\tablehead{
 & \multicolumn{5}{c}{Parameter ($\Lambda$CDM)}\\
Data set & $\Omega_{\rm c}h^2$ & $\Omega_{\rm m}$ & $\sigma_8$ & $h$ & $\sigma_8 (\Omega_{\rm m}/0.27)^{0.3}$
}
\startdata
Without ACT Cluster Data & \\
\quad WMAP7 & 0.111 $\pm$ 0.006 & 0.266 $\pm$ 0.029 & 0.801 $\pm$ 0.030 & 0.710 $\pm$ 0.025 & 0.797 $\pm$ 0.053\\
\quad WMAP7 + BAO + H0 & 0.112 $\pm$ 0.003 & 0.272 $\pm$ 0.016 & 0.809 $\pm$ 0.024 & 0.704 $\pm$ 0.014 & 0.811 $\pm$ 0.034\\
Fixed Scaling Relations (\S \ref{sec:parameters_fixed}) & \\
\quad BBN + H0 + \ACTCL{(B12)} & 0.115 $\pm$ 0.024 & 0.252 $\pm$ 0.047 & 0.872 $\pm$ 0.065 & 0.741 $\pm$ 0.036 & 0.848 $\pm$ 0.032\\
\quad WMAP7 + \ACTCL{(UPP)} & 0.107 $\pm$ 0.002 & 0.250 $\pm$ 0.012 & 0.786 $\pm$ 0.013 & 0.720 $\pm$ 0.015 & 0.768 $\pm$ 0.015\\
\quad WMAP7 + \ACTCL{(B12)} & 0.114 $\pm$ 0.002 & 0.285 $\pm$ 0.014 & 0.824 $\pm$ 0.014 & 0.693 $\pm$ 0.015 & 0.837 $\pm$ 0.017\\
\quad WMAP7 + \ACTCL{(Non)} & 0.117 $\pm$ 0.002 & 0.303 $\pm$ 0.016 & 0.839 $\pm$ 0.014 & 0.680 $\pm$ 0.015 & 0.869 $\pm$ 0.018\\
Dynamical Mass Constraints (\S \ref{sec:parameters_dyn}) & \\
\quad BBN + H0 + \ACTCL{(dyn)} & 0.141 $\pm$ 0.042 & 0.301 $\pm$ 0.082 & 0.975 $\pm$ 0.108 & 0.737 $\pm$ 0.037 & 0.999 $\pm$ 0.130\\
\quad WMAP7 + \ACTCL{(dyn)} & 0.115 $\pm$ 0.004 & 0.292 $\pm$ 0.025 & 0.829 $\pm$ 0.024 & 0.688 $\pm$ 0.021 & 0.848 $\pm$ 0.042\\
\quad WMAP7 + \ACTCL{(dyn)} + BAO + H0 & 0.114 $\pm$ 0.003 & 0.282 $\pm$ 0.016 & 0.829 $\pm$ 0.022 & 0.696 $\pm$ 0.013 & 0.840 $\pm$ 0.031
\enddata
\tablecomments{
Numbers indicate the mean and standard deviation of the marginalized
posterior distribution.  \ACTCL{} results for scaling relations based
on the Universal Pressure Profile (UPP), and the B12 and
\emph{Nonthermal20} (Non) models do not include marginalization over
scaling relation uncertainties (Section~\ref{sec:parameters_fixed}).
\ACTCL{(dyn)} results use Southern and Equatorial cluster data,
including dynamical mass measurements for the Southern clusters
(Section~\ref{sec:parameters_dyn}).
}
\label{tab:params_lcdm}\end{deluxetable*}

In a \wCDM{} model (which differs from the \LCDM~model in that the equation of
state parameter $w$ may deviate from $-1$), cluster counts are sensitive to
the effect of dark energy on the expansion rate of the recent universe.  As
shown in Table~\ref{tab:params_wcdmx}, the ACT cluster data provides slightly
improved constraints on $\sigma_8$ relative to WMAP7 alone; but the true power
of the cluster data is to break degeneracies between $\sigma_8$, $\Omegam$,
and $w$.  We present composite parameters defined by
$\sigma_8(\Omegam/0.27)^{0.4}$ and $w(\Omegam/0.27)$ to express these
improvements.

\begin{figure}
\includegraphics[width=8.5cm]{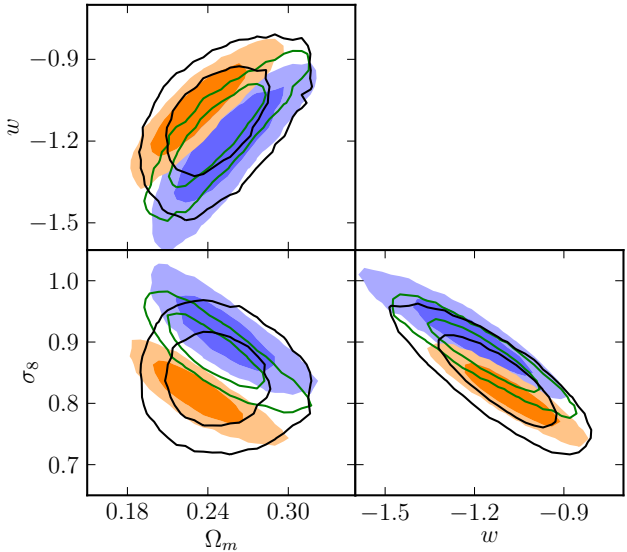}
\caption{Constraints on \wCDM{} cosmological parameters from WMAP7+H0 (solid
  black lines), and WMAP7+\ACTCL{}+H0 for three scaling relations (B12 scaling
  relation is green lines; UPP is orange contours; Nonthermal20 is violet
  contours.)}
\label{fig:params_wcdm}
\end{figure}

In Figure~\ref{fig:params_wcdm}, we show 2-d marginalized constraints for
$\Omegam, \sigma_8$, and $w$.  Note that, for this plot only, we have included
the H0 prior, to partially break the degeneracy between these three parameters
and to emphasize the role that normalization of the SZ scaling relation plays
in this space.

\begin{deluxetable*}{l c c c c}\tablecaption{Cosmological parameter constraints for the flat $w$CDM model.}

\tablehead{
 & \multicolumn{4}{c}{Parameter ($w$CDM)}\\
Data set & $\Omega_{\rm c}h^2$ & $\sigma_8$ & $\sigma_8 (\Omega_{\rm m}/0.27)^{0.4}$ & $w(\Omega_{\rm m}/0.27)$
}
\startdata
Without ACT Cluster Data & \\
\quad WMAP7 & 0.111 $\pm$ 0.006 & 0.832 $\pm$ 0.134 & 0.790 $\pm$ 0.065 & \phantom{-}-0.95 $\pm$ 0.13\\
\quad WMAP7 + SNe & 0.111 $\pm$ 0.006 & 0.791 $\pm$ 0.042 & 0.798 $\pm$ 0.060 & \phantom{-}-0.99 $\pm$ 0.11\\
Fixed Scaling Relations (\S \ref{sec:parameters_fixed}) & \\
\quad WMAP7 + ACTcl(UPP) & 0.108 $\pm$ 0.002 & 0.854 $\pm$ 0.106 & 0.766 $\pm$ 0.018 & \phantom{-}-0.90 $\pm$ 0.06\\
\quad WMAP7 + ACTcl(B12) & 0.115 $\pm$ 0.003 & 0.915 $\pm$ 0.111 & 0.849 $\pm$ 0.021 & \phantom{-}-1.04 $\pm$ 0.07\\
\quad WMAP7 + ACTcl(Non) & 0.119 $\pm$ 0.003 & 0.915 $\pm$ 0.116 & 0.887 $\pm$ 0.023 & \phantom{-}-1.11 $\pm$ 0.08\\
Dynamical Mass Constraints (\S \ref{sec:parameters_dyn}) & \\
\quad WMAP7 + ACTcl(dyn) & 0.116 $\pm$ 0.005 & 0.921 $\pm$ 0.108 & 0.851 $\pm$ 0.052 & \phantom{-}-1.05 $\pm$ 0.11\\
\quad WMAP7 + ACTcl(dyn) + SNe & 0.115 $\pm$ 0.004 & 0.835 $\pm$ 0.034 & 0.858 $\pm$ 0.049 & \phantom{-}-1.08 $\pm$ 0.10
\enddata
\tablecomments{Cluster data provide important constraints in the space of $\sigma_8$,
$\Omegam$ and $w$.  \ACTCL{} results are presented for scaling
relations based on the Universal Pressure Profile (UPP), and the B12
and \emph{Nonthermal20} (Non) models; these do not include
marginalization over scaling relation uncertainties.  Results
constrained using dynamical mass data (dyn) include a full
marginalization over SZ scaling relation parameters.
}
\label{tab:params_wcdmx}\end{deluxetable*}

\begin{deluxetable*}{l c c c c c}\tablecaption{Cosmological parameter constraints for the flat $w$CDM model,
for various combinations of WMAP7, ACT cluster data (with scaling
relation constrained using dynamical mass data), and Type Ia
Supernovae results.}

\tablehead{
 & \multicolumn{5}{c}{Parameter ($w$CDM)}\\
Data set & $\Omega_{\rm c}h^2$ & $\Omega_{\rm m}$ & $\sigma_8$ & $h$ & $w$
}
\startdata
Without ACT Cluster Data & \\
\quad WMAP7 & \phantom{-}0.111 $\pm$ 0.006 & \phantom{-}0.259 $\pm$ 0.096 & \phantom{-}0.832 $\pm$ 0.134 & \phantom{-}0.753 $\pm$ 0.131 & -1.117 $\pm$ 0.394\\
\quad WMAP7 + SNe & \phantom{-}0.111 $\pm$ 0.006 & \phantom{-}0.276 $\pm$ 0.020 & \phantom{-}0.791 $\pm$ 0.042 & \phantom{-}0.697 $\pm$ 0.016 & -0.969 $\pm$ 0.054\\
Dynamical Mass Constraints & \\
\quad WMAP7 + ACTcl(dyn) & \phantom{-}0.116 $\pm$ 0.005 & \phantom{-}0.237 $\pm$ 0.080 & \phantom{-}0.921 $\pm$ 0.108 & \phantom{-}0.792 $\pm$ 0.119 & -1.306 $\pm$ 0.356\\
\quad WMAP7 + ACTcl(dyn) + SNe & \phantom{-}0.115 $\pm$ 0.004 & \phantom{-}0.289 $\pm$ 0.017 & \phantom{-}0.835 $\pm$ 0.034 & \phantom{-}0.691 $\pm$ 0.014 & -1.011 $\pm$ 0.052
\enddata
\label{tab:params_wcdm}\end{deluxetable*}

Overall, we find no disagreement between WMAP7 and the ACT cluster data for
any of the three model-based scaling relations considered.  While the three
scaling relations produce results that almost completely span the range of
$\Omega_m$ and $\sigma_8$ preferred by WMAP7, a better understanding of
scaling relation parameters can provide significant improvements in parameter
constraints given even a relatively small cluster sample.  This is addressed
in the next section.

\subsection{Parameter Constraints from Dynamical Mass Data}
\label{sec:parameters_dyn}

As an alternative to fixing the SZ scaling relation parameters based on
models, in this section we perform a cosmological analysis using the ACT
Southern and Equatorial cluster samples, including the dynamical mass
information for the Southern clusters from \cite{sifon/etal:inprep}.  The two
samples are included as separate contributions to the likelihood.  In this
analysis, the dynamical masses directly inform the cosmology, while the
selection function is understood in terms of the observed SZ signal, through
the modeling of the cluster signal with the UPP.  In this framework the
scaling relation parameters will be naturally constrained to be consistent
with the observed sample sizes and with the $\ytilde$ measurements of the
clusters that also have dynamical mass measurements.

For the Equatorial clusters we apply the same sample selection criteria used
in Section~\ref{sec:parameters_fixed}, and thus include the same 15 clusters.
As before, these clusters contribute to the likelihood through their observed
redshifts and $\ytilde$ measurements.

For the Southern cluster sample, we obtain $\ytilde$ measurements from the
three-season \arone{} maps as described in the Appendix.  For the cosmological
analysis we restrict the Southern sample based on a signal to noise ratio
threshold of $\ytilde / \delta \ytilde > 5.7$ and an observed redshift
requirement of $0.315 < z < 1.4$.  The $\ytilde / \delta \ytilde$ threshold is
high enough to exclude new, unconfirmed candidates in our analysis of the
three-season Southern maps.  The lower redshift bound restricts the sample to
the clusters for which \cite{sifon/etal:inprep} have measured dynamical
masses.  This yields a sample of seven clusters, which are identified in
Table~\ref{tab:clusters_south}.  Of the nine clusters used in
\cite{sehgal/etal:2011}, our sample includes the five clusters at $z > 0.315$.
The sample includes the exceptional cluster ACT-CL J0102$-$4915; the inclusion
or exclusion of this cluster does not change the cosmological parameter
constraints significantly.  The contribution from the Southern clusters to the
likelihood is in the form of equation (\ref{eqn:Lthetapsic}).

The data set consisting of this combination of Equatorial SZ data and Southern
SZ and dynamical mass data is denoted \ACTCL{(dyn)}.  These data are combined
with other data sets in an MCMC approach to parameter estimation as described
in Section~\ref{sec:parameters_fixed}, except that now we allow the SZ scaling
relation parameters to vary, assuming flat priors over ranges $1.7 < A_m <
0.9$, $-1 < B < 3$, $-2 < C < 2$ and $0 < \sigma_{\rm int}^2 <2$.  These
priors are not intended to be informative, but rather to permit the scaling
relation parameters to range freely over values supported by the data.  For
the dynamical mass bias parameter, we apply a Gaussian prior corresponding to
$1/\bdyn = 1.00 \pm 0.15$,\footnote{This description of the prior is an
  artifact of our initial implementation of the likelihood, where the
  parameter describing the bias corresponded to $1/\bdyn$.} motivated by the
results of \cite{evrard/etal:2008} as described in Section~\ref{sec:caldyn}.

\subsubsection{\LCDM{} Constraints}

\begin{figure}
\includegraphics[width=8.5cm]{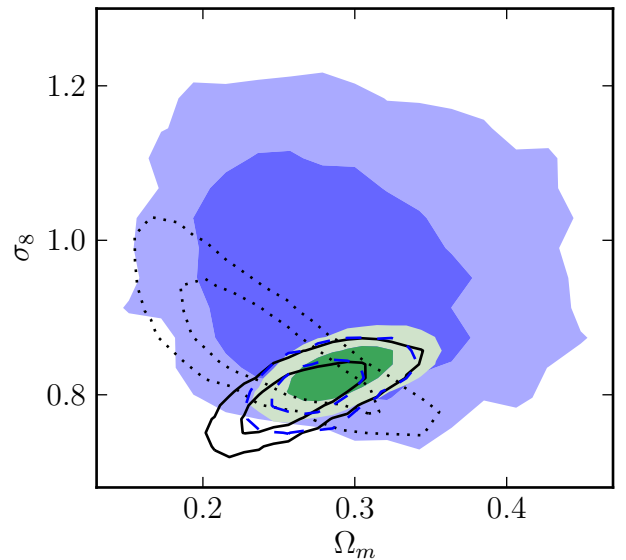}
\caption{Constraints on \LCDM{} cosmological parameters from Equatorial and
  Southern clusters.  Results from \ACTCL{(dyn)}+BBN+H0 (violet contours), and
  WMAP7+\ACTCL{(dyn)} (green contours), which both include full
  marginalization over scaling relation and dynamical mass bias parameters,
  may be compared to WMAP alone (solid black lines).  Dotted line shows
  constraints for \ACTCL{}+BBN+H0, using the same cluster sample but with the
  scaling relation fixed to the central values obtained from the dynamical
  mass fit of Section~\ref{sec:caldyn}; note the similarity to contours in
  Figure~\ref{fig:params_solo2} obtained for Equatorial SZ data with B12 fixed
  scaling relation parameters.  Dashed blue line shows WMAP7+\ACTCL{(dyn)},
  with full marginalization over scaling relation parameters but with $\bdyn$
  fixed to 1.33.}
\label{fig:params_solo4}
\end{figure}

The effect of the increased freedom in the scaling relation parameters may be
seen in Figure~\ref{fig:params_solo4}, which shows the confidence regions on
$\sigma_8$ and $\Omegam$, in a \LCDM{} model, from ACT cluster data combined
with BBN and H0.  The scaling relation parameters are not well constrained
without some prior information, so for this chain only we include a Gaussian
prior on the redshift evolution corresponding to $C = 0.0 \pm 0.5$, based on
the B12 model fits.  Compared to results for fixed scaling relation
parameters, the distribution of acceptable $\sigma_8$ and $\Omegam$ is broader
and skewed, at fixed $\Omegam$, towards high $\sigma_8$.

\begin{figure}
\includegraphics[width=8.5cm]{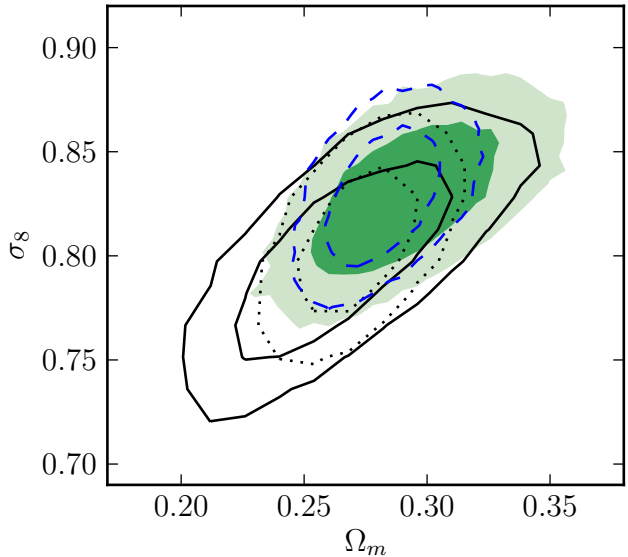}
\caption{Constraints on \LCDM{} cosmological parameters from the combined
  Southern and Equatorial cluster samples, including dynamical mass
  measurements for the Southern clusters and full marginalization over scaling
  relation parameters.  WMAP7 and WMAP7+\ACTCL{(dyn)} are identified as in
  Figure~\ref{fig:params_solo4} (solid black line and green contours,
  respectively).  Also shown are WMAP7+BAO+H0 (dotted black line) and
  WMAP7+\ACTCL{(dyn)}+BAO+H0 (dashed blue lines).}
\label{fig:params_solo3}
\end{figure}

In a \LCDM{} model, the addition of \ACTCL{(dyn)} data improves the
constraints on $\sigma_8$, $\Omegam$, and $h$ relative to WMAP7 alone by
factors of 0.8 to 0.9, as can be seen in Figure~\ref{fig:params_solo4}.
WMAP7+\ACTCL{(dyn)} prefers slightly larger values of $\sigma_8$ than does
WMAP7+BAO+H0.  For the composite parameter $\sigma_8(\Omegam/0.27)^{0.3}$, the
combination of WMAP7+BAO+H0+\ACTCL{(dyn)} improves the uncertainty by a factor
of 0.6 compared to WMAP7.  Parameter values are presented in
Table~\ref{tab:params_lcdm} and confidence regions for $\sigma_8$ and
$\Omegam$ are plotted in Figure~\ref{fig:params_solo3}.

Within the WMAP7+\ACTCL{(dyn)} chain for \LCDM{} the dynamical mass bias
parameter is pushed to $\bdyn = 1.12 \pm 0.17$ (corresponding to $1/\bdyn =
0.91 \pm 0.12$), a substantial change given the prior on $\bdyn$.  To explore
the consequences of a large systematic bias in the dynamical mass
measurements, we study a \LCDM{} chain run with fixed $\bdyn = 1.33$.  We find
that $\sigma_8$ and $\Omegam$ move towards the central values preferred by
WMAP7, with parameter uncertainty slightly reduced.  The confidence contours
associated with this chain can be seen in Figure~\ref{fig:params_solo4}.

In the chains presented we have not included any intrinsic scatter in the
relationship between dynamical mass and halo mass, because the measurement
errors on the masses are already at the 20-50\% level, and because we do not
use dynamical mass in the sample selection criteria.  However, large levels of
intrinsic scatter in $\mdyn$, or correlations between $\mdyn$ and $\ytr$ as
mass proxies will affect the derived constraints to some degree.  In chains
that include a 30\% scatter in the dynamical mass relative to halo mass, the
central values of $\sigma_8$, $\Omegam$, and $\sigma_8(\Omegam/0.3)^{0.27}$
decrease by up to 20\% of their quoted uncertainties.  Adding positive
correlation to the scatter in $\mdyn$ and $\ytr$ lowers the preferred
parameter values further.  Running WMAP7+\ACTCL{(dyn)} with an additional
constraint that the two proxies scatter with
correlation coefficient $\rho = 0.5$, we obtain constraints $\sigma_8 = 0.820
\pm 0.025$, $\Omegam = 0.284 \pm 0.025$, and $\sigma_8(\Omegam/0.3)^{0.27} =
0.832 \pm 0.042$.  The addition of these two effects changes the central
parameter constraints by roughly 40\% of the quoted uncertainty; limits on the
intrinsic scatter and its correlation between proxies will be important in
higher precision studies.

\subsubsection{Neutrino Mass Constraints}

\begin{figure}
\includegraphics[width=8.5cm]{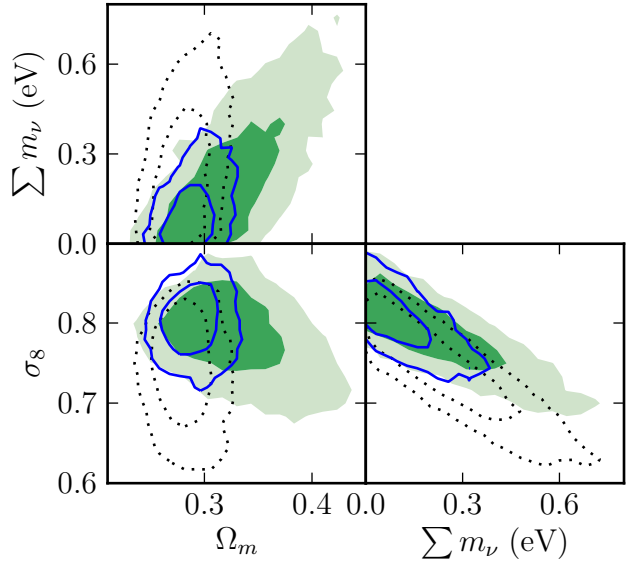}
\caption{Constraints within an extension to \LCDM{} that allows for non-zero
  neutrino density.  The data sets shown are WMAP7+BAO+H0 (dotted black
  lines), WMAP7+\ACTCL{(dyn)} (green contours), and WMAP7+\ACTCL{(dyn)}+BAO+H0
  (solid blue lines).  The total number of relativistic species is fixed to
  $N_{\rm eff} = 3.046$. }
\label{fig:params_sumnu}
\end{figure}

As an extension to \LCDM{}, we also run chains where the cosmic mass density
of neutrinos is allowed to vary.  Constraints are interpreted in terms of the
sum of the neutrino mass species according to the relation $\Omega_\nu h^2 =
\sum_\nu m_\nu / (93~\rm{eV})$.  Combining the ACT cluster data with WMAP7 and
BAO+H0 leads to significant improvements in this constraint, as shown in
Figure~\ref{fig:params_sumnu} and Table~\ref{tab:params_sumnu}.  For
WMAP7+BAO+H0+\ACTCL{(dyn)} we obtain an upper limit, at 95\% confidence, of
$\sum m_\nu < 0.29~\rm{eV}$.  The improvement in this constraint is driven by
the preference of the ACT cluster data for values of $\sigma_8$ and $\Omegam$
that are in the upper range of those consistent with WMAP.  Interpretations of
this preference are discussed below.

\begin{deluxetable*}{l c c c c}\tablecaption{Cosmological parameter constraints for $\Lambda$CDM, extended with one
additional parameter for non-zero neutrino density.}

\tablehead{
 & \multicolumn{4}{c}{Parameter ($\Lambda$CDM + $\sum m_\nu$)}\\
Data set & $\Omega_{\rm m}$ & $\sigma_8$ & $h$ & $\sum m_\nu$ (eV)\\
 &  &  &  & 95\% CL
}
\startdata
\quad WMAP7 + BAO + H0 & 0.282 $\pm$ 0.018 & 0.742 $\pm$ 0.053 & 0.693 $\pm$ 0.016 & $< 0.58$\\
\quad WMAP7 + \ACTCL{(dyn)} & 0.325 $\pm$ 0.041 & 0.787 $\pm$ 0.041 & 0.663 $\pm$ 0.029 & $< 0.57$\\
\quad WMAP7 + \ACTCL{(dyn)} + BAO + H0 & 0.289 $\pm$ 0.018 & 0.802 $\pm$ 0.031 & 0.690 $\pm$ 0.015 & $< 0.29$
\enddata
\tablecomments{The cluster data
greatly assist in breaking the degeneracy between $\sigma_8$,
$\Omegam$, and the neutrino density (as parametrized by $\sum
m_\nu$).  \ACTCL{(dyn)} results use South and Equatorial cluster
data, including dynamical mass measurements for the Southern clusters
(Section~\ref{sec:parameters_dyn}).
}
\label{tab:params_sumnu}\end{deluxetable*}

\subsubsection{\wCDM{} Constraints}

\begin{figure}
\includegraphics[width=8.5cm]{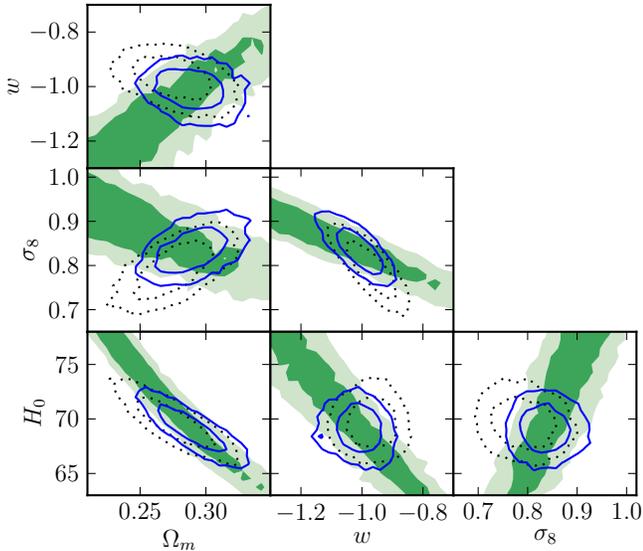}
\caption{Constraints on \wCDM{} cosmological parameters from the combined
  Southern and Equatorial cluster samples, with scaling relation parameters
  constrained based on dynamical mass measurements.  Data sets shown are
  WMAP7+SNe (dotted black lines), WMAP7+\ACTCL{(dyn)} (green contours), and
  WMAP7+\ACTCL{(dyn)}+SNe (solid blue lines).  The units of $H_0$ are ${\rm
    km~s}^{-1}~{\rm Mpc}^{-1}$.  }
\label{fig:params_wcdm_sn}
\end{figure}

Within a \wCDM{} model, we consider the \ACTCL{(dyn)} data in combination with
WMAP7 and with SNe (which provides important, complementary constraints on the
recent cosmic expansion history).  The WMAP7+\ACTCL{(dyn)} are consistent with
WMAP7+SNe, but with a preference (as was found in \LCDM{}) for slightly higher
values of $\sigma_8$ and $\Omegam$.  The importance of cluster information is
demonstrated by the improvement, over both WMAP7 and WMAP7+SNe, in the
composite parameter $\sigma_8(\Omegam/0.27)^{0.4}$ (see
Table~\ref{tab:params_wcdmx}).  As a result, the main impact of adding the
cluster data to either WMAP7 or WMAP7+SNe is to reduce the uncertainties in
$\sigma_8$ and $\Omegam$ by factors of $\approx 0.8$.  The \wCDM{} parameter
constraints for each combination of \ACTCL{(dyn)} and SNe with WMAP7 are
presented in Table~\ref{tab:params_wcdm}, with marginalized 2-d confidence
regions shown in Figure~\ref{fig:params_wcdm_sn}.

The slight preference of \ACTCL{(dyn)} for higher values of $\sigma_8$ and
$\Omegam$ than are preferred by WMAP7+SNe alone also induces a shift in the
posterior distribution for $w$.  The value of the composite parameter
$w(\Omegam/0.27)$ decreases by almost one standard deviation when
\ACTCL{(dyn)} are added to WMAP7 + SNe.

\subsubsection{Scaling Relation Constraints}

\begin{figure}
\includegraphics[width=8.5cm]{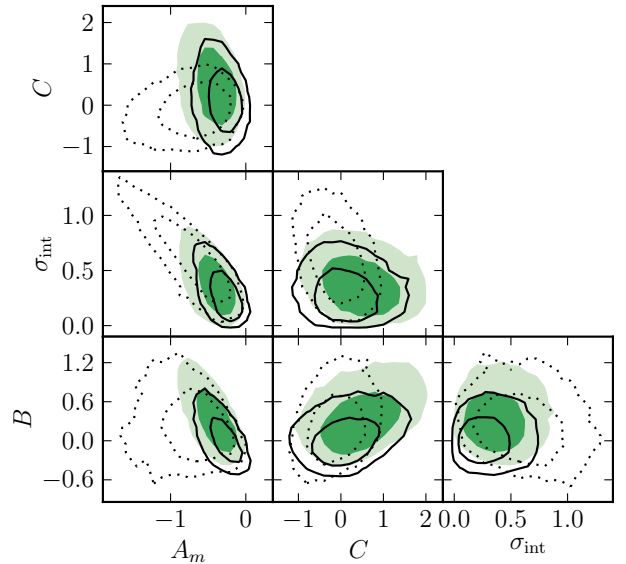}
\caption{Constraints on SZ scaling relation parameters from the combined
  Southern and Equatorial cluster samples, \ACTCL{(dyn)}, constrained based on
  dynamical mass measurements in a cosmological MCMC.  Green contours are for
  WMAP7+\ACTCL{(dyn)} chain; black solid lines are for WMAP7+\ACTCL{(dyn)} but
  with J0102$-$4915 excluded from the Southern sample.  Dotted line shows
  constraints for \ACTCL{(dyn)}+BBN+H0 (i.e., without CMB information), but
  with a Gaussian prior on $C$ of 0.0 $\pm$ 0.5.}
\label{fig:params_scaling}
\end{figure}

The marginalized constraints on the SZ scaling relation parameters derived for
the \ACTCL{(dyn)} chains are presented in Table~\ref{tab:abcfits}.  The
parameters indicating deviations from self-similar scaling with mass and
redshift ($B$ and $C$) are each consistent with 0, and consistent with the
fits to all models.  The intrinsic scatter, $\sigma_{\rm int}$ is somewhat
higher than the 20\% seen in the model results, but it is not well-constrained
by these data.  Furthermore, since the data span a fairly restricted range of
masses, there is significant covariance between $A_m, B$, and $\sigma_{\rm
  int}$.  The value of $A$ is provided for $\Mpivot = 7\times
10^{14}~\hinv\Msun$, chosen to produce negligible covariance between $A$ and
$B$.  While the cosmological results change only slightly when J0102$-$4915 is
excluded from the Southern cluster sample, we note that scaling relation
parameters are somewhat more affected, and in particular that the slope
parameter $B$ drops from $0.36 \pm 0.36$ to $0.06 \pm 0.27$ and the scatter
$\sigma_{\rm int}$ drops from $0.42 \pm 0.19$ to $0.33 \pm 0.17$.  The 2-d
confidence regions for scaling relation parameters are shown in
Figure~\ref{fig:params_scaling}.

\subsubsection{Discussion}

We note that the constraints on cosmological parameters are due mostly to the
inclusion of the dynamical mass measurements for the Southern clusters, rather
than due to the SZ data of the larger Equatorial sample.  However, while the
removal of the Equatorial sample from the likelihood computation produces no
change in the cosmological parameter constraints, it leads to a significant
weakening in the SZ scaling relation parameter constraints.  This is because
the Equatorial sample constrains the scaling relation parameters to values
that predict sample selection functions consistent with the Equatorial sample
size.  By including the Equatorial SZ measurements in the likelihood, we
obtain simultaneous, self-consistent constraints on cosmological parameters
and the SZ scaling relation parameters.  Despite the low weight of the SZ
information in the cosmological parameter constraints, it remains true that
the sample is SZ selected and thus approximately mass-limited over a broad
range of redshifts.

Immediate improvement in the cosmological constraining power can be obtained
through improved calibration of the SZ mass relation.  This is underway in the
form of a campaign to collect dynamical mass data for the Equatorial sample
(Sif\'on et al, in prep.).  The current analysis would also benefit from an
improved understanding of any systematic biases in the measurement of halo
mass using galaxy velocity dispersions.  In addition to the dynamical mass
data, we are pursuing weak lensing \citep[e.g., ][]{miyatake/etal:2013},
X-ray \citep[e.g., ][]{menanteau/etal:2012}, and additional SZ \citep[e.g.,
][]{reese/etal:2012} measurements to improve constraints on the SZ--mass
scaling relation from the ACT cluster sample.  All such mass measurements can
be easily included in our formulation.  The sample of clusters appropriate for
our approach to the cosmological analysis will grow as targeted follow-up on
the Equatorial field candidates is completed to lower S/N ratios.

While this work presents a new way to quantify the SZ-mass relation for
$10^{15}~\Msun$ clusters, we note that it is but one component of a growing
web of observation that tie optical and SZ data together. Recently the Planck
team presented results \citep{planck/intermediateXI:prep} on the SZ emission
from SDSS galaxies that extended the SZ-mass relation down to $10^{13}~\Msun$
systems and showed that the gas properties of dark matter haloes are similar
to those in massive clusters.

\section{Conclusion}

The ACT Equatorial maps at \arone{}, along with an optical and infrared
confirmation campaign, have yielded a sample of 68 confirmed, SZ selected
galaxy clusters in the redshift range from approximately 0.1 to 1.4, in an
area of 504 square degrees.  Inside the 270 square degree overlap with SDSS
Stripe 82, and assuming the scaling relation parameters associated with the
Universal Pressure Profile of A10, the sample is estimated to be 90\% complete
above a mass of $M_{500c} \approx 4.5 \times 10^{14}~\hinv\Msun$.  (The
completeness level is dependent on the normalization of the SZ--mass scaling
relation; a similar estimate using the most likely scaling relation parameters
from a full cosmological MCMC that incorporates dynamical mass information
gives a 90\% completeness mass of $\approx 9 \times 10^{14}~\hinv\Msun$.)

In order to use the SZ signal to determine cluster mass and to constrain
cosmological parameters, we have developed a framework that predicts the
cluster amplitude in a matched-filtered map based on cluster mass and
redshift.  The approach naturally handles varying noise levels in a map, and
can be adapted easily to accommodate alternative cluster pressure profiles and
scaling relations.

The framework is based on the Universal Pressure Profile of A10, but we obtain
alternative normalizations through an analysis of several cluster physics
models.  While the normalization obtained directly from the UPP is consistent
with X-ray masses from the MCXC, the calibration to the B12 model is more
consistent with dynamical mass measurements of the ACT Southern cluster
sample.  From this we conclude that it is necessary to consider a broad range
of scaling relation parameters when using the cluster data to constrain
cosmological parameters.

We have presented constraints in \LCDM{} and \wCDM{} models using the ACT
Equatorial cluster sample for several fixed normalizations of the SZ scaling
relations.  In each case the cosmological constraints are consistent with
WMAP.  The results obtained for fixed scaling relations demonstrate the
potential for such a cluster sample to provide important new cosmological
information, even for modest cluster samples, provided the systematics of the
SZ--mass calibration can be better understood.

We have also demonstrated cosmological constraints based on the combination of
SZ measurements and dynamical mass data, in which the four parameters of the
SZ scaling relation are calibrated simultaneously with cosmological
parameters.  The results provide significant constraints that are
complementary to CMB, BAO, and Type Ia Supernovae data.  The scaling relation
parameters obtained in this analysis are consistent with the B12 model and
inconsistent with the normalization arising directly from the UPP.

Our results are consistent with a study of the SZ signal from the ACT Southern
cluster sample \citep{sehgal/etal:2011}, as well as the $\sigma_8$ constraints
from the skewness analysis of the ACT Equatorial \arone{} maps.
\citep{wilson/etal:2012}.

While the ACT SZ cluster data prefer matter density parameters that are at the
upper end of those supported by WMAP7 data, our results are consistent with
cluster studies that incorporate a variety of other mass proxies.  They are
also in agreement with the recent ACT angular power spectrum results (Sievers
et al, in prep.), which also favor the upper limits of $\sigma_8$ and
$\Omegam$ permitted by WMAP7.

The combined X-ray, $f_{\rm gas}$, Supernovae, BAO, and CMB results of
\citet{mantz/etal:2010} produce $\sigma_8 = 0.80 \pm 0.02$ ($0.79 \pm 0.03$)
and $\Omegam = 0.257 \pm 0.015$ ($0.272 \pm 0.016$) in \LCDM{} (\wCDM{});
these results are marginally consistent with but lower than the results of our
analysis.  For the \ACTCL{(dyn)}+H0+BBN run in \LCDM{}, we compute
$\sigma_8(\Omegam/0.25)^{0.47} = 0.87 \pm 0.04$, which is higher than, but
consistent with, the value of $0.813 \pm 0.027$ obtained for the X-ray cluster
study of \citet{vikhlinin/etal:2009}.

Using SPT cluster SZ and X-ray mass information, \cite{reichardt/etal:2013}
obtain $\sigma_8$ and $\Omegam$ measurements in a combined WMAP7, high-$\ell$
power spectrum, and clusters analysis that lie slightly below the central
values preferred by CMB measurements alone (including Sievers et al, in
prep.).  From their Figure~5 we also estimate that their cluster data in
combination with H0 and BBN (i.e., without CMB information) produce a
composite parameter constraint of $\sigma_8(\Omegam/0.27)^{0.3} = 0.77 \pm
0.05$, lower by roughly 1.3-$\sigma$ than either our WMAP7+\ACTCL{(dyn)} or
BBN+H0+\ACTCL{(dyn)} results.  Our results are in general agreement with SPT
despite several differences in the SZ signal interpretation and mass
calibration, which we summarize here.

SPT makes use of cluster signal simulations, analogous to the B12 models used
here, in order to interpret an observable based on signal to noise ratio.  In
contrast, our analysis relies instead on the assumption of a simple relation
between mass and the cluster pressure profile.  We test the \PBAA{} approach
on simulated maps based on models of cluster physics, but we do not use these
models to place priors on the scaling relation parameters.  The SPT mass
calibration is ultimately derived from measurements of $Y_{\rm X}$, defined as
the product of the core-excised X-ray temperature and the cluster gas mass
\citep{kravtsov/vikhlinin/nagai:2006}.  The use of $Y_{\rm X}$ as a mass proxy
has, in turn, been calibrated to weak lensing masses
\citep{vikhlinin/etal:2009b}.  The slight preference of our data for higher
values of $\sigma_8$ and $\Omegam$ (for both the \ACTCL{(dyn)} and
\ACTCL{(B12)} studies) compared to SPT may indicate a complicated relationship
between the various mass proxies.


This work was supported by the U.S. National Science Foundation through awards
AST-0408698 and AST-0965625 for the ACT project, as well as awards PHY-0855887
and PHY-1214379. Funding was also provided by Princeton University, the
University of Pennsylvania, and a Canada Foundation for Innovation (CFI) award
to UBC. ACT operates in the Parque Astron\'omico Atacama in northern Chile
under the auspices of the Comisi\'on Nacional de Investigaci\'on Cient\'ifica
y Tecnol\'ogica de Chile (CONICYT). Computations were performed on the GPC
supercomputer at the SciNet HPC Consortium. SciNet is funded by the CFI under
the auspices of Compute Canada, the Government of Ontario, the Ontario
Research Fund -- Research Excellence; and the University of Toronto.

\bibliographystyle{apj}

\LongTables 

\renewcommand\mytitle{Confirmed galaxy clusters in the ACT Equatorial region.}
\renewcommand\mycaption{ Coordinates are in the J2000 standard equinox.
  Redshifts for which uncertainties are quoted are photometric; all others are
  spectroscopic (full details can be found in \citealt{menanteau/etal:2013}).
  The Region (Reg.) column indicates whether the cluster lies within the
  coverage of SDSS Stripe 82 or in the shallower coverage of SDSS DR8; an
  asterisk denotes clusters used for the cosmological analysis of
  Section~\ref{sec:cosmology}.  The signal to noise ratio and filter scale
  $\thetaD$ are provided for the matched filter template that yielded the most
  significant detection.  The uncorrected central Compton parameter $\ytilde$
  is obtained from maps filtered with the matched filter having $\thetaD =
  \thetaFixed$ (see Section~\ref{sec:fixed_scale}).}

\newpage
\clearpage
\begin{deluxetable}{c c c c c c c r@{ $\pm$ }l l}\tablecaption{\mytitle}

\tablehead{
ACT ID & R.A. & Dec. & Redshift & Reg. & $S/N$ & $\thetaD$ & \multicolumn{2}{c}{$\ytilde$} & Alternate ID (ref.)\\
 & $(^\circ)$ & $(^\circ)$ &  &  &  & $(\mathrm{arcmin})$ & \multicolumn{2}{c}{$(10^{-4})$} & 
}
\startdata
ACT--CL J0008.1$+$0201 & \phantom{00}2.0418 & $\phantom{-}2.0204$ & 0.36 $\pm$ 0.04 & DR8\phantom{*} & \phantom{0}4.7 & \phantom{0}4.71 & 0.96 & 0.21 & WHL J000810.4+020112 (1)\\
ACT--CL J0012.0$-$0046 & \phantom{00}3.0152 & $-0.7693$ & 1.36 $\pm$ 0.06 & S82\phantom{*} & \phantom{0}5.3 & 16.47 & 0.91 & 0.18 & \\
ACT--CL J0014.9$-$0057 & \phantom{00}3.7276 & $-0.9502$ & 0.533 & S82* & \phantom{0}7.8 & \phantom{0}3.53 & 1.34 & 0.18 & GMB11 J003.71362-00.94838 (2)\\
ACT--CL J0017.6$-$0051 & \phantom{00}4.4138 & $-0.8580$ & 0.211 & S82\phantom{*} & \phantom{0}4.2 & \phantom{0}4.71 & 0.73 & 0.17 & SDSS CE J004.414726-00.876164 (3)\\
ACT--CL J0018.2$-$0022 & \phantom{00}4.5623 & $-0.3795$ & 0.75 $\pm$ 0.04 & S82\phantom{*} & \phantom{0}4.4 & \phantom{0}4.71 & 0.74 & 0.17 & \\
ACT--CL J0022.2$-$0036 & \phantom{00}5.5553 & $-0.6050$ & 0.805 & S82* & \phantom{0}9.8 & \phantom{0}1.18 & 1.35 & 0.16 & WHL J002213.0-003634 (1)\\
ACT--CL J0026.2$+$0120 & \phantom{00}6.5699 & $\phantom{-}1.3367$ & 0.65 $\pm$ 0.04 & DR8\phantom{*} & \phantom{0}6.3 & \phantom{0}4.71 & 0.99 & 0.16 & \\
ACT--CL J0044.4$+$0113 & \phantom{0}11.1076 & $\phantom{-}1.2221$ & 1.11 $\pm$ 0.03 & S82\phantom{*} & \phantom{0}5.5 & \phantom{0}1.18 & 0.70 & 0.15 & \\
ACT--CL J0045.2$-$0152 & \phantom{0}11.3051 & $-1.8827$ & 0.545 & DR8\phantom{*} & \phantom{0}7.5 & \phantom{0}3.53 & 1.31 & 0.18 & WHL J004512.5-015232 (1)\\
ACT--CL J0051.1$+$0055 & \phantom{0}12.7875 & $\phantom{-}0.9323$ & 0.69 $\pm$ 0.03 & S82\phantom{*} & \phantom{0}4.2 & \phantom{0}1.18 & 0.53 & 0.15 & WHL J005112.9+005555 (1)\\
ACT--CL J0058.0$+$0030 & \phantom{0}14.5189 & $\phantom{-}0.5106$ & 0.76 $\pm$ 0.02 & S82\phantom{*} & \phantom{0}5.0 & \phantom{0}2.35 & 0.72 & 0.15 & \\
ACT--CL J0059.1$-$0049 & \phantom{0}14.7855 & $-0.8326$ & 0.786 & S82* & \phantom{0}8.4 & \phantom{0}2.35 & 1.24 & 0.15 & \\
ACT--CL J0104.8$+$0002 & \phantom{0}16.2195 & $\phantom{-}0.0495$ & 0.277 & S82\phantom{*} & \phantom{0}4.3 & 12.94 & 0.62 & 0.15 & SDSS CE J016.232412+00.058164 (3)\\
ACT--CL J0119.9$+$0055 & \phantom{0}19.9971 & $\phantom{-}0.9193$ & 0.72 $\pm$ 0.03 & S82\phantom{*} & \phantom{0}5.0 & \phantom{0}3.53 & 0.73 & 0.15 & \\
ACT--CL J0127.2$+$0020 & \phantom{0}21.8227 & $\phantom{-}0.3468$ & 0.379 & S82\phantom{*} & \phantom{0}5.1 & \phantom{0}2.35 & 0.72 & 0.15 & SDSS CE J021.826914+00.344883 (3)\\
ACT--CL J0139.3$-$0128 & \phantom{0}24.8407 & $-1.4769$ & 0.70 $\pm$ 0.03 & DR8\phantom{*} & \phantom{0}4.3 & \phantom{0}1.18 & 0.54 & 0.17 & \\
ACT--CL J0152.7$+$0100 & \phantom{0}28.1764 & $\phantom{-}1.0059$ & 0.230 & S82* & \phantom{0}9.0 & \phantom{0}3.53 & 1.30 & 0.15 & Abell 267 (4)\\
ACT--CL J0156.4$-$0123 & \phantom{0}29.1008 & $-1.3879$ & 0.45 $\pm$ 0.04 & DR8\phantom{*} & \phantom{0}5.2 & \phantom{0}2.35 & 0.67 & 0.15 & WHL J015624.3-012317 (1)\\
ACT--CL J0206.2$-$0114 & \phantom{0}31.5567 & $-1.2428$ & 0.676 & S82* & \phantom{0}6.9 & \phantom{0}2.35 & 0.94 & 0.14 & \\
ACT--CL J0215.4$+$0030 & \phantom{0}33.8699 & $\phantom{-}0.5091$ & 0.865 & S82* & \phantom{0}5.5 & \phantom{0}1.18 & 0.75 & 0.15 & \\
ACT--CL J0218.2$-$0041 & \phantom{0}34.5626 & $-0.6883$ & 0.672 & S82* & \phantom{0}5.8 & \phantom{0}2.35 & 0.82 & 0.15 & GMB11 J034.56995-00.69963 (2)\\
ACT--CL J0219.8$+$0022 & \phantom{0}34.9533 & $\phantom{-}0.3755$ & 0.537 & S82\phantom{*} & \phantom{0}4.7 & \phantom{0}2.35 & 0.66 & 0.15 & GMB11 J034.94761+00.35956 (2)\\
ACT--CL J0219.9$+$0129 & \phantom{0}34.9759 & $\phantom{-}1.4973$ & 0.35 $\pm$ 0.02 & DR8\phantom{*} & \phantom{0}4.9 & \phantom{0}2.35 & 0.62 & 0.14 & NSCS J021954+013102 (5)\\
ACT--CL J0221.5$-$0012 & \phantom{0}35.3925 & $-0.2063$ & 0.589 & S82\phantom{*} & \phantom{0}4.0 & \phantom{0}1.18 & 0.33 & 0.15 & GMB11 J035.40587-00.21967 (2)\\
ACT--CL J0223.1$-$0056 & \phantom{0}35.7939 & $-0.9466$ & 0.663 & S82* & \phantom{0}5.8 & \phantom{0}2.35 & 0.84 & 0.15 & GMB11 J035.79247-00.95712 (2)\\
ACT--CL J0228.5$+$0030 & \phantom{0}37.1250 & $\phantom{-}0.5033$ & 0.72 $\pm$ 0.02 & S82\phantom{*} & \phantom{0}4.0 & \phantom{0}1.18 & 0.56 & 0.15 & GMB11 J037.11459+00.52965 (2)\\
ACT--CL J0230.9$-$0024 & \phantom{0}37.7273 & $-0.4043$ & 0.44 $\pm$ 0.03 & S82\phantom{*} & \phantom{0}4.2 & \phantom{0}8.24 & 0.63 & 0.15 & WHL J023055.3-002549 (6)\\
ACT--CL J0239.8$-$0134 & \phantom{0}39.9718 & $-1.5758$ & 0.375 & DR8\phantom{*} & \phantom{0}8.8 & \phantom{0}4.71 & 1.61 & 0.18 & Abell 370 (4)\\
ACT--CL J0240.0$+$0116 & \phantom{0}40.0102 & $\phantom{-}1.2693$ & 0.62 $\pm$ 0.03 & DR8\phantom{*} & \phantom{0}4.8 & \phantom{0}4.71 & 0.70 & 0.15 & WHL J024001.7+011606 (1)\\
ACT--CL J0241.2$-$0018 & \phantom{0}40.3129 & $-0.3109$ & 0.684 & S82\phantom{*} & \phantom{0}5.1 & \phantom{0}1.18 & 0.59 & 0.15 & WHL J024115.5-001841 (1)\\
ACT--CL J0245.8$-$0042 & \phantom{0}41.4645 & $-0.7013$ & 0.179 & S82\phantom{*} & \phantom{0}4.1 & 10.59 & 0.61 & 0.15 & Abell 381 (4)\\
ACT--CL J0250.1$+$0008 & \phantom{0}42.5370 & $\phantom{-}0.1403$ & 0.78 $\pm$ 0.03 & S82\phantom{*} & \phantom{0}4.5 & \phantom{0}2.35 & 0.62 & 0.15 & \\
ACT--CL J0256.5$+$0006 & \phantom{0}44.1354 & $\phantom{-}0.1049$ & 0.363 & S82* & \phantom{0}5.4 & \phantom{0}7.06 & 0.82 & 0.15 & SDSS CE J044.143375+00.105766 (3)\\
ACT--CL J0301.1$-$0110 & \phantom{0}45.2925 & $-1.1716$ & 0.53 $\pm$ 0.04 & S82\phantom{*} & \phantom{0}4.2 & \phantom{0}2.35 & 0.51 & 0.15 & GMB11 J045.30649-01.17805 (2)\\
ACT--CL J0301.6$+$0155 & \phantom{0}45.4158 & $\phantom{-}1.9219$ & 0.167 & DR8\phantom{*} & \phantom{0}5.8 & \phantom{0}4.71 & 1.12 & 0.20 & RXC J0301.6+0155 (7)\\
ACT--CL J0303.3$+$0155 & \phantom{0}45.8343 & $\phantom{-}1.9214$ & 0.153 & DR8\phantom{*} & \phantom{0}5.2 & \phantom{0}7.06 & 1.00 & 0.20 & Abell 409 (4)\\
ACT--CL J0308.1$+$0103 & \phantom{0}47.0481 & $\phantom{-}1.0607$ & 0.633 & S82\phantom{*} & \phantom{0}4.8 & \phantom{0}1.18 & 0.60 & 0.15 & GMB11 J047.03754+01.04350 (2)\\
ACT--CL J0320.4$+$0032 & \phantom{0}50.1239 & $\phantom{-}0.5399$ & 0.384 & S82\phantom{*} & \phantom{0}4.9 & \phantom{0}3.53 & 0.72 & 0.15 & SDSS CE J050.120594+00.533045 (3)\\
ACT--CL J0326.8$-$0043 & \phantom{0}51.7075 & $-0.7312$ & 0.448 & S82* & \phantom{0}9.1 & \phantom{0}1.18 & 1.24 & 0.15 & GMBCG J051.70814-00.73104 (8)\\
ACT--CL J0336.9$-$0110 & \phantom{0}54.2438 & $-1.1705$ & 1.32 $\pm$ 0.05 & S82\phantom{*} & \phantom{0}4.8 & \phantom{0}3.53 & 0.68 & 0.14 & \\
ACT--CL J0342.0$+$0105 & \phantom{0}55.5008 & $\phantom{-}1.0873$ & 1.07 $\pm$ 0.06 & S82* & \phantom{0}5.9 & \phantom{0}4.71 & 0.89 & 0.15 & \\
ACT--CL J0342.7$-$0017 & \phantom{0}55.6845 & $-0.2899$ & 0.310 & S82\phantom{*} & \phantom{0}4.6 & \phantom{0}5.88 & 0.70 & 0.15 & SDSS CE J055.683678-00.286974 (3)\\
ACT--CL J0348.6$+$0029 & \phantom{0}57.1612 & $\phantom{-}0.4892$ & 0.297 & S82\phantom{*} & \phantom{0}5.0 & \phantom{0}2.35 & 0.69 & 0.15 & WHL J034837.9+002900 (6)\\
ACT--CL J0348.6$-$0028 & \phantom{0}57.1605 & $-0.4681$ & 0.345 & S82\phantom{*} & \phantom{0}4.7 & \phantom{0}2.35 & 0.67 & 0.15 & WHL J034841.5-002807 (6)\\
ACT--CL J2025.2$+$0030 & 306.3006 & $\phantom{-}0.5130$ & 0.34 $\pm$ 0.02 & DR8\phantom{*} & \phantom{0}6.4 & \phantom{0}9.41 & 1.05 & 0.17 & WHL J202512.8+003134 (1)\\
ACT--CL J2050.5$-$0055 & 312.6264 & $-0.9311$ & 0.622 & S82* & \phantom{0}5.6 & \phantom{0}1.18 & 0.83 & 0.16 & GMB11 J312.62475-00.92697 (2)\\
ACT--CL J2050.7$+$0123 & 312.6814 & $\phantom{-}1.3857$ & 0.333 & DR8\phantom{*} & \phantom{0}7.4 & \phantom{0}4.71 & 1.16 & 0.16 & RXC J2050.7+0123 (7)\\
ACT--CL J2051.1$+$0056 & 312.7935 & $\phantom{-}0.9488$ & 0.333 & S82\phantom{*} & \phantom{0}4.1 & \phantom{0}1.18 & 0.62 & 0.17 & WHL J205111.1+005646 (6)\\
ACT--CL J2051.1$+$0215 & 312.7885 & $\phantom{-}2.2628$ & 0.321 & DR8\phantom{*} & \phantom{0}5.2 & \phantom{0}5.88 & 1.36 & 0.26 & RXC J2051.1+0216 (7)\\
ACT--CL J2055.4$+$0105 & 313.8581 & $\phantom{-}1.0985$ & 0.408 & S82\phantom{*} & \phantom{0}4.9 & \phantom{0}3.53 & 0.77 & 0.16 & WHL J205526.6+010511 (6)\\
ACT--CL J2058.8$+$0123 & 314.7234 & $\phantom{-}1.3836$ & 0.32 $\pm$ 0.02 & DR8\phantom{*} & \phantom{0}8.3 & 10.59 & 1.25 & 0.15 & WHL J205853.1+012411 (1)\\
ACT--CL J2128.4$+$0135 & 322.1036 & $\phantom{-}1.5996$ & 0.385 & DR8\phantom{*} & \phantom{0}7.3 & \phantom{0}5.88 & 1.34 & 0.18 & WHL J212823.4+013536 (1)\\
ACT--CL J2129.6$+$0005 & 322.4186 & $\phantom{-}0.0891$ & 0.234 & S82* & \phantom{0}8.0 & \phantom{0}1.18 & 1.23 & 0.17 & RXC J2129.6+0005 (9)\\
ACT--CL J2130.1$+$0045 & 322.5367 & $\phantom{-}0.7590$ & 0.71 $\pm$ 0.04 & S82\phantom{*} & \phantom{0}4.4 & \phantom{0}4.71 & 0.74 & 0.17 & \\
ACT--CL J2135.1$-$0102 & 323.7907 & $-1.0396$ & 0.33 $\pm$ 0.01 & S82\phantom{*} & \phantom{0}4.1 & \phantom{0}8.24 & 0.68 & 0.17 & WHL J213512.1-010258 (6)\\
ACT--CL J2135.2$+$0125 & 323.8151 & $\phantom{-}1.4247$ & 0.231 & DR8\phantom{*} & \phantom{0}9.3 & \phantom{0}4.71 & 1.47 & 0.16 & Abell 2355 (4)\\
ACT--CL J2135.7$+$0009 & 323.9310 & $\phantom{-}0.1568$ & 0.118 & S82\phantom{*} & \phantom{0}4.0 & 10.59 & 0.68 & 0.17 & Abell 2356 (4)\\
ACT--CL J2152.9$-$0114 & 328.2375 & $-1.2458$ & 0.69 $\pm$ 0.02 & S82\phantom{*} & \phantom{0}4.4 & \phantom{0}3.53 & 0.70 & 0.17 & \\
ACT--CL J2154.5$-$0049 & 328.6319 & $-0.8197$ & 0.488 & S82* & \phantom{0}5.9 & \phantom{0}3.53 & 0.95 & 0.17 & WHL J215432.2-004905 (6)\\
ACT--CL J2156.1$+$0123 & 329.0407 & $\phantom{-}1.3857$ & 0.224 & DR8\phantom{*} & \phantom{0}6.0 & \phantom{0}7.06 & 0.95 & 0.16 & Abell 2397 (4)\\
ACT--CL J2220.7$-$0042 & 335.1922 & $-0.7095$ & 0.57 $\pm$ 0.03 & S82\phantom{*} & \phantom{0}4.0 & \phantom{0}1.18 & 0.63 & 0.18 & GMB11 J335.19871-00.69024 (2)\\
ACT--CL J2229.2$-$0004 & 337.3042 & $-0.0743$ & 0.61 $\pm$ 0.05 & S82\phantom{*} & \phantom{0}4.0 & 15.29 & 0.66 & 0.17 & \\
ACT--CL J2253.3$-$0031 & 343.3432 & $-0.5280$ & 0.54 $\pm$ 0.01 & S82\phantom{*} & \phantom{0}4.0 & \phantom{0}2.35 & 0.64 & 0.17 & \\
ACT--CL J2302.5$+$0002 & 345.6427 & $\phantom{-}0.0419$ & 0.520 & S82\phantom{*} & \phantom{0}4.9 & \phantom{0}4.71 & 0.82 & 0.17 & WHL J230235.1+000234 (6)\\
ACT--CL J2307.6$+$0130 & 346.9176 & $\phantom{-}1.5161$ & 0.36 $\pm$ 0.02 & DR8\phantom{*} & \phantom{0}6.1 & \phantom{0}2.35 & 0.95 & 0.17 & WHL J230739.9+013056 (1)\\
ACT--CL J2327.4$-$0204 & 351.8660 & $-2.0777$ & 0.705 & DR8\phantom{*} & 13.1 & \phantom{0}3.53 & 2.65 & 0.21 & RCS2 J2327.4$-$0204 (10)\\
ACT--CL J2337.6$+$0016 & 354.4156 & $\phantom{-}0.2690$ & 0.275 & S82* & \phantom{0}8.2 & \phantom{0}7.06 & 1.43 & 0.18 & Abell 2631 (4)\\
ACT--CL J2351.7$+$0009 & 357.9349 & $\phantom{-}0.1538$ & 0.99 $\pm$ 0.03 & S82\phantom{*} & \phantom{0}4.7 & \phantom{0}2.35 & 0.89 & 0.21 & 
\enddata
\tablecomments{\mycaption References: (1)~\citet{wen/etal:2012}; (2)~\citet{geach/etal:2011}; (3)~\citet{goto/etal:2002}; (4)~\citet{abell/1958}; (5)~\citet{lopes/etal:2004}; (6)~\citet{wen/etal:2009}; (7)~\citet{bohringer/etal:2000}; (8)~\citet{hao/etal:2010}; (9)~\citet{ebeling/etal:1998}; (10)~\citet{gralla/etal:2011}.}
\label{tab:clusters}\end{deluxetable}

\renewcommand\mytitle{SZ-derived mass estimates for ACT clusters.}
\renewcommand\mycaption{Mass estimates obtained
  as described in Section~\ref{sec:corrected_masses}.  The $\ytilde$-$M$
  scaling relation is fixed to either the UPP result, the fit to the B12
  model, the fit to Nonthermal20 model (superscript ``non''), or the fit to
  dynamical masses of \cite{sifon/etal:inprep} (superscript ``dyn'').  Masses
  and SZ quantities are computed from the cluster's uncorrected central
  Compton parameter $\ytilde$ and redshift measurements, including correction
  for mass function bias.  The cluster scale $\theta_{500}$, value of the bias
  function $Q(m,z)$, and integrated Compton parameter $Y_{500}$ are those
  inferred from the UPP scaling relation.  Uncertainties do not include
  uncertainty in the scaling relation parameters, but do include the effects
  of intrinsic scatter and the measurement uncertainty.  SZ and mass errors
  are highly correlated.}

\begin{deluxetable}{c r@{ $\pm$ }l r@{ $\pm$ }l r@{ $\pm$ }l r@{ $\pm$ }l r@{ $\pm$ }l r@{ $\pm$ }l r@{ $\pm$ }l}\tablecaption{\mytitle}

\tablehead{
ID & \multicolumn{2}{c}{$\theta_{500}$} & \multicolumn{2}{c}{$Q(m,z)$} & \multicolumn{2}{c}{$Y_{500}$} & \multicolumn{2}{c}{$M_{500c}^\text{UPP}$} & \multicolumn{2}{c}{$M_{500c}^\text{B12}$} & \multicolumn{2}{c}{$M_{500c}^\text{non}$} & \multicolumn{2}{c}{$M_{500c}^\text{dyn}$}\\
 & \multicolumn{2}{c}{(arcmin)} & \multicolumn{2}{c}{} & \multicolumn{2}{c}{$(10^{-4}~\mathrm{arcmin}^2)$} & \multicolumn{2}{c}{$(10^{14} \hinv \Msun)$} & \multicolumn{2}{c}{$(10^{14} \hinv \Msun)$} & \multicolumn{2}{c}{$(10^{14} \hinv \Msun)$} & \multicolumn{2}{c}{$(10^{14} \hinv \Msun)$}
}
\startdata
ACT-CL J0008.1$+$0201 & 3.3 & 0.3 & 0.78 & 0.05 & 4.0 & 1.6 & 3.9 & 1.1 & 5.2 & 1.6 & 6.2 & 2.0 & 4.9 & 2.0\\
ACT-CL J0012.0$-$0046 & 1.2 & 0.1 & 0.28 & 0.03 & 1.4 & 0.5 & 3.0 & 0.8 & 3.9 & 1.1 & 3.7 & 1.0 & 3.3 & 1.2\\
ACT-CL J0014.9$-$0057 & 2.8 & 0.1 & 0.69 & 0.02 & 5.0 & 0.9 & 5.7 & 1.1 & 7.6 & 1.4 & 9.0 & 1.7 & 8.2 & 1.9\\
ACT-CL J0017.6$-$0051 & 4.6 & 0.5 & 0.93 & 0.04 & 5.0 & 2.3 & 2.9 & 1.0 & 3.9 & 1.4 & 4.8 & 1.8 & 3.4 & 1.6\\
ACT-CL J0018.2$-$0022 & 1.8 & 0.2 & 0.45 & 0.05 & 1.5 & 0.6 & 3.1 & 0.9 & 4.0 & 1.3 & 4.4 & 1.4 & 3.6 & 1.4\\
ACT-CL J0022.2$-$0036 & 2.1 & 0.1 & 0.52 & 0.02 & 3.8 & 0.6 & 5.5 & 0.9 & 7.3 & 1.2 & 8.0 & 1.4 & 7.7 & 1.6\\
ACT-CL J0026.2$+$0120 & 2.2 & 0.1 & 0.56 & 0.03 & 2.8 & 0.6 & 4.4 & 0.9 & 5.8 & 1.2 & 6.6 & 1.3 & 5.8 & 1.5\\
ACT-CL J0044.4$+$0113 & 1.3 & 0.1 & 0.32 & 0.04 & 1.1 & 0.4 & 2.7 & 0.8 & 3.5 & 1.0 & 3.5 & 1.1 & 3.0 & 1.1\\
ACT-CL J0045.2$-$0152 & 2.8 & 0.1 & 0.68 & 0.02 & 4.8 & 0.9 & 5.6 & 1.1 & 7.5 & 1.4 & 8.8 & 1.6 & 8.0 & 1.9\\
ACT-CL J0051.1$+$0055 & 1.7 & 0.2 & 0.42 & 0.06 & 0.9 & 0.5 & 2.2 & 0.8 & 2.8 & 1.1 & 3.0 & 1.3 & 2.2 & 1.1\\
ACT-CL J0058.0$+$0030 & 1.8 & 0.1 & 0.45 & 0.04 & 1.5 & 0.5 & 3.2 & 0.8 & 4.1 & 1.1 & 4.5 & 1.3 & 3.7 & 1.3\\
ACT-CL J0059.1$-$0049 & 2.1 & 0.1 & 0.53 & 0.02 & 3.5 & 0.6 & 5.2 & 0.9 & 6.9 & 1.2 & 7.6 & 1.3 & 7.2 & 1.5\\
ACT-CL J0104.8$+$0002 & 3.5 & 0.4 & 0.81 & 0.06 & 2.8 & 1.3 & 2.6 & 0.9 & 3.5 & 1.2 & 4.2 & 1.5 & 3.0 & 1.4\\
ACT-CL J0119.9$+$0055 & 1.9 & 0.1 & 0.47 & 0.04 & 1.7 & 0.5 & 3.3 & 0.8 & 4.3 & 1.1 & 4.7 & 1.3 & 3.9 & 1.3\\
ACT-CL J0127.2$+$0020 & 3.0 & 0.2 & 0.73 & 0.04 & 2.8 & 0.9 & 3.3 & 0.9 & 4.4 & 1.2 & 5.3 & 1.4 & 4.1 & 1.4\\
ACT-CL J0139.3$-$0128 & 1.6 & 0.2 & 0.40 & 0.06 & 0.8 & 0.5 & 2.1 & 0.9 & 2.6 & 1.1 & 2.8 & 1.2 & 2.1 & 1.1\\
ACT-CL J0152.7$+$0100 & 5.4 & 0.2 & 0.98 & 0.01 & 13.0 & 2.3 & 5.7 & 1.1 & 7.9 & 1.6 & 9.7 & 1.9 & 8.7 & 2.3\\
ACT-CL J0156.4$-$0123 & 2.6 & 0.2 & 0.64 & 0.05 & 2.1 & 0.8 & 3.1 & 0.9 & 4.0 & 1.2 & 4.7 & 1.4 & 3.6 & 1.4\\
ACT-CL J0206.2$-$0114 & 2.2 & 0.1 & 0.54 & 0.02 & 2.6 & 0.6 & 4.3 & 0.8 & 5.7 & 1.1 & 6.4 & 1.2 & 5.7 & 1.4\\
ACT-CL J0215.4$+$0030 & 1.9 & 0.1 & 0.47 & 0.03 & 1.8 & 0.5 & 3.5 & 0.8 & 4.5 & 1.1 & 5.0 & 1.2 & 4.2 & 1.3\\
ACT-CL J0218.2$-$0041 & 2.1 & 0.1 & 0.54 & 0.03 & 2.2 & 0.6 & 3.8 & 0.8 & 4.9 & 1.1 & 5.6 & 1.3 & 4.7 & 1.3\\
ACT-CL J0219.8$+$0022 & 2.2 & 0.2 & 0.57 & 0.05 & 1.7 & 0.7 & 3.0 & 0.9 & 3.9 & 1.2 & 4.5 & 1.4 & 3.5 & 1.4\\
ACT-CL J0219.9$+$0129 & 3.0 & 0.3 & 0.73 & 0.05 & 2.3 & 0.9 & 2.8 & 0.8 & 3.7 & 1.1 & 4.5 & 1.4 & 3.3 & 1.3\\
ACT-CL J0221.5$-$0012 & 1.6 & 0.2 & 0.40 & 0.06 & 0.5 & 0.3 & 1.4 & 0.6 & 1.8 & 0.8 & 2.0 & 0.9 & 1.4 & 0.7\\
ACT-CL J0223.1$-$0056 & 2.1 & 0.1 & 0.53 & 0.03 & 2.2 & 0.6 & 3.8 & 0.8 & 5.0 & 1.1 & 5.7 & 1.3 & 4.8 & 1.4\\
ACT-CL J0228.5$+$0030 & 1.7 & 0.2 & 0.42 & 0.06 & 1.0 & 0.5 & 2.4 & 0.8 & 3.0 & 1.1 & 3.3 & 1.3 & 2.5 & 1.2\\
ACT-CL J0230.9$-$0024 & 2.5 & 0.3 & 0.64 & 0.06 & 1.9 & 0.8 & 2.8 & 0.9 & 3.7 & 1.2 & 4.3 & 1.4 & 3.2 & 1.4\\
ACT-CL J0239.8$-$0134 & 3.9 & 0.1 & 0.85 & 0.01 & 9.4 & 1.6 & 6.7 & 1.3 & 9.1 & 1.7 & 11.1 & 2.0 & 10.3 & 2.5\\
ACT-CL J0240.0$+$0116 & 2.1 & 0.2 & 0.53 & 0.04 & 1.8 & 0.6 & 3.3 & 0.8 & 4.3 & 1.1 & 4.8 & 1.3 & 3.9 & 1.3\\
ACT-CL J0241.2$-$0018 & 1.8 & 0.2 & 0.44 & 0.06 & 1.1 & 0.5 & 2.5 & 0.9 & 3.2 & 1.2 & 3.6 & 1.3 & 2.7 & 1.2\\
ACT-CL J0245.8$-$0042 & 5.0 & 0.6 & 0.95 & 0.04 & 4.9 & 2.4 & 2.5 & 0.9 & 3.4 & 1.3 & 4.1 & 1.6 & 2.8 & 1.4\\
ACT-CL J0250.1$+$0008 & 1.7 & 0.2 & 0.41 & 0.05 & 1.2 & 0.5 & 2.7 & 0.8 & 3.5 & 1.1 & 3.7 & 1.2 & 3.0 & 1.2\\
ACT-CL J0256.5$+$0006 & 3.1 & 0.2 & 0.76 & 0.03 & 3.4 & 1.0 & 3.8 & 0.9 & 5.0 & 1.2 & 6.1 & 1.4 & 4.8 & 1.5\\
ACT-CL J0301.1$-$0110 & 2.0 & 0.3 & 0.51 & 0.07 & 1.1 & 0.6 & 2.2 & 0.8 & 2.8 & 1.1 & 3.2 & 1.3 & 2.3 & 1.2\\
ACT-CL J0301.6$+$0155 & 6.7 & 0.4 & 1.01 & 0.01 & 15.7 & 5.0 & 4.7 & 1.2 & 6.6 & 1.8 & 8.1 & 2.1 & 6.7 & 2.4\\
ACT-CL J0303.3$+$0155 & 6.9 & 0.6 & 1.01 & 0.01 & 14.9 & 5.5 & 4.2 & 1.2 & 5.8 & 1.8 & 7.1 & 2.1 & 5.6 & 2.3\\
ACT-CL J0308.1$+$0103 & 1.9 & 0.2 & 0.48 & 0.05 & 1.3 & 0.6 & 2.7 & 0.8 & 3.4 & 1.1 & 3.8 & 1.3 & 3.0 & 1.3\\
ACT-CL J0320.4$+$0032 & 3.0 & 0.2 & 0.72 & 0.04 & 2.8 & 0.9 & 3.3 & 0.9 & 4.4 & 1.2 & 5.3 & 1.4 & 4.1 & 1.4\\
ACT-CL J0326.8$-$0043 & 3.1 & 0.1 & 0.75 & 0.02 & 5.5 & 0.9 & 5.5 & 1.0 & 7.4 & 1.4 & 8.9 & 1.6 & 7.9 & 1.9\\
ACT-CL J0336.9$-$0110 & 1.2 & 0.1 & 0.27 & 0.03 & 1.0 & 0.4 & 2.5 & 0.7 & 3.2 & 0.9 & 3.1 & 0.9 & 2.7 & 1.0\\
ACT-CL J0342.0$+$0105 & 1.5 & 0.1 & 0.37 & 0.02 & 1.8 & 0.4 & 3.7 & 0.7 & 4.8 & 1.0 & 4.8 & 1.0 & 4.5 & 1.1\\
ACT-CL J0342.7$-$0017 & 3.4 & 0.3 & 0.80 & 0.04 & 3.2 & 1.2 & 3.1 & 0.9 & 4.1 & 1.2 & 5.0 & 1.5 & 3.7 & 1.4\\
ACT-CL J0348.6$+$0029 & 3.5 & 0.3 & 0.81 & 0.04 & 3.3 & 1.2 & 3.1 & 0.9 & 4.1 & 1.2 & 5.0 & 1.5 & 3.7 & 1.4\\
ACT-CL J0348.6$-$0028 & 3.1 & 0.3 & 0.75 & 0.05 & 2.6 & 1.0 & 3.0 & 0.9 & 4.0 & 1.2 & 4.8 & 1.5 & 3.5 & 1.4\\
ACT-CL J2025.2$+$0030 & 3.6 & 0.2 & 0.83 & 0.02 & 5.5 & 1.3 & 4.6 & 1.0 & 6.3 & 1.4 & 7.7 & 1.6 & 6.4 & 1.8\\
ACT-CL J2050.5$-$0055 & 2.2 & 0.1 & 0.56 & 0.04 & 2.2 & 0.7 & 3.8 & 0.9 & 4.9 & 1.2 & 5.6 & 1.4 & 4.7 & 1.4\\
ACT-CL J2050.7$+$0123 & 3.9 & 0.2 & 0.85 & 0.02 & 6.7 & 1.4 & 5.1 & 1.0 & 7.0 & 1.4 & 8.5 & 1.7 & 7.4 & 1.9\\
ACT-CL J2051.1$+$0056 & 3.0 & 0.4 & 0.73 & 0.07 & 2.1 & 1.1 & 2.5 & 0.9 & 3.3 & 1.3 & 4.0 & 1.6 & 2.8 & 1.4\\
ACT-CL J2051.1$+$0215 & 4.0 & 0.3 & 0.87 & 0.03 & 7.6 & 2.5 & 5.3 & 1.4 & 7.1 & 1.9 & 8.6 & 2.4 & 7.2 & 2.6\\
ACT-CL J2055.4$+$0105 & 2.9 & 0.2 & 0.70 & 0.04 & 2.8 & 1.0 & 3.5 & 0.9 & 4.6 & 1.3 & 5.5 & 1.5 & 4.3 & 1.5\\
ACT-CL J2058.8$+$0123 & 4.1 & 0.2 & 0.88 & 0.02 & 7.8 & 1.4 & 5.5 & 1.1 & 7.5 & 1.5 & 9.2 & 1.8 & 8.0 & 2.0\\
ACT-CL J2128.4$+$0135 & 3.5 & 0.1 & 0.81 & 0.02 & 6.8 & 1.4 & 5.7 & 1.1 & 7.7 & 1.6 & 9.4 & 1.8 & 8.3 & 2.1\\
ACT-CL J2129.6$+$0005 & 5.2 & 0.2 & 0.97 & 0.01 & 11.4 & 2.4 & 5.3 & 1.1 & 7.3 & 1.6 & 9.1 & 1.9 & 7.9 & 2.2\\
ACT-CL J2130.1$+$0045 & 1.9 & 0.2 & 0.47 & 0.05 & 1.6 & 0.6 & 3.2 & 0.9 & 4.1 & 1.2 & 4.5 & 1.4 & 3.7 & 1.4\\
ACT-CL J2135.1$-$0102 & 3.1 & 0.4 & 0.75 & 0.06 & 2.6 & 1.2 & 2.8 & 1.0 & 3.7 & 1.3 & 4.5 & 1.6 & 3.2 & 1.5\\
ACT-CL J2135.2$+$0125 & 5.3 & 0.2 & 0.97 & 0.01 & 14.2 & 2.4 & 6.3 & 1.2 & 8.7 & 1.7 & 10.8 & 2.1 & 9.9 & 2.6\\
ACT-CL J2135.7$+$0009 & 7.3 & 1.0 & 1.01 & 0.01 & 10.8 & 6.3 & 2.6 & 1.1 & 3.6 & 1.6 & 4.3 & 1.9 & 2.9 & 1.8\\
ACT-CL J2152.9$-$0114 & 1.9 & 0.2 & 0.47 & 0.05 & 1.5 & 0.6 & 3.0 & 0.9 & 3.9 & 1.3 & 4.2 & 1.4 & 3.4 & 1.4\\
ACT-CL J2154.5$-$0049 & 2.7 & 0.2 & 0.68 & 0.03 & 3.4 & 0.9 & 4.3 & 0.9 & 5.7 & 1.3 & 6.8 & 1.5 & 5.7 & 1.6\\
ACT-CL J2156.1$+$0123 & 4.9 & 0.3 & 0.95 & 0.02 & 7.9 & 2.2 & 4.1 & 1.0 & 5.7 & 1.4 & 7.0 & 1.7 & 5.7 & 1.8\\
ACT-CL J2220.7$-$0042 & 2.0 & 0.3 & 0.50 & 0.07 & 1.2 & 0.7 & 2.5 & 1.0 & 3.1 & 1.3 & 3.5 & 1.5 & 2.6 & 1.4\\
ACT-CL J2229.2$-$0004 & 2.0 & 0.2 & 0.49 & 0.06 & 1.3 & 0.7 & 2.7 & 1.0 & 3.4 & 1.3 & 3.8 & 1.5 & 2.9 & 1.4\\
ACT-CL J2253.3$-$0031 & 2.1 & 0.3 & 0.54 & 0.06 & 1.5 & 0.7 & 2.7 & 0.9 & 3.5 & 1.3 & 4.0 & 1.5 & 3.0 & 1.4\\
ACT-CL J2302.5$+$0002 & 2.5 & 0.2 & 0.62 & 0.04 & 2.5 & 0.8 & 3.7 & 0.9 & 4.8 & 1.3 & 5.6 & 1.5 & 4.6 & 1.5\\
ACT-CL J2307.6$+$0130 & 3.4 & 0.2 & 0.80 & 0.03 & 4.5 & 1.2 & 4.2 & 1.0 & 5.7 & 1.3 & 6.9 & 1.6 & 5.7 & 1.7\\
ACT-CL J2327.4$-$0204 & 2.8 & 0.1 & 0.67 & 0.01 & 10.1 & 1.0 & 9.4 & 1.5 & 12.5 & 2.0 & 14.3 & 2.2 & 14.9 & 3.0\\
ACT-CL J2337.6$+$0016 & 4.8 & 0.2 & 0.94 & 0.01 & 11.5 & 2.2 & 6.1 & 1.2 & 8.4 & 1.7 & 10.3 & 2.0 & 9.3 & 2.5\\
ACT-CL J2351.7$+$0009 & 1.5 & 0.2 & 0.36 & 0.05 & 1.5 & 0.7 & 3.2 & 1.0 & 4.0 & 1.4 & 4.1 & 1.5 & 3.4 & 1.6
\enddata
\tablecomments{\mycaption}
\label{tab:masses}\end{deluxetable}

\clearpage

\appendix
\section{Analysis of ACT Southern clusters}

In this appendix we apply the methodology of Sections~\ref{sec:fixed_scale}
and \ref{sec:corrected_masses} to estimate masses and SZ quantities for the 23
clusters from the original ACT Southern cluster sample, described in
\citet[hereafter M11]{marriage/etal:2011} and \cite{menanteau/etal:2010}.  For
this analysis, we use updated maps that include additional data acquired in
this field during the 2009 and 2010 observing seasons.

In Table~\ref{tab:clusters_south} we summarize the basic properties of the
Southern cluster sample, and include our measurement of the uncorrected
central temperature decrement taken from maps filtered with the $\thetaD =
\thetaFixed$ UPP matched filter.  While this filter incorporates the same
cluster profile that was used in the analysis of the Equatorial sample, the
filter itself is slightly different due to the different noise properties in
the Southern maps.  In Table~\ref{tab:masses_south} we present cluster mass
and SZ quantity estimates.

There is substantial overlap between the ACT Southern field and the cluster
samples presented by the SPT collaboration.  The M11 sample includes six of
the clusters analyzed in \citet[hereafter W11]{williamson/etal:2011}, and five
of the clusters in \citet[hereafter R12]{reichardt/etal:2013}.  In W11 a
scaling relation based on the SPT signal to noise ratio and calibrated to the
models of \citet{shaw/etal:2010} is used to obtain SZ based mass estimates.
In R12, masses are presented that are derived from a combination of SZ and
X-ray measurements, though X-ray measurements are expected to dominate because
of the smaller uncertainty in the X-ray scaling relation.  (Four of the five
ACT clusters appearing in R12 have X-ray mass measurements.)

\begin{figure}
\begin{center}
\includegraphics[width=8.5cm]{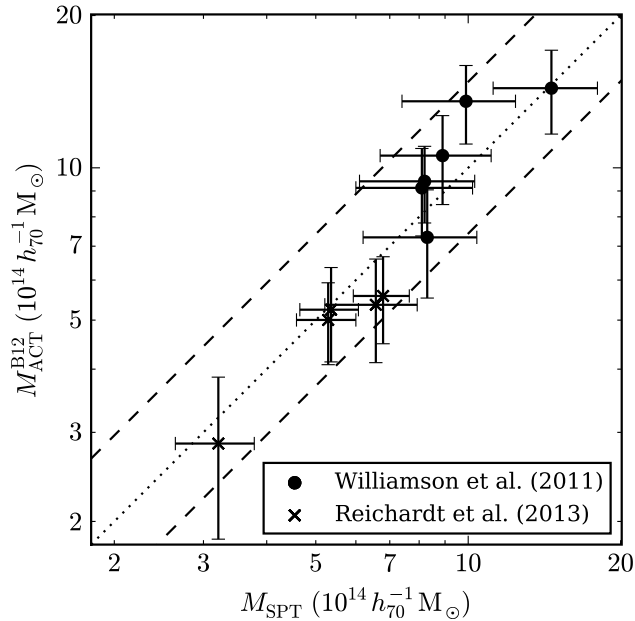}
\caption{Comparison of ACT SZ based masses (from B12 scaling relation) to
  masses from the South Pole Telescope.  Both are $M_{500c}$.  ACT masses are
  computed using the B12 scaling relation and may be found in
  Table~\ref{tab:masses_south}.  SPT masses are taken from
  \citet{williamson/etal:2011} (mass is computed from SZ signal based on
  scaling relation calibrated to \citet{shaw/etal:2010} models; only
  statistical uncertainty is included in error bars), and
  \citet{reichardt/etal:2013} (mass is computed from combination of $Y_{\rm
    X}$ and SZ measurements).  The weighted mean mass ratio for the 11
  clusters is $0.99 \pm 0.06$, though there is evidence of a systematic
  difference between the mass calibration in the two SPT catalogs.  Dotted
  line traces equality of SPT and ACT masses for the B12 scaling relation.
  Dashed lines trace approximate loci of agreement between SPT masses and ACT
  masses based on the UPP (upper line) and Nonthermal20 (lower line) scaling
  relations.  Uncertainties in the ACT and SPT measurements for each cluster
  are likely to be partially correlated, since both use properties of cluster
  gas to infer a total mass.}
\label{fig:south_spt}
\end{center}
\end{figure}

We find the SPT masses to be in good agreement with our masses based on the
B12 model scaling relation parameters, with weighted mean mass ratio
$M^{\rm{B12}}/M^{\rm{SPT}} = 0.99 \pm 0.06$.  We note, however, a difference
in the mean ratio for each of the two SPT catalogs.  For the clusters in R12
the mass ratio is $0.88 \pm 0.09$, while for the higher mass, lower redshift
sample of W11 the ratio is $1.10 \pm 0.09$.  In Figure~\ref{fig:south_spt} we
show the masses.  We note that this difference in mass ratio is consistent
with the observation in W11 that their average SZ determined masses are
smaller than $Y_{\rm X}$ based masses by a factor of $0.78 \pm 0.06$.  In
contrast, the R12 masses for four of the five common clusters include input
from observations of $Y_{\rm X}$, which is likely to dominate over the
contribution from SZ data.

\renewcommand\mytitle{Uncorrected central Compton parameters for the clusters
  from \cite{marriage/etal:2011}.}

\renewcommand\mycaption{Clusters used for cosmological and scaling relation
  constraints in Section~\ref{sec:cosmology} are marked with an asterisk.
  Redshifts are all spectroscopic, obtained from literature as described in
  \cite{menanteau/etal:2010} or as presented in \cite{sifon/etal:inprep}.  We
  provide the S/N of detection as presented in \cite{marriage/etal:2011}, for
  reference.  Uncorrected central Compton parameters $\ytilde$ are obtained
  from the application of a matched filter with $\thetaD = \thetaFixed$ (see
  Section~\ref{sec:fixed_scale}). Alternate IDs are provided for clusters
  detected before the initial release of the Southern sample, and for any
  cluster appearing in current SPT results. }

\begin{deluxetable}{c c c c r@{.}l r@{ $\pm$ }l l}\tablecaption{\mytitle}

\tablehead{
ACT ID & R.A. & Dec. & Redshift & \multicolumn{2}{c}{$S/N$} & \multicolumn{2}{c}{$\ytilde$} & Alternate ID (ref.)\\
 & $(^\circ)$ & $(^\circ)$ &  & \multicolumn{2}{c}{} & \multicolumn{2}{c}{$(10^{-4})$} & 
}
\startdata
ACT--CL J0102$-$4915* & \phantom{0}15.7208 & $-49.2553$ & 0.870 & 9&0 & 3.51 & 0.43 & SPT-CL J0102-4915 (1)\\
ACT--CL J0145$-$5301\phantom{*} & \phantom{0}26.2458 & $-53.0169$ & 0.118 & 4&0 & 0.86 & 0.19 & Abell 2941 (2)\\
ACT--CL J0215$-$5212\phantom{*} & \phantom{0}33.8250 & $-52.2083$ & 0.480 & 4&9 & 0.79 & 0.18 & \\
ACT--CL J0217$-$5245\phantom{*} & \phantom{0}34.2958 & $-52.7556$ & 0.343 & 4&1 & 0.79 & 0.18 & RXC J0217.2-5244 (3)\\
ACT--CL J0232$-$5257\phantom{*} & \phantom{0}38.1875 & $-52.9522$ & 0.556 & 4&7 & 0.61 & 0.17 & \\
ACT--CL J0235$-$5121\phantom{*} & \phantom{0}38.9667 & $-51.3544$ & 0.278 & 6&2 & 0.98 & 0.19 & \\
ACT--CL J0237$-$4939\phantom{*} & \phantom{0}39.2625 & $-49.6575$ & 0.334 & 3&9 & 0.93 & 0.26 & \\
ACT--CL J0245$-$5302\phantom{*} & \phantom{0}41.3875 & $-53.0344$ & 0.300 & 9&1 & 1.57 & 0.17 & Abell S0295 (4)\\
  &   &   &   & \multicolumn{2}{c}{ } & \multicolumn{2}{c}{ } & SPT-CL J0245-5302 (1)\\
ACT--CL J0304$-$4921\phantom{*} & \phantom{0}46.0625 & $-49.3617$ & 0.392 & 3&9 & 1.52 & 0.32 & \\
ACT--CL J0330$-$5227* & \phantom{0}52.7250 & $-52.4678$ & 0.442 & 6&1 & 1.23 & 0.18 & Abell 3128 NE (5)\\
ACT--CL J0346$-$5438\phantom{*} & \phantom{0}56.7125 & $-54.6483$ & 0.530 & 4&4 & 1.05 & 0.22 & \\
ACT--CL J0438$-$5419* & \phantom{0}69.5792 & $-54.3181$ & 0.421 & 8&0 & 1.62 & 0.13 & SPT-CL J0438-5419 (1)\\
ACT--CL J0509$-$5341* & \phantom{0}77.3375 & $-53.7014$ & 0.461 & 4&8 & 0.82 & 0.14 & SPT-CL J0509-5342 (6)\\
ACT--CL J0516$-$5430\phantom{*} & \phantom{0}79.1250 & $-54.5083$ & 0.294 & 4&6 & 0.87 & 0.15 & Abell S0520 (4)\\
  &   &   &   & \multicolumn{2}{c}{ } & \multicolumn{2}{c}{ } & SPT-CL J0516-5430 (6)\\
ACT--CL J0528$-$5259\phantom{*} & \phantom{0}82.0125 & $-52.9981$ & 0.768 & 3&1 & 0.50 & 0.13 & SPT-CL J0528-5300 (6)\\
ACT--CL J0546$-$5345* & \phantom{0}86.6542 & $-53.7589$ & 1.066 & 6&5 & 0.92 & 0.14 & SPT-CL J0546-5345 (6)\\
ACT--CL J0559$-$5249* & \phantom{0}89.9292 & $-52.8203$ & 0.609 & 5&1 & 0.89 & 0.14 & SPT-CL J0559-5249 (6)\\
ACT--CL J0616$-$5227* & \phantom{0}94.1500 & $-52.4597$ & 0.684 & 5&9 & 1.00 & 0.15 & \\
ACT--CL J0638$-$5358\phantom{*} & \phantom{0}99.6917 & $-53.9792$ & 0.222 & 10&0 & 1.77 & 0.15 & Abell S0592 (4)\\
  &   &   &   & \multicolumn{2}{c}{ } & \multicolumn{2}{c}{ } & SPT-CL J0638-5358 (1)\\
ACT--CL J0641$-$4949\phantom{*} & 100.3958 & $-49.8089$ & 0.146 & 4&7 & 0.58 & 0.26 & Abell 3402 (2)\\
ACT--CL J0645$-$5413\phantom{*} & 101.3750 & $-54.2275$ & 0.167 & 7&1 & 1.19 & 0.17 & Abell 3404 (2)\\
  &   &   &   & \multicolumn{2}{c}{ } & \multicolumn{2}{c}{ } & SPT-CL J0645-5413 (1)\\
ACT--CL J0658$-$5557\phantom{*} & 104.6250 & $-55.9511$ & 0.296 & 11&5 & 2.65 & 0.21 & 1E0657-56/Bullet (7)\\
  &   &   &   & \multicolumn{2}{c}{ } & \multicolumn{2}{c}{ } & SPT-CL J0658-5556 (1)\\
ACT--CL J0707$-$5522\phantom{*} & 106.8042 & $-55.3800$ & 0.296 & 3&3 & 0.54 & 0.21 & 
\enddata
\tablecomments{\mycaption References: (1)~\citet{williamson/etal:2011}; (2)~\citet{abell/1958}; (3)~\citet{bohringer/etal:2004}; (4)~\citet{abell/corwin/olowin:1989}; (5)~\citet{werner/etal:2007}; (6)~\citet{reichardt/etal:2013}; (7)~\citet{tucker/etal:1995}.}
\label{tab:clusters_south}\end{deluxetable}

\renewcommand\mytitle{SZ-derived mass estimates for the ACT Southern cluster
  sample of \cite{marriage/etal:2011}.}

\renewcommand\mycaption{Columns are as described in Table~\ref{tab:masses}.}

\begin{deluxetable}{c r@{ $\pm$ }l r@{ $\pm$ }l r@{ $\pm$ }l r@{ $\pm$ }l r@{ $\pm$ }l r@{ $\pm$ }l r@{ $\pm$ }l}\tablecaption{\mytitle}

\tablehead{
ID & \multicolumn{2}{c}{$\theta_{500}$} & \multicolumn{2}{c}{$Q(m,z)$} & \multicolumn{2}{c}{$Y_{500}$} & \multicolumn{2}{c}{$M_{500c}^\text{UPP}$} & \multicolumn{2}{c}{$M_{500c}^\text{B12}$} & \multicolumn{2}{c}{$M_{500c}^\text{non}$} & \multicolumn{2}{c}{$M_{500c}^\text{dyn}$}\\
 & \multicolumn{2}{c}{(arcmin)} & \multicolumn{2}{c}{} & \multicolumn{2}{c}{$(10^{-4}~\mathrm{arcmin}^2)$} & \multicolumn{2}{c}{$(10^{14} h_{70}^{-1} M_\odot)$} & \multicolumn{2}{c}{$(10^{14} h_{70}^{-1} M_\odot)$} & \multicolumn{2}{c}{$(10^{14} h_{70}^{-1} M_\odot)$} & \multicolumn{2}{c}{$(10^{14} h_{70}^{-1} M_\odot)$}
}
\startdata
ACT-CL J0102$-$4915 & 2.5 & 0.1 & 0.60 & 0.02 & 11.5 & 2.0 & 10.5 & 1.8 & 13.8 & 2.4 & 15.0 & 2.6 & 16.0 & 3.5\\
ACT-CL J0145$-$5301 & 8.1 & 0.9 & 1.02 & 0.01 & 16.9 & 8.2 & 3.5 & 1.2 & 4.8 & 1.8 & 5.7 & 2.2 & 4.2 & 2.2\\
ACT-CL J0215$-$5212 & 2.5 & 0.2 & 0.62 & 0.05 & 2.4 & 1.0 & 3.5 & 1.0 & 4.5 & 1.4 & 5.2 & 1.7 & 4.1 & 1.7\\
ACT-CL J0217$-$5245 & 3.2 & 0.3 & 0.75 & 0.05 & 3.2 & 1.3 & 3.4 & 1.0 & 4.4 & 1.4 & 5.3 & 1.8 & 4.0 & 1.7\\
ACT-CL J0232$-$5257 & 2.0 & 0.3 & 0.49 & 0.07 & 1.2 & 0.7 & 2.5 & 1.0 & 3.1 & 1.3 & 3.5 & 1.5 & 2.6 & 1.4\\
ACT-CL J0235$-$5121 & 4.1 & 0.3 & 0.87 & 0.03 & 5.9 & 2.0 & 4.1 & 1.1 & 5.6 & 1.5 & 6.8 & 1.8 & 5.4 & 1.9\\
ACT-CL J0237$-$4939 & 3.2 & 0.5 & 0.74 & 0.09 & 3.1 & 2.1 & 3.1 & 1.4 & 3.9 & 1.9 & 4.6 & 2.4 & 3.2 & 2.1\\
ACT-CL J0245$-$5302 & 4.6 & 0.2 & 0.92 & 0.01 & 12.1 & 2.0 & 6.7 & 1.3 & 9.2 & 1.8 & 11.3 & 2.1 & 10.4 & 2.6\\
ACT-CL J0304$-$4921 & 3.5 & 0.3 & 0.79 & 0.05 & 6.8 & 2.6 & 5.7 & 1.6 & 7.4 & 2.3 & 8.7 & 2.8 & 7.3 & 3.1\\
ACT-CL J0330$-$5227 & 3.2 & 0.1 & 0.74 & 0.02 & 5.4 & 1.2 & 5.4 & 1.1 & 7.3 & 1.5 & 8.7 & 1.7 & 7.7 & 2.0\\
ACT-CL J0346$-$5438 & 2.6 & 0.2 & 0.62 & 0.04 & 3.3 & 1.1 & 4.4 & 1.2 & 5.7 & 1.6 & 6.6 & 1.9 & 5.5 & 2.0\\
ACT-CL J0438$-$5419 & 3.6 & 0.1 & 0.80 & 0.01 & 8.8 & 1.0 & 7.0 & 1.2 & 9.5 & 1.6 & 11.5 & 1.9 & 10.9 & 2.3\\
ACT-CL J0509$-$5341 & 2.8 & 0.1 & 0.66 & 0.03 & 3.1 & 0.7 & 4.0 & 0.8 & 5.3 & 1.1 & 6.3 & 1.3 & 5.2 & 1.4\\
ACT-CL J0516$-$5430 & 3.9 & 0.2 & 0.85 & 0.03 & 5.1 & 1.3 & 4.0 & 0.9 & 5.4 & 1.2 & 6.7 & 1.5 & 5.3 & 1.6\\
ACT-CL J0528$-$5259 & 1.6 & 0.2 & 0.37 & 0.05 & 0.9 & 0.4 & 2.3 & 0.8 & 2.9 & 1.0 & 3.1 & 1.1 & 2.4 & 1.1\\
ACT-CL J0546$-$5345 & 1.6 & 0.1 & 0.36 & 0.02 & 2.0 & 0.4 & 3.9 & 0.7 & 5.1 & 0.9 & 5.2 & 1.0 & 4.9 & 1.1\\
ACT-CL J0559$-$5249 & 2.3 & 0.1 & 0.56 & 0.02 & 2.8 & 0.6 & 4.3 & 0.8 & 5.7 & 1.1 & 6.5 & 1.3 & 5.6 & 1.4\\
ACT-CL J0616$-$5227 & 2.2 & 0.1 & 0.53 & 0.02 & 3.0 & 0.6 & 4.6 & 0.9 & 6.1 & 1.2 & 6.9 & 1.3 & 6.2 & 1.5\\
ACT-CL J0638$-$5358 & 6.1 & 0.2 & 1.00 & 0.01 & 22.4 & 3.0 & 7.5 & 1.4 & 10.5 & 2.0 & 13.0 & 2.4 & 12.5 & 3.2\\
ACT-CL J0641$-$4949 & 4.8 & 1.0 & 0.93 & 0.06 & 2.9 & 3.2 & 1.4 & 0.9 & 1.8 & 1.2 & 2.2 & 1.4 & 1.3 & 1.1\\
ACT-CL J0645$-$5413 & 6.9 & 0.3 & 1.02 & 0.01 & 18.2 & 4.3 & 5.1 & 1.2 & 7.3 & 1.7 & 8.9 & 2.0 & 7.7 & 2.4\\
ACT-CL J0658$-$5557 & 5.4 & 0.1 & 0.97 & 0.01 & 26.6 & 3.2 & 10.3 & 1.9 & 14.3 & 2.6 & 17.6 & 3.1 & 18.2 & 4.3\\
ACT-CL J0707$-$5522 & 2.8 & 0.5 & 0.69 & 0.09 & 1.4 & 1.1 & 1.7 & 0.9 & 2.2 & 1.2 & 2.6 & 1.4 & 1.7 & 1.1
\enddata
\tablecomments{\mycaption}
\label{tab:masses_south}\end{deluxetable}

\end{document}